\documentclass[]{aa}
\usepackage{natbib,twoopt}
\usepackage{amsmath}

\usepackage{color}
\usepackage[breaklinks=true]{hyperref} 
\bibpunct{(}{)}{;}{a}{}{,}             
\makeatletter
  \newcommandtwoopt{\citeads}[3][][]{\href{http://adsabs.harvard.edu/abs/#3}%
    {\def\hyper@linkstart##1##2{}%
     \let\hyper@linkend\@empty\citealp[#1][#2]{#3}}}
  \newcommandtwoopt{\citepads}[3][][]{\href{http://adsabs.harvard.edu/abs/#3}%
    {\def\hyper@linkstart##1##2{}%
     \let\hyper@linkend\@empty\citep[#1][#2]{#3}}}
  \newcommandtwoopt{\citetads}[3][][]{\href{http://adsabs.harvard.edu/abs/#3}%
    {\def\hyper@linkstart##1##2{}%
     \let\hyper@linkend\@empty\citet[#1][#2]{#3}}}
  \newcommandtwoopt{\citeyearads}[3][][]%
    {\href{http://adsabs.harvard.edu/abs/#3}
    {\def\hyper@linkstart##1##2{}%
     \let\hyper@linkend\@empty\citeyear[#1][#2]{#3}}}
\makeatother

\usepackage{graphicx}
\usepackage{txfonts}
\begin{document}

   \title{Probing star formation and ISM properties using galaxy disk inclination III}
   \subtitle{Evolution in dust opacity and clumpiness between redshift 0.0 < z< 0.7 constrained from UV to NIR}
   
\authorrunning{S.A. van der Giessen et al.}

   \author{S.A. van der Giessen
          \inst{1,2,3}
          \and S. K. Leslie \inst{1}
		  \and B. Groves \inst{4,5}
		  \and J. A. Hodge \inst{1}
		  \and C. C. Popescu \inst{6}
         \and M. T. Sargent \inst{7, 8}
         \and E. Schinnerer \inst{9}
		 \and R. J. Tuffs \inst{10}
          }
\institute{Leiden Observatory, Leiden University, PO Box 9513, NL-2300 RA Leiden, the Netherlands
\and Sterrenkundig Observatorium, Ghent University, Krijgslaan 281 - S9, 9000 Gent, Belgium
\and Dept. Fisica Teorica y del Cosmos, Universidad de Granada, Spain
\and International Centre for Radio Astronomy Research, The University of Western Australia, 7 Fairway, Crawley, WA 6009, Australia
\and Research School of Astronomy and Astrophysics, Australian National University, Mt Stromlo Observatory, Weston Creek, ACT 2611, Australia
\and Jeremiah Horrocks Institute, University of Central Lancashire, Preston PR1 2HE
\and Astronomy Centre, Department of Physics and Astronomy, University of Sussex, Brighton BN1 9QH, UK
\and International Space Science Institute (ISSI), Hallerstrasse 6, CH-3012 Bern, Switzerland
\and Max-Planck-Institut f\"{u}r Astronomie, K\"{o}nigstuhl 17, D-69117, Heidelberg, Germany
\and Max-Plank Institute for Nuclear Physics (MPIK), Saupfercheckweg 1, 69117 Heidelberg, Germany\\
\email{stefan.stefananthonyvandergiessen@ugent.be, leslie@strw.leidenuniv.nl}
             }

   \date{Version 4 November 2021}

\abstract{ 
   Attenuation by dust severely impacts our ability to obtain unbiased observations of galaxies, especially as the amount and wavelength dependence of the attenuation varies with the stellar mass $M_{*}$, inclination $i$, and other galaxy properties. In this study, we used the attenuation - inclination models in ultraviolet (UV), optical, and near-infrared (NIR) bands designed by Tuffs and collaborators to investigate the average global dust properties in galaxies as a function of $M_{*}$, the stellar mass surface density $\mu_{*}$, the star-formation rate $SFR$, the specific star-formation rate $sSFR$, the star-formation main-sequence offset $dMS$, and the star-formation rate surface density $\Sigma_{SFR}$ at redshifts $z \sim 0$ and $z \sim 0.7$. We used star-forming galaxies from the Sloan Digital Sky Survey (SDSS; $\sim$ 20000) and Galaxy And Mass Assembly (GAMA; $\sim$ 2000) to form our low-z sample at $0.04 < z < 0.1$ and star-forming galaxies from Cosmological Evolution Survey (COSMOS; $\sim$ 2000) for the sample at $0.6 <z < 0.8$. We found that galaxies at $z \sim 0.7$ have a higher optical depth $\tau_{B}^{f}$ and clumpiness $F$ than galaxies at $z \sim 0$. The increase in $F$ hints that the stars of $z \sim 0.7$ galaxies are less likely to escape their birth cloud, which might indicate that the birth clouds are larger. We also found that $\tau_{B}^{f}$ increases with $M_{*}$ and $\mu_{*}$, independent of the sample and therefore redshift. We found no clear trends in $\tau_{B}^{f}$ or $F$ with the $SFR$, which could imply that the dust mass distribution is independent of the $SFR$. In turn, this would imply that the balance of dust formation and destruction is independent of the $SFR$. Based on an analysis of the inclination dependence of the Balmer decrement, we found that reproducing the Balmer line emission requires not only a completely optically thick dust component associated with star-forming regions, as in the standard model, but an extra component of an optically thin dust within the birth clouds.
This new component implies the existence of dust inside HII regions that attenuates the Balmer emission before it escapes through gaps in the birth cloud and we found it is more important in high-mass galaxies. These results will inform our understanding of dust formation and dust geometry in star-forming galaxies across redshift.
}

   \maketitle
%

\section{Introduction}\label{Sec:Intro}
Astronomers use the emission of galaxies to determine their properties such as the stellar mass $M_{*}$ and the star-formation rate $SFR$. However, dust within the galactic disk and star-forming regions of late-type galaxies scatters and absorbs the ultraviolet (UV), optical, and near-infrared (NIR) radiation, making the intrinsic stellar emission difficult to constrain. The absorption and scattering of light from extended objects is known as dust attenuation. Dust attenuation depends on both the dust properties \citep[e.g.,][]{Dra07} and relative geometry of stars and dust \citep[e.g.,][]{Calzetti1997, Char00, T04, Pop11} and is often mathematically indicated by an optical depth $\tau$ or a attenuation slope $\beta$. The attenuation is wavelength dependent and often expressed as follows:
\begin{equation}\label{eq:calz}
    F_{\lambda} \approx \lambda^{\beta},
\end{equation}
where $\lambda$ is the wavelength, $F_{\lambda}$ is the flux per wavelength, and $\beta$ is the attenuation slope \citep{Cal96}. 

The effect of the dust distribution on attenuation can be studied using the orientation of a galaxy to the line of sight. If there were no dust, the luminosity of a galaxy would be the same for all inclinations, with only the surface brightness changing with inclination as the apparent area of the galaxy changes. However, observations show the rate at which the surface brightness increases is lower than expected because light must travel through more dust at higher galaxy inclinations \citep[e.g.,][]{Hol58}.
The change in surface brightness with inclination gives us information about the dust geometry in galaxies that can help constrain the amount of dust needed to obtain dust-corrected magnitudes. \citet{Sar10} made use of galaxy inclination to investigate the difference in attenuation as a function of redshift. They took measurements from low-z galaxies in the Sloan Digital Sky Survey \citep[SDSS,][]{Ab09} and galaxies at $z\sim0.7$ in the Cosmological Evolution Survey \citep[COSMOS,][]{Scar07}. They artificially redshifted the SDSS galaxies to z$\sim$0.7 and compared the surface brightness $-$ inclination relation between the SDSS and COSMOS galaxies. They found the surface brightness$-$inclination relation for the COSMOS galaxies was flatter than seen for SDSS, suggesting that distant disk galaxies have a higher optical depth or a different spatial distribution of dust.

Attenuation can be modeled using radiative transfer equations that simulate interactions between photons and dust. Dust properties such as size, dust grain type, and spatial distribution relative to the stars, can be controlled in these models, allowing us to investigate their influence on attenuation.
Combining the results of models with the constraints from observations allows us to study the global dust properties of galaxies.

For example, \citet{Chev15} used four different models relying on different radiative transfer codes to investigate the average attenuation properties of the interstellar medium and birth clouds: \citet{Sil98} which relies on GRAphite and SILicate (GRASIL), \citet{T04} which relies on the code from \citet{Kyl87}, \citet{Pier04} which relies on DustI Radiative Transfer, Yeah! \citep[DIRTY,][]{Gor01}, and \citet{Jon10} which relies on SUNRISE \citep{Jon06}. Despite their differences, these models all showed similar attenuation curves at different galaxy inclinations, allowing them to be combined. \citet{Chev15} applied this combined model to calculate the dust attenuation of galaxies selected from the SDSS survey following \citet{Wild11}, separating the attenuation effects of the diffuse dust and the clumpy dust based on when they would attenuate the star during stellar birth and stellar migration from the birth cloud. They found the attenuation tends to increase with the bulge-to-total ratio, while the face-on optical depth remains the same. They also found attenuation tends to change the photometry up to 0.3-0.4 mag at optical wavelengths and 0.1 mag at near-infrared wavelengths.

Given the four models compared by \cite{Chev15} provided consistent results, we choose to focus on a single model in this study, the \citet{T04} model. Several studies have obtained consistent average optical depths of large samples of low-z star-forming galaxies using only the \citet{T04} model. \citet{Dri07} applied the \citet{T04} model to r-band measurements of a sample of low-z galaxies in the Millennium Galaxy Catalog to obtain the average face-on B-band optical depth $\tau_{B}^{f}$, finding $\tau_{B}^{f} = 3.8_{-0.7}^{+0.7}$. \citet{Andrae12} applied the same model on r-band measurements in low-z galaxies from the Galaxy And Mass Assembly (GAMA) survey, obtaining $\tau_{B}^{f} = 4.0_{-0.5}^{+0.5}$.

\citet{Les18} determined attenuation using the \citet{T04} model with FUV emission from low-z SDSS galaxies and $\sim$ 0.7 galaxies in the COSMOS field. They calculated the FUV attenuation using the expected emission from the star-formation rate main-sequence and compared the attenuation - inclination relation with the \citet{T04} models to retrieve $\tau_{B}^{f}$ and the likelihood of emission being absorbed by opaque star-forming regions, given as the clumpiness $F$. They obtained $\tau_{B}^{f} = 3.95_{-0.15}^{+0.16}$ and $F = 0.09_{-0.02}^{+0.02}$ for the galaxies in SDSS and $\tau_{B}^{f} = 3.5_{-1.8}^{+1.1}$  and $F = 0.55_{-0.04}^{+0.06}$ for galaxies in COSMOS. The factor-of-five increase in the fraction of attenuation in optically thick star-forming regions ($F$) agrees with the increase in molecular gas mass fraction of galaxies with redshift out to $z<1$. The authors concluded that the interstellar medium of galaxies is more clumpy at higher redshift. \cite{Les18} only used one UV band to constrain $F$, which could make the fitting sensitive to detection biases.
\citet{Les18b} investigated different methods of dust corrections for FUV data used to calculate SFRs for $z \sim 0 $ and $z \sim 0.7$ galaxies. They tested three correction methods using the UV-slope $\beta$ and converting $\beta$ to attenuation following \citet{Boquien12}, using the attenuation - inclination relation from \citet{T04}, and using mid-infrared (MIR) hybrid corrections \citep{Ken09, Hao11, Cat15}. Only the UV-slope correction failed to remove the inclination dependence for $z \sim 0 $ galaxies. The resulting SFRs for $z \sim 0.7$ galaxies using the UV-slope correction lie on average below the star-formation main-sequence (SF MS) by $\sim 0.44$ dex. The smaller inclination dependence of FUV attenuation in the higher redshift galaxies occurs because the clumpy component $F$ dominates the attenuation (as found by \citealt{Les18}). They suggested that the clumpy component and the diffuse dust disk have different UV-attenuation laws as was proposed by other studies \citep[e.g.,][]{DaCu08, Wild11, Chev15}.

This study expands the work done in \citet{Les18} by fitting attenuation-inclination models for multiple photometric bands and comparing the best-fit parameters at $z\sim0$ and $z\sim0.7$. We refer to the inclination using $1 -\cos(i)$, where $1 -\cos(i) = 0$ is face-on, and $1 -\cos(i) = 1$ is edge-on. We use UV, optical, and NIR band measurements for star-forming galaxies with redshift 0.0 $< z <$ 0.1 in the SDSS and GAMA fields and 0.6 $< z <$ 0.8 in the COSMOS field and explain the selection process in Section \ref{Sec:Data}. In Section \ref{Sec:Methods}, we describe the \citet{T04} model and how we use it to retrieve the dust parameters. In Section \ref{Sec:Results}, we show the dependence of fitted dust parameters with galaxy properties, and compare the results between low-z galaxies and galaxies at higher redshift. The properties investigated are $M_{*}$, the stellar mass surface density $\mu_{*}$, $SFR$, the $SFR$ per unit $M_{*}$ referred to as specific star-formation rate $sSFR$, the distance from the star-forming main-sequence $dMS$, and the $SFR$ surface density $\Sigma_{\mathrm{SFR}}$. We also investigate the $H\alpha/H\beta$ - inclination relation for the low-z galaxies and what set of dust parameters are required for the model to match the observations. In Section \ref{Sec:Discussion}, we discuss the results, what they imply, and how they compare to previous studies of the attenuation curve. We present our conclusions in Section \ref{Sec:Conclusion}. For any cosmology related calculations, we assume a flat $\mathrm{\Lambda CDM}$ universe with $H_{0} = 70 km\cdot s^{-1}\cdot pc$, and $\Omega_{m} = 0.3$.

\section{Data and sample selection}\label{Sec:Data}
We have compiled data from galaxies in two different redshift regimes: galaxies at redshift $0.04 < z < 0.1$ from the SDSS and GAMA surveys, and galaxies at $0.6 < z < 0.8$ from the COSMOS survey. We select galaxies at $0.04 < z < 0.1$ correspond to being observed when the universe was $t \sim 13.5Gyr$ old, which cover $\sim$5\% of the age of the universe, whereas galaxies at $0.6 < z < 0.8$  correspond to being observed when the universe was $t \sim 7Gyr$ old, which cover $\sim$8\%. Our selections allow us to study how dust properties have changed over $\approx$6.5 Gyr. The low redshift lower boundary ensures that the 3" SDSS fiber covers at least 20\% of the galaxy \citep{Kew06}, ensuring our Balmer line ratios used in Section \ref{Sec:FittingBalmer} come from a representative region of the galaxy. We set the upper boundary ($z < 0.1$) to minimize galaxy evolution effects following \citet{Les18}. The $0.6 < z < 0.8$ redshift boundaries ensure we can retrieve similar rest-frame wavelength coverage as our low redshift sample \citep{Kampczyk2007}. We convert the photometry of all three samples to $z=0$ rest-frame photometry using k-corrections. The next three subsections will cover our retrieval of literature photometric data (UV, optical, NIR, and IR), $M_{*}$, inclination, Sérsic index $n$, and $SFR$ measurements. Section \ref{Sec:Sample} explains how we select well-matched samples of star-forming galaxies.

For our study, we focus on late-type, disk-dominated galaxies for which the \cite{T04} model was developed. Inclination and bulge-to-total ratio ($B/T$) are two parameters used in the \cite{T04} model. We derive inclinations from the galaxy axis ratios and use the $n$ as a proxy for $B/T$, derived from single-component Sérsic model fits on rest-frame g-band imaging across the three surveys (see Appendix \ref{Ap:GalProp} for the GAMA and COSMOS $n$ calculations).

\subsection{COSMOS}\label{Sec:COSMOS}
We select galaxies with redshifts 0.6 $\mathrm{<}$ z $\mathrm{<}$ 0.8 from the COSMOS Zurich Structure and Morphology Catalog (ZSMC, \citet{Scar07}, available on \href{https://irsa.ipac.caltech.edu/data/COSMOS/tables/}{IRSA}\footnote{\url{https://irsa.ipac.caltech.edu/data/COSMOS/tables/}}) in a similar manner to \citet{Sar10}. We match the selected sample with the COSMOS2015 catalog \citep{Lai16}. 
\paragraph{Photometry:} The COSMOS2015 catalog contains precise PSF-matched rest-frame photometry, photometric redshifts, $M_{*}$, and $SFR$ of more than half a million COSMOS sources. We use the measured GALEX NUV data as rest-frame GALEX FUV data because at redshift $z = 0.7$ the GALEX NUV band aligns with the redshift $z \sim 0.1$ GALEX FUV band \citep{Zam07}. For the rest-frame GALEX NUV, we use k-corrected values from the COSMOS2015 catalog, where they derived the fluxes using a spectral energy distribution (SED) that fit the detections of bands close to the NUV. We also use k-corrected photometry from the Subaru B, V, and I bands for the optical coverage and UVISTA J and K bands for the NIR coverage. We use the Spitzer Multi-band Imaging Photometer for Spitzer (MIPS) 24$\mathrm{\mu m}$ observations \citep{San07,Floc09} to calculate infrared luminosities (see Section \ref{Sec:Methods}, Eq. \ref{eq:TIR}).

\paragraph{Properties:} The COSMOS2015 photometric redshifts are calculated with the code LePHARE \citep{Arnout02, Il06} using 3" aperture fluxes and a similar method to \citet{Il13}.  
The $M_{*}$ of the galaxies is also calculated with LePHARE, creating a library of synthetic spectra generated from the stellar population synthesis model of \citet{Bruz03} assuming a Chabrier initial mass function \citep[IMF][]{Chabrier2003}. Throughout this study, we convert $M_{*}$ and $SFR$ to match the Kroupa IMF \citep{Kroupa2001} following \citet{Zahid2012}. The 90\% $M_{*}$ completeness limit for star-forming galaxies in the redshift range of our sample ($0.6 < z < 0.8$) is $10^{9} \mathrm{M}_{\odot}$ in the UVISTA Deep field \citep{Lai16}. The $SFR$ is constrained using a set of 12 templates using the \citet{Bruz03} models and optical data. For this work, we use the $SFR$ corresponding to the maximum likelihood. 

\paragraph{Morphology:} The morphology is measured using the images from the HST Advanced Camera for Surveys (ACS) F814W I-band \citep{Koek07} with a pixel scale of 0.05" pixel$^{-1}$ and a resolution of $\sim$ 0.1" \citep{Scar07}. All galaxies in the sample with a magnitude I $<$ 22.5 have been modeled, as described in \citet{Sargent2007}, with a single component Sérsic profile using the Galaxy IMage2D package \citep[GIM2D;][]{Marl98, Sim02}. GIM2D seeks the best-fit values for the total flux, half-light radius $r_{1/2}$, the position angle, the central position of the galaxy, the residual background level, and the ellipticity $e = 1 - b/a$, with a and b being the semi-major and semi-minor axes of the brightness distribution, respectively.

\subsection{SDSS}\label{Sec:SDSS}
We select galaxies with redshifts $0.04 < z < 0.1$ from the spectroscopic catalog of SDSS Data Release 7 (DR7, \citealt{Ab09}).
\paragraph{Photometry:}
The magnitudes of the FUV and NUV bands are provided by the GALEX Medium-depth Imaging Survey (MIS) by \citet{Bianchi11}, who matched their catalogs of unique UV sources to the seventh SDSS data release. For these GALEX bands, we find the k-correction using the IDL code kcorrect\_v4\_2 \citep{Blanton07} and the SDSS photometry.
We draw the photometry in the optical and the near-infrared bands from the New York University (NYU) k-corrected catalog \citep{Blanton05}. They matched the SDSS ugriz photometry to the Two Micron All Sky Survey (2MASS) JHK photometry \citep{Skrut06}. Both SDSS and 2MASS photometry are used to calculate the k-corrections using kcorrect\_v4\_2 \citep{Blanton05}. 
We add $\mathrm{H\alpha}$ and $\mathrm{H\beta}$ lines from the spectroscopic Max Planck for Astrophysics and Johns Hopkins University (MPA-JHU) catalog \citep{Brinchmann2004}. We use the 12$\mathrm{\mu m}$ corrected fluxes from \citet{Cha15} from the Wide-field Infrared Survey Explorer (WISE) survey to calculate the total infrared luminosities.

\paragraph{Properties:} As mentioned, we use spectroscopic redshifts from the spectroscopic catalog of SDSS Data Release 7 (DR7, \citealt{Ab09}).
The total stellar masses were calculated by the MPA-JHU team from SDSS ugriz galaxy photometry using the model grids of \citet{Kauf03} assuming a Kroupa IMF.
We retrieve the $SFR$ from \citet{Sal18}. They computed the SFR using Code Investigating GALaxy Emission \citep[CIGALE;][]{Nol09, Boquien19}, performing SED fitting of the UV and optical photometry available in the construction of GALEX-SDSS-WISE Legacy Catalog version 1 \citep[GSWLC-1;][]{Sal16}. We choose these SFRs, as they align within 1 $\mathrm{\sigma}$ with the SFR used in the GAMA survey.

\paragraph{Morphology:} We add bulge+disk decomposition data from \citet{Sim11}. \citet{Sim11} performed 2D point-spread-functions-convolved model fitting in both the g- and r-band, giving the best-fit parameters of the galaxy profile. They set the sensitivity limit of the sample to $\mu_{1/2}$ = 23.0 mag $\mathrm{arcsec^{-2}}$ for completeness and our average g-band resolution is 1.7". For this work, we use the SDSS g-band GIM2D single Sérsic profile fits as the g-band allows for comparison with the ACS I band imaging samples at z$\sim$0.7. 

\subsection{GAMA}\label{Sec:GAMA}
In addition to the galaxies in SDSS, we also select galaxies from the GAMA Data Release 3 \citep{Dri09, Baldry18}. The GAMA input catalog is based on data taken from spectroscopic SDSS and UKIRT Infrared Deep Sky Survey. GAMA is a deeper and more complete survey compared to SDSS, with an r$ < 19.8$ magnitude limit for 98.5\% redshift completeness \citep{Baldry18}, compared to SDSS with an r$ < 17.7$ magnitude for 94\% completeness limit \citep{Strauss2002}.

\paragraph{Photometry:} The FUV and NUV photometry is obtained from GALEX \citep{Lisk15}, the optical from VLT Survey Telescope (VST) Kilo-Degree Survey (KiDS) survey \citep{Jong13}, the near-infrared from Visible and Infrared Survey for Telescope for Astronomy (VISTA) Kilo-degree Infrared Galaxy (VIKING) public survey \citep{Edge13}, and the spectra from Anglo-Australian Telescope \citep[AAT,][]{Lisk15}. We use the updated photometric data in GALEX FUV, NUV, SDSS g, r, and i, and 2MASS J and K bands from \citet{Bell20}. The k-corrections of the galaxies are calculated with kcorrect\_v4\_2 using magnitudes from the ApMatchedCat \citep{Love12}. The $H\alpha$ and $H\beta$ fluxes are derived from Gaussian fitting following \citet{Gor17}. We use IR data of GAMA DR3 from the WISE survey \citep{Clu14} at 12$\mathrm{\mu m}$.

\paragraph{Properties:} We use the spectroscopic redshifts from the GAMA survey described in \citet{Lisk15}.
We make use of the stellar masses in the table StellarMassLamdar, which used matched-aperture photometry from the GAMA DR 3 catalog ApMatchedCat \citep{Hill11, Lisk15, Dri15} as the input and derived the $M_{*}$ using the code LAMBDAR \citep{Wri16}. This code derives unblended photometry across the optical to NIR bands used for the fitting. Because the masses were derived with aperture photometry, we need to take into account aperture corrections using the formula:
\begin{equation}
    \log(M_{*, total}) = \log(M_{*, before}) + \log(\mathrm{fluxscale}),
\end{equation}
where $M_{*, before}$ and $M_{*, total}$ are the stellar mass from the LAMBDAR code before and after applying the correction respectively, and $fluxscale$ is the documented flux ratio between the r-band flux from the LAMBDAR photometry and the Sérsic profile fitting. The fitting assumes a Chabrier IMF, which we convert to Kroupa IMF.
The SFR used for the GAMA sample is calculated using MAGPHYS \citep{DaCu08} adopting templates from \citet{Bruz03} and comparing SEDs to the resulting photometry of LambdarCat, taking into account the energy balance between dust absorption and dust emission. We select the SFRs found in the GaussFitSimple data table \citep{Gor17}.

\paragraph{Morphology:} The half-light radii and axes ratios of the galaxies are documented in SersicCatSDSS \citep{Kel12}, resulting from single Sérsic profile fits performed on the g-band images from the SDSS catalog.

\subsection{Sample selection}\label{Sec:Sample}
Ideally, we would like to select samples of star-forming galaxies with similar physical properties to ensure that only the variation in dust attenuation causes differences in galaxy emission between the three catalogs. Such a selection would allow us to analyze the dust properties of galaxy populations at different redshifts.
In the following, we describe our selection of star-forming galaxies (Sec \ref{Sec:SFselec}) and our selections on $M_{*}$ and $n$ (Sec \ref{Sec:PhysSelec}).

\subsubsection{Star-formation selection}\label{Sec:SFselec}
First, we classify galaxies based on their colors and use color cuts to select star-forming galaxies. We use the criteria for the NUV-r and r-J colors following \citet{Il13}.
For GAMA and SDSS, we also use the 3" aperture optical spectra to select only star-forming galaxies based on their emission line ratios using the \citet{Kew06} diagnostic curves. We also select star-forming galaxies by their WISE photometry following the criteria from \citet{Assef18}, where the mid-IR selection identifies AGN with 90\% reliability.
Because we do not have the same mid-IR data for the COSMOS sample, we use a MIR-selection based on \citet{Don12} using SPLASH photometry to remove IR AGNs. We also remove galaxies in COSMOS with AGN based on X-ray emission \citep{Mar16}:
\begin{equation}
    L_{xray} > 10^{42} \mathrm{erg\cdot s^{-1}},
\end{equation}
with $L_{xray}$ as the 2 - 10 keV X-ray luminosity, taken from the Chandra COSMOS-Legacy survey \citep{Civ16, Mar16}.

\subsubsection{Physical properties and morphology}\label{Sec:PhysSelec}
We select galaxies within a range of physical properties. Because these physical properties are derived from photometry, we have to take into account possible detection biases. The  inclination of disk galaxies impacts detection biases in two conflicting ways. More inclined galaxies will appear brighter due to their emission being spread over a smaller apparent surface area. Conversely, more inclined galaxies will also have lower apparent luminosities, especially at shorter wavelengths, due to the increase in attenuation. The combination of these effects will influence the photometry of our sample and may result in extrinsic trends between the inclination and photometry-derived galaxy properties.
inclination dependent biases in physical properties in our analysis are observed in $M_{*}$, $r_{1/2}$, and $n$. To remove these dependencies, we calculate the median values of the physical properties for bins of inclinations and model the following relation through the data based on the method of \citet{Les18}:
\begin{equation}\label{eq:lin_exp}
    x = a(1 - \cos(i))^{b} + c,
\end{equation}
with $x$ being one of the physical properties and $a$, $b$, and $c$ are the fitting coefficients of the relations. These coefficients are given in Table \ref{tab:in_bias}. The results from our fitting vary slightly from \citet{Les18} as our selections differ, but the trends agree qualitatively. We find a slight positive correlation for $M_{*}$ and size with inclination and a negative correlation for $n$.

\begin{table}[t]
\centering
\caption{Best-fit values for the inclination bias in the $M_{*}$, $r_{1/2}$, and $n$, using Eq. \ref{eq:lin_exp}. We note that GAMA seems to have an inclination independent mass $\log(M_{*}[\mathrm{M}_{\odot}])$, causing large uncertainties in the inclination dependent fitting parameters $a$ and $b$.}
\label{tab:in_bias}
\resizebox{\columnwidth}{!}{%
\begin{tabular}{ll|l|l|l}
\hline\hline
       &           & a                  & b                    & c                  \\ \hline
       & $\log(M_{*}[\mathrm{M}_{\odot}])$ & 0.13$\pm$ 0.14        & 0.26$\pm$0.42           & 10.28$\pm$0.15        \\
SDSS   & $r_{1/2}$ (kpc)       & 3.78$\pm$0.16         & 3.35$\pm$0.25           & 4.47$\pm$0.049        \\
       & $n$         & -0.49$\pm$0.029       & 2.94$\pm$0.32           & 3.43$\pm$0.034       \\ \hline
       & $\log(M_{*}[\mathrm{M}_{\odot}])$ & $2.84\pm120$ & $82.2\pm143$ & $9.86\pm0.015$ \\
GAMA  & $r_{1/2}$ (kpc)       & 3.79$\pm$0.15         & 5.16$\pm$0.37           & 3.72$\pm$0.042         \\
       & $n$         & -0.35$\pm$0.047       & 2.73$\pm$0.82           & 1.00$\pm$0.024        \\ \hline
       & $\log(M_{*}[\mathrm{M}_{\odot}])$ & 0.25$\pm$0.067        & 4.56$\pm$2.09           & 9.97$\pm$0.018        \\
COSMOS & $r_{1/2}$ (kpc)       & 8.07$\pm$0.36        & 3.75$\pm$0.31           & 3.71$\pm$0.12        \\
       & $n$         & -0.58$\pm$0.057       & 3.54$\pm$0.67           & 0.92$\pm$0.02        \\
       \hline
\end{tabular}%
}
\end{table}
We remove the inclination dependence found from the fitting before making cuts in physical properties by subtracting the inclination dependent part of Eq. (\ref{eq:lin_exp}) from the data. This inclination dependency of physical properties likely means there is a detection bias with inclination in the photometry. For our analysis, we compensate for the bias in photometry using a technique called importance sampling. Following \citet{Chev15}, we use importance sampling to weigh the galaxies based on the physical properties ($z$, $\log(M_{*}[\mathrm{M}_{\odot}])$, $r_{1/2}$, and $n$) and their relation with inclination. We use these weights for calculating unbiased averages of photometric data in bins of inclination to find the unbiased attenuation-inclination curve. For a more detailed explanation on importance sampling, see Appendix \ref{Ap:ImpSamp}. \\

\begin{figure*}[!ht]
    \centering
    \includegraphics[width=\linewidth]{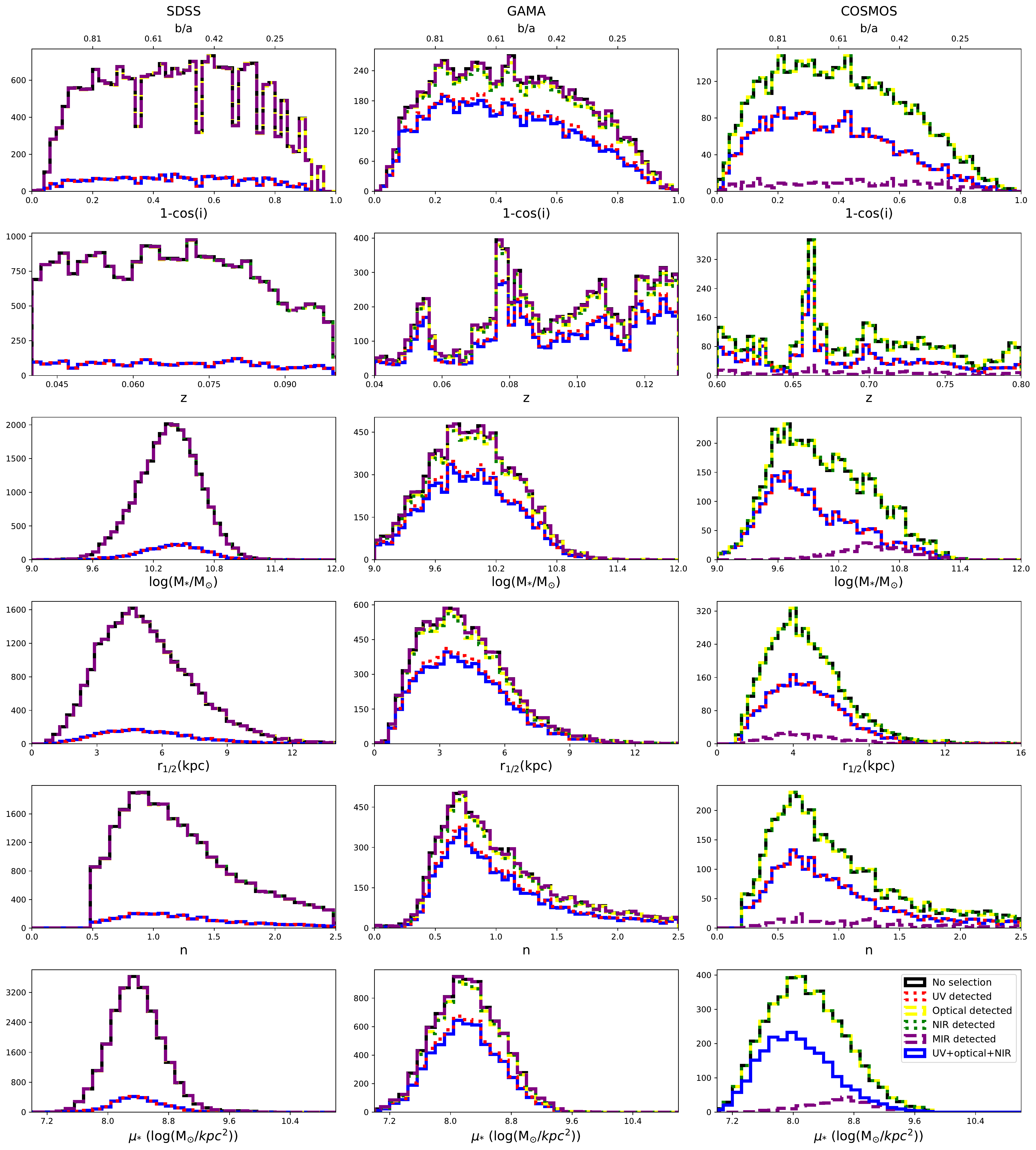}
    \caption{Distributions of the inclination corrected properties of star-forming galaxies in the three galaxy datasets SDSS, GAMA, and COSMOS, based on the different rest-frame detection criteria. The criteria are: (1) the galaxy is detected in at least one band (black), (2) the galaxy is detected in at least one UV band, one optical, one NIR, and one MIR band with S/N > 3 (blue), (3) the galaxy is detected in at least one UV band with S/N > 3 (red), (4) the galaxy is detected in at least one optical band with S/N > 3 (yellow), (5) the galaxy is detected in at least one NIR band with S/N > 3 (green), (6) the galaxy is detected in at least one MIR band with S/N > 3 (purple). The properties are, from top to bottom, inclination $1-\cos(i)$, redshift $z$, stellar mass $\log(M_{*})$, half-light radius $r_{1/2}$, Sérsic index $n$ and stellar mass surface density $\mu_{*}$. In general, the UV selection affects the number of galaxies detected in all samples. Most of the differences in selections are not visible, as the distributions overlap with each other. There is a bias toward more massive galaxies when requiring a MIR-detection in COSMOS.}
    \label{fig:optical_bias}
\end{figure*}

We investigate the potential detection bias by selecting galaxies based on their S/N (S/N > 3.0) in the UV, optical, NIR, and MIR bands used for our analysis. We show these observational selections in Figure \ref{fig:optical_bias}.
Figure \ref{fig:optical_bias} shows the physical property distributions of the three datasets. The properties shown are inclination $1-\cos(i)$, $z$, $\log(M_{*})$, $r_{1/2}$, $n$ and $\mu_{*}$. $\mu_{*}$ is given by:
\begin{equation}\label{eq:dens}
    \mu_{*} = \frac{M_{*}}{2r_{1/2}^{2}}.
\end{equation} 
The samples generally do not show a bias in a property when requiring a detection at a particular wavelength. The COSMOS sample is the only exception, where using MIR-selections would bias our sample towards higher-mass galaxies. We do not restrict our galaxies based on being detected in a specific band due to this lack of bias, resulting in a larger sample of galaxies to use for our analysis. 

From the inclination corrected galaxy property distributions, we determine the cuts in physical properties needed to select disk-dominated galaxies that only differ in their redshift distribution between the three samples. We chose the cuts to ensure that the three samples are complete and have similar physical property distributions. We only select in $M_{*}$ and $n$, with the following criteria:
\begin{equation*}
    10.0 < \log\left(\frac{M_{*}}{\mathrm{M}_{\odot}}\right) < 12.0,
\end{equation*}
\begin{equation*}
    0.0 < n < 2.5.
\end{equation*}

\begin{figure}[t]
    \centering
    \includegraphics[width = \columnwidth]{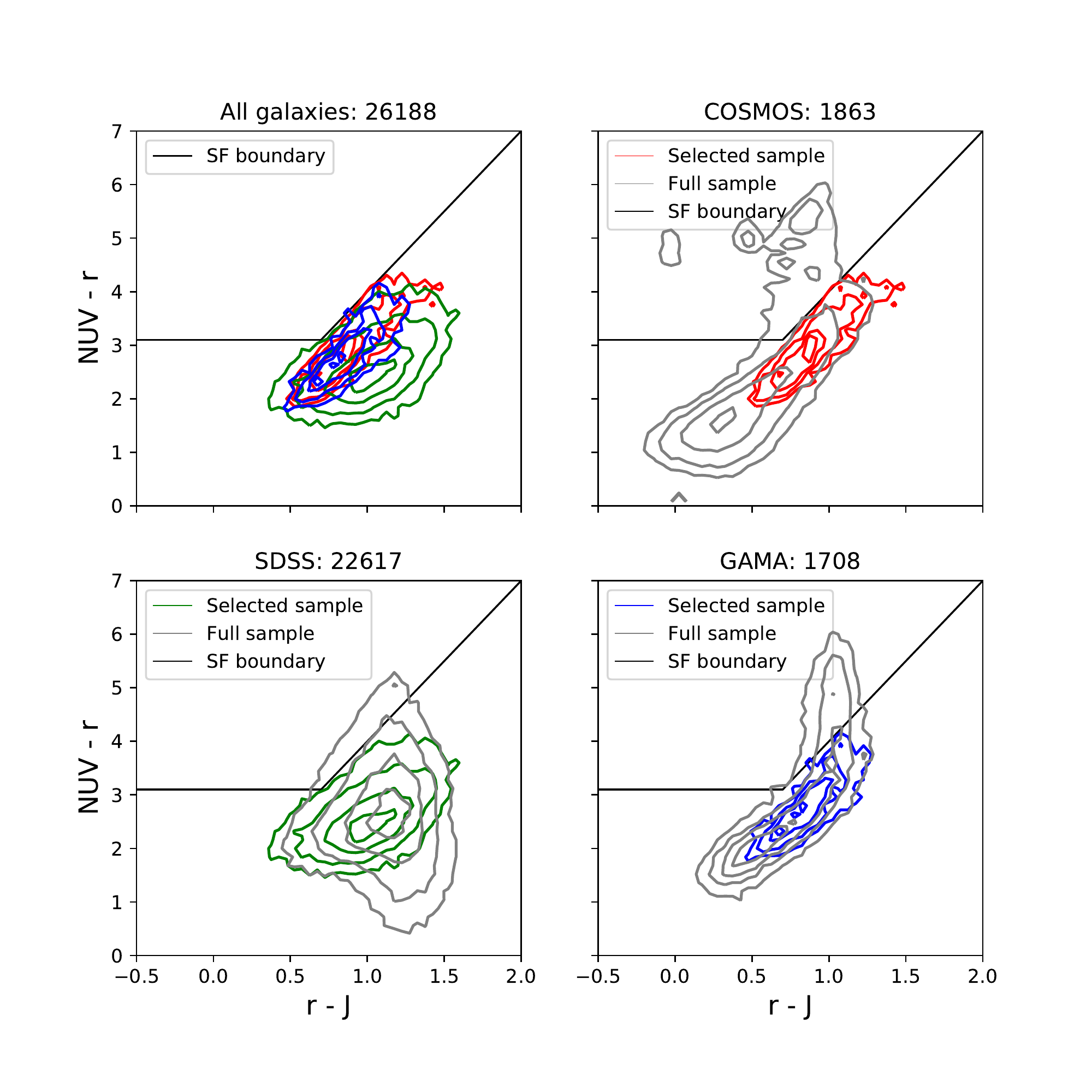}
    \caption{Rest-frame NUV - r and r - J color diagram of galaxies used before (gray) and after (colored) applying the selection in star-formation classification and physical properties $\log(M_{*})$ and $n$, including AGN removal. We select star-forming galaxies to lie below the black line.}
    \label{fig:SFR_UV2}
\end{figure}

Figure \ref{fig:SFR_UV2} shows the NUV-r and r-J color-color diagram of the sample after applying the physical selections.
The main purpose of this figure is to show that galaxies from the different surveys align well, especially the GAMA and COSMOS surveys. \\
We have $\sim 2000$ galaxies for COSMOS and GAMA and $\sim 20 000$ galaxies for SDSS after applying all the selections.

\section{Methods}\label{Sec:Methods}
\begin{figure*}
    \centering
    \includegraphics[width= \linewidth, trim = 0cm 6cm 0cm 1cm clip]{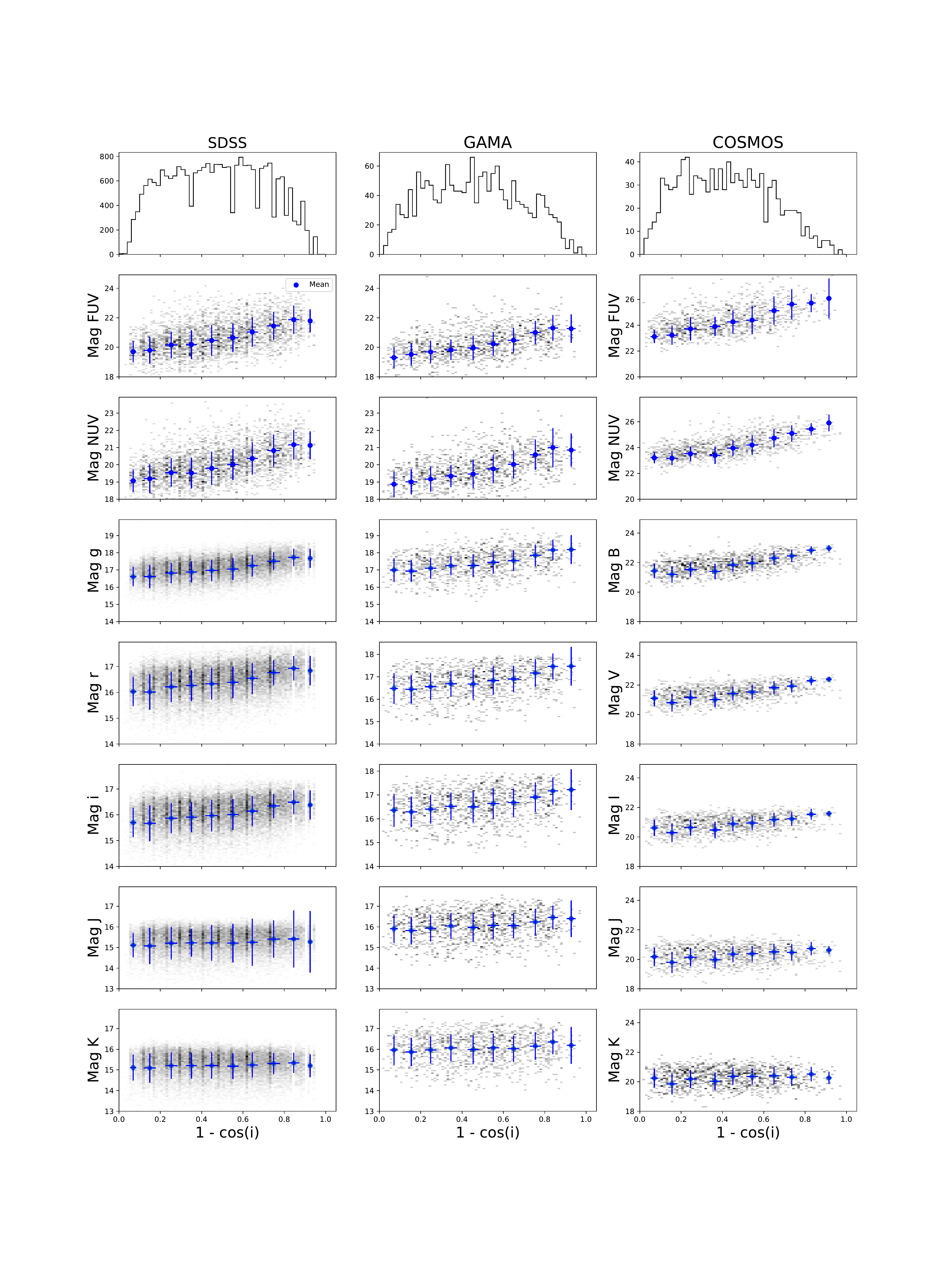}
    \caption{Magnitude-inclination relations. The top row shows the inclination distribution of all selected star-forming galaxies in the three samples. The other panels show the magnitude-inclination relation for SDSS, GAMA, and COSMOS in rest-frame UV, optical, and NIR bands. The grayscale is a 2D histogram of all the galaxies with a signal-to-noise ratio > 3 in the respective band and the blue points show the mean values with the 32-68th percentile as errors in 10 bins of inclination using the importance sampling weights. We note that our sample selection does not require a galaxy to be detected in all photometric bands.}
    \label{fig:mag_inc_SGC}
\end{figure*}

After applying the selections, we analyze the magnitudes of the galaxies in different bands as a function of their inclination, as shown in Figure \ref{fig:mag_inc_SGC}. We describe the magnitude - inclination relations using the \citet[][hereafter T04]{T04} attenuation-inclination model and fit for the model parameters, making assumptions about the unattenuated emission in the GALEX UV-bands.

T04 designed a model to predict the attenuation at different UV and optical wavelengths and inclination for a star-forming galaxy with parametrised size and structures using radiative transfer modeling of UV, optical, NIR, FIR, and submm-bands multiwavelength observations of low redshift star-forming galaxies \citep[e.g.,][]{Pop00}.
In the T04 model, a galaxy consists of two major component types: diffuse components describe the distributions of stars and dust on the scales of the galactic disks and the clumpy component describes these distributions within the star-forming regions. Specifically, the T04 model contains a dustless stellar bulge, a stellar disk harboring old stellar populations and a diffuse dust disk, a thin stellar disk harboring young stellar populations and the clumpy component, and  a diffuse thin dust disk spatially correlated with the thin stellar disk. The disks and bulge were described by different exponential and de Vacouleur distributions, respectively. 
T04 calculated intrinsic and attenuated images of a model galaxy in eight optical and nine ultraviolet (UV) bands for the bulge, disk, and thin disk from which they derive attenuation-inclination relations. We use the updated attenuation - inclination models described in \citet{Pop11}. The updated model uses the \citet{Wein01} and \citet{Dra07} dust models, which include a mixture of silicate, graphite, and PAH molecules.
T04 fitted the attenuation of the diffuse component in galaxies at each wavelength for different $\tau_{B}^{f}$ as a function of inclination with polynomial functions:
\begin{equation}\label{Tuffs_model_power}
    \Delta m(\tau_{B}^{f}) = \Sigma_{j = 0}^{k} a_{j}(\tau_{B})[1 - \cos(i)]^{j},
\end{equation}
where $\Delta m$ is the difference in magnitudes between the dusty images and the intrinsic images, $i$ the inclination angle, $a_{j}$ a set of coefficients for each $\tau_{B}$, and $k$ the maximum fitted power of the polynomial. T04 used $k = 4$ for the bulge component and $k = 5$ for the disk and thin disk. The $\tau_{B}^{f}$ used here is measured at the center of the galaxy and assumes that the dust follows an exponential profile, supported by studies of galaxies in the local Universe \citep[e.g.,][]{Alt98, Bian07, Muno09, Hunt15, Casa17}. However, when looking at our own Milky Way \citep{Popescu2017, Natale21} or the nearby M33 \citep{Thir20}, it was found that the dust distribution is exponential down to an inner radius, and then it decreases towards the center. So in this respect, the parameter $\tau_{B}^{f}$ should be thought of an effective $\tau$, describing the global distribution of dust.
The attenuation of all the components combined is given as:
\begin{equation}\label{eq:main}
    \begin{split}
    \Delta m_{\lambda}(\tau_{B}^{f},i)
    &= 2.5\log(r_{\lambda}^{disk}10^{\frac{\Delta m_{\lambda^{disk}}(\tau_{B}^{f},i)}{2.5}} \\
    &+ \frac{1 - r_{\lambda}^{disk} - r_{\lambda}^{bulge}}{1 - \mathrm{Ff_{\lambda}}}10^{\frac{\Delta m_{\lambda^{tdisk}}(\tau_{B}^{f},i)}{2.5}}\\ 
    &+ r_{\lambda}^{bulge}10^{\frac{\Delta m_{\lambda}^{bulge}(\tau_{B}^{f},i)}{2.5}}). \\
    \end{split}
\end{equation}
This is the general expression for the attenuation for the disk $\Delta m_{\lambda^{disk}}(\tau_{B}^{f},i)$, thin disk $\Delta m_{\lambda^{tdisk}}(\tau_{B}^{f},i)$, and the bulge $\Delta m_{\lambda^{bulge}}(\tau_{B}^{f},i)$. The attenuation per component depends on the wavelength $\lambda$, $\tau_{B}^{f}$, flux fractions of the bulge-to-total ratio $r_{\lambda}^{bulge}$ and thin disk-to-total ratio $r_{\lambda}^{disk}$, wavelength-independent clumpiness $F$, and wavelength-dependent conversion factor of clumpiness to attenuation $f_{\lambda}$, and the inclination angle $i$. For the clumpy component, we assume that there is no cloud fragmentation due to feedback, limiting F to be $<$ 0.61 (see T04, \citet{Pop11} for more details). The wavelength dependence of $f_{\lambda}$ only comes from the escape fraction of different types of stars. Stars with lower masses and redder emission escape further from their parent cloud in their lifetimes compared to higher mass bluer stars. Therefore, the light of redder stars is generally less attenuated by their parent clouds. In T04, they derive the wavelength-dependent values $f_{\lambda}$ in their Appendix A, summarized in T04 Table A.1.
Because the stellar components are separated based on old and young stellar population, it is possible to rewrite Eq. (\ref{eq:main}) by assuming that all UV emission comes from the young stellar disk, also refered to as the thin disk, and the optical to NIR attenuation comes from the old stellar disk, also refered to as the disk, and bulge. These assumptions mean that we can express the bulge and disk flux fractions such that:
\begin{equation}
    r_{\lambda}^{disk} + r_{\lambda}^{bulge} = 1.
\end{equation}
As the young stellar population dominates the UV-emission, the general expression of the T04 model is written for the UV range as:
\begin{equation}\label{eq:T_UV}
    \Delta m_{\lambda}^{UV} = \Delta m_{\lambda}^{tdisk} - 2.5\log(1 - Ff_{\lambda}).
\end{equation}
Because the old stellar population dominates the optical and NIR emission, the expressions for the optical and NIR range can be rewritten as:
\begin{equation}\label{eq:T_opt}
    \Delta m_{\lambda}^{optical} = 2.5\log((1 - r_{\lambda}^{bulge})10^{\frac{\Delta m_{\lambda}^{disk}}{2.5}} + r_{\lambda}^{bulge}10^{\frac{\Delta m_{\lambda}^{bulge}}{2.5}}),
\end{equation}
with the attenuation of the components varying with $\tau_{B}$, $\lambda$, and $1-\cos(i)$.

Adopting the above expressions allows us to fit for the parameters $\tau_{B}^{f}$, $F$, and hereby called bulge fraction $r_{bulge}$. Since the T04 models do not give explicit predictions for the SDSS ugriz bands, but the wavelengths corresponding to the BVIJK bands, we use interpolations to derive the attenuation values at the desired wavelengths. We separate the magnitude - inclination data into ten separate bins with an equal width in inclination or $1 - cos(i)$. Then, we need to derive the attenuation from our observations by normalizing the data.\\
Since we do not know the intrinsic emission in the optical and NIR bands, we cannot use the models directly in these bands. Instead, we normalize the data and the models by a near-face-on average magnitude, as it is the least affected by attenuation. The optical and NIR data and models are normalized to the value in the second inclination bin to avoid the low number of detections in the first bin.\\
However, we do need to estimate the intrinsic emission in the UV bands to fit for $F$ describing the inclination independent offset between the attenuated and intrinsic emission.
\citet{Les18} used the star-formation main-sequence and the SFR-UV conversion from \citet{Ken12} to estimate the intrinsic UV luminosity, assuming that the sample is dominated by main-sequence galaxies. We show the results obtained using this method in Appendix \ref{Ap:IntrUV}.  
But in this work, we aim to investigate trends as a function of galaxy parameters such as distance from the main-sequence and therefore choose to use a MIR-based correction for each galaxy. We normalize the FUV and NUV data by assuming dust-corrected FUV and NUV emission following \citet{Hao11}:
\begin{equation}
L_{FUV, corrected} = L_{FUV, observed} + 0.46L_{TIR},
\end{equation}
\begin{equation}
L_{NUV, corrected} = L_{NUV, observed} + 0.27L_{TIR},
\end{equation}
with the total infrared luminosity derived following \citet{Cluver2017}:
\begin{equation}\label{eq:TIR}
\log(L_{TIR}) = 0.889 \log(L_{12 \mu m}) + 2.21,
\end{equation}
with $L_{12 \mu m}$ the luminosity at 12 micron in solar luminosities. For SDSS and GAMA, we use the WISE3 flux, and for COSMOS, we use the MIPS 24$\mathrm{\mu m}$ flux and make small k-corrections using the \citet{Wuyts2008} SED template.

In these ten inclination bins, we calculate the mean magnitudes of the galaxies applying weights found using importance sampling (Appendix \ref{Ap:ImpSamp}). The calculated means of bins two up to nine (avoiding possible detection biases in the first and last bins affecting the results) are compared to the model value at the mean inclination per bin per waveband using the MCMC python package emcee.py \citep{For13} with the following likelihood function:
\begin{equation}\label{eq:LikeT}
    L = -\Sigma_{k}\Sigma_{i} \left(\frac{y_{k}(x_{i}) - y_{model,k}(x_{i})}{\sigma_{i}}\right)^{2},
\end{equation}
where $y$ is the mean value in inclination bin $i$ and band $k$, $\mathrm{y_{model}}$ the T04 model at mean inclination per bin $x_{i}$, and $\sigma_i$ the uncertainty in the magnitude of the bands for which we use the sample standard deviation. We find the best-fit parameters by selecting the 50th percentile in the sampler distribution and the uncertainties by selecting the 32nd and 68th percentiles.\\
We assume uniform priors for $\tau_{B}$ and $F$ over the range of the models. We limit $\tau_{B}^{f}$ to be between $ 0 < \tau_{B}^{f} < 8$, as the T04 models were calculated over this range and the model may not hold for higher values. We limit $F$ to be between $0 < F < 0.61$, as the multiplication of $F$ with the wavelength-dependent parameter $f_{\lambda}$ of Eq. (\ref{eq:T_UV}) needs to be smaller than 1 to get finite numbers in a logarithm. We interpolate the wavelength-dependent parameter $f_{\lambda}$ from T04 Table A.1  and obtain $f_{FUV} = 1.361$ for the GALEX FUV band and $f_{NUV} = 0.839$ for the GALEX NUV band. Therefore, we know that the maximum value $F$ can have is constrained by the $f_{FUV}$, corresponding to $F = \frac{1}{1.361} = 0.61$. We use the observed or inferred bulge-to-total ratio as a prior for $r_{bulge}$ by fitting a skewed Gaussian distribution to the observations, assuming that the variation with optical wavelength is negligible. For SDSS, we use the $B/T$ ratios from the g-band bulge-disk decomposition with an $n=4$ bulge from \citet{Sim11}. For GAMA and COSMOS, we do not have access to $B/T$ ratios for the full sample. Instead, we train a model using scikit learn described in Appendix \ref{Ap:GalProp} to predict the $B/T$ using $n$, $M_{*}$, and color. For GAMA, we train the model using the GAMA galaxies that are also found in the SDSS sample and the g-r color, whereas for COSMOS, we use the CANDELS $B/T$ ratios from \citet{Haus13} and the B-R color.\\
We fit all three samples separately, and by comparing SDSS and GAMA with each other, we verify whether the results for $z\sim0.1$ galaxies are robust and not dependent on the sample. 
The fitting is sensitive to the chosen boundaries of the priors. In Appendix \ref{Ap:Model}, we show how much the boundaries influence the overall results.

\section{Results}\label{Sec:Results}
We apply the fitting regime described in Section \ref{Sec:Methods} to the selected samples from Section \ref{Sec:Data}. First, we fit the selected data with the T04 model to study how the results for best-fit $\tau_{B}^{f}$, $F$, and $r_{bulge}$ depend on redshift. Then, we investigate the fitting parameter dependence on inferred galaxy parameters, namely $M_{*}$, $\mu_{*}$, $SFR$, $sSFR$, $dMS$, and $\Sigma_{SFR}$. In Section \ref{Sec:FittingBalmer}, we explore the dependence of the Balmer ratio $\mathrm{H\alpha/H\beta}$ on the inclination and $M_{*}$ in the SDSS and GAMA samples.

\subsection{T04 model fits to SDSS, GAMA, and COSMOS}\label{Sec:FittingZ}
\begin{figure}[!ht]
    \centering
    \includegraphics[width = \columnwidth, trim = 0cm 1cm 0cm 0.5cm clip]{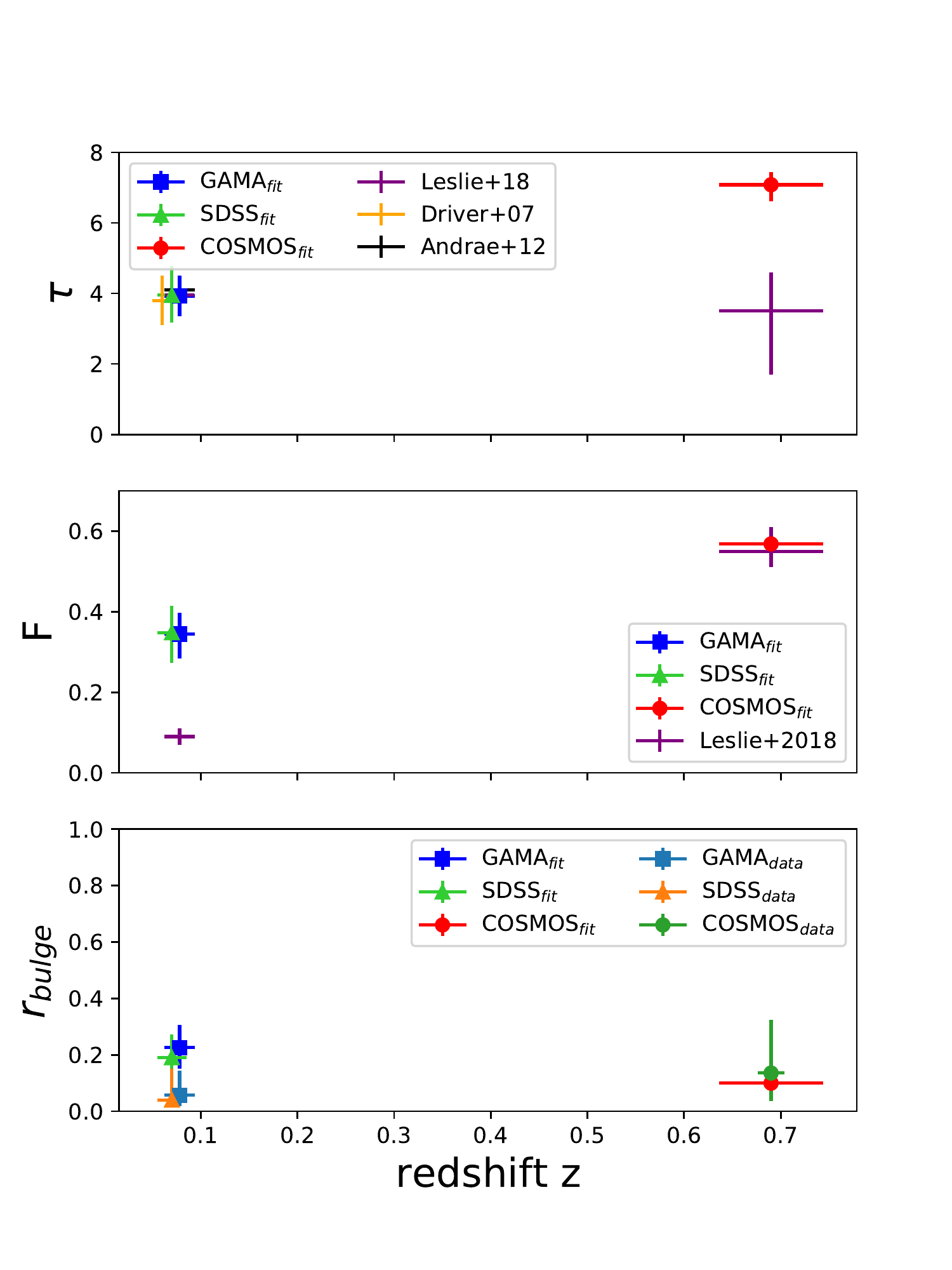}
    \caption{Results of fitting the T04 model for galaxies in the GAMA, SDSS, and COSMOS datasets. The fitted parameters are the $\tau_{B}^{f}$ (top), $F$ (middle), and $r_{bulge}$ (bottom). We compare the fitted values of $\tau_{B}^{f}$ and $F$ to the ones found in \citet{Les18}, \citet{Dri07}, and \citet{Andrae12}, and the fitted values for $r_{bulge}$ to the average and 32-68th percentile of the bulge-to-total distribution of the selected samples, labeled with the subscript "data". Our best-fit values are in line with the literature values. We see that $\tau_{B}^{f}$ and $F$ increase with redshift.}
    \label{fig:gal_z}
\end{figure}

We first take all the selected star-forming galaxies, compare their magnitude-inclination relations in seven bands from FUV - K, with the T04 model, and find the best-fit parameters for the SDSS, GAMA, and COSMOS data sets. The results are given in Figure \ref{fig:gal_z} and Table \ref{tab:gal_z}.

\begin{table}[b]
\centering
\caption{Median redshift $z_{med}$ and the best-fit values for the different datasets.}
\label{tab:gal_z}
\resizebox{\columnwidth}{!}{%
\begin{tabular}{l|l|l|l|l}
\hline\hline
       & $z_{med}$ & $\tau_{B}^{f}$            & $F$              & $r_{bulge}$     \\ \hline
SDSS   & $0.05^{+0.08}_{-0.07}$ & $3.95^{+0.82}_{-0.77}$ & $0.35^{+0.08}_{-0.07}$ & $0.19^{+0.08}_{-0.07}$  \\
GAMA   & $0.078^{+0.015}_{-0.015}$& $3.93^{+0.58}_{-0.58}$ & $0.34^{+0.05}_{-0.06}$ & $0.23^{+0.08}_{-0.07}$  \\
COSMOS & $0.69^{+0.054}_{-0.054}$  & $7.08^{+0.36}_{-0.46}$ & $0.57^{+0.02}_{-0.02}$ & $0.10^{+0.05}_{-0.04}$\\
\hline
\end{tabular}%
}
\end{table}

Table \ref{tab:gal_z} shows the best-fit values and their uncertainties for $\tau_{B}^{f}$, $F$ and $r_{bulge}$. The fitted values for $\tau_{B}^{f}$ of the low redshift samples are slightly higher than what was found in \citet{Les18}, \citet{Dri07}, and \citet{Andrae12}. This difference is explained by the differences between the original T04 model published in \citet{T04}, used in the reference studies, and the updated version published in \citet{Pop11} used for our results. \citet{Pop11} mention that for similar attenuation-inclination curves, the new models will have a $10\%$ higher $\tau_{B}^{f}$, which explains why our results have higher $\tau_{B}^{f}$. The difference in $F$ between our results and \citet{Les18} is due to the difference in methods. If we were to follow the methods described in \citet{Les18}, we would obtain the same results. We investigate the effects of the assumed intrinsic UV emission in Appendix \ref{Ap:IntrUV} and the number of wavelengths used in Appendix \ref{Ap:Model}.

Comparing our results of the $0.0 < z < 0.1$ galaxies from SDSS and GAMA with the $0.6 < z < 0.8$ galaxies from COSMOS shows that the $\tau_{B}^{f}$ and $F$ increase with redshift, whereas the bulge fraction decreases. Our fitting results at $z\sim$0.7 are again higher than \citet{Les18}, and the main difference is that they only found an increase in $F$ with redshift but no significant increase in $\tau_{B}^{f}$. However, an increase in $\tau_{B}^{f}$ with redshift was suggested in \citet{Sar10}, as they found a flatter B-band surface brightness - inclination relation for pure disk galaxies in COSMOS. If the surface brightness - inclination curve is flatter, it means that more light gets attenuated with inclination. Considering that \citet{Sar10} only looked at pure disk galaxies, we can ignore the effects of $r_{bulge}$ and therefore, a steeper attenuation-inclination relation implies a higher $\tau_{B}^{f}$ for their sample of COSMOS galaxies at $z\sim0.7$, in qualitative agreement with our result.

\begin{figure}[!ht]
    \centering
    \includegraphics[width = \columnwidth, trim = 0cm 1cm 0cm 0cm clip]{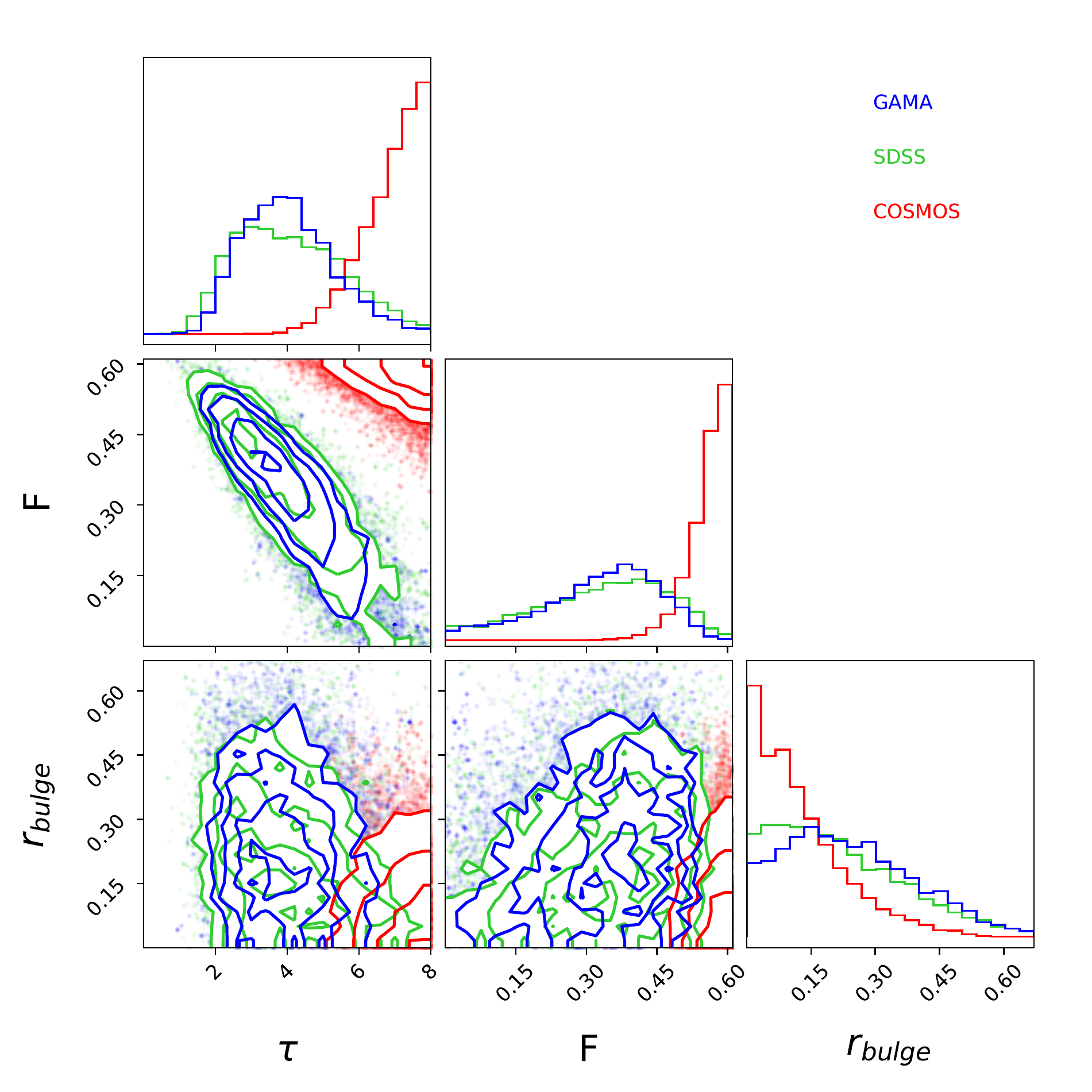}
    \caption{Results of the MCMC sampling and the best-fit parameter distributions using the T04 attenuation-inclination model. We show the fitting results for galaxies in SDSS (green), GAMA (blue), and COSMOS (red) datasets. We compare the fitted values for $r_{bulge}$ to the average and 32-68th percentile of the bulge-to-total distribution of the selected samples, labeled as data.}
    \label{fig:corner_z}
\end{figure}
In Figure \ref{fig:corner_z}, we see the distribution of samplers of the MCMC fitting, illustrating how the fitted parameters are dependent on each other. We see in the figure panels that the distributions of $\tau_{B}^{f}$ and $F$ are narrow for the SDSS and GAMA sample, where the $1\sigma$ value is less than 30$\%$. The distribution of $r_{bulge}$ is wider. We also see that $\tau_{B}^{f}$ and $F$ are highly covariant. The COSMOS sample is best fit by values at the extreme ends of the T04 model, and as such, our uncertainties are likely underestimated.

\subsection{Variation of T04 model parameters with galaxy properties}\label{Sec:FittingBins}
\begin{figure}[t]
    \centering
    \includegraphics[width = \columnwidth, trim = 0cm 0cm 0cm 0cm clip]{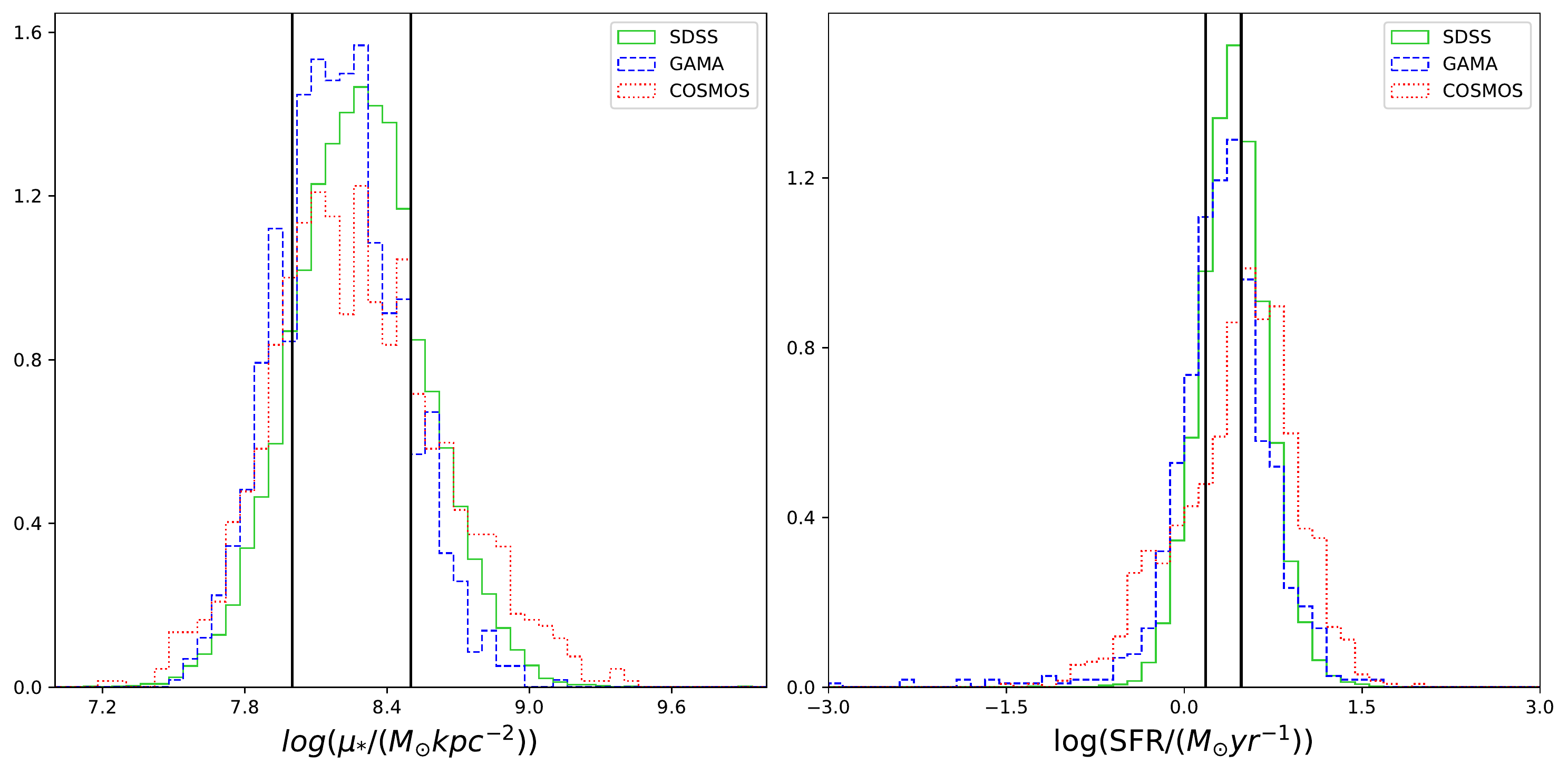}
    \caption{Distributions of our selected star-forming galaxy samples in SDSS, GAMA, and COSMOS. The figure shows the distribution in $\mu_{*}$ (left panel), and $SFR$ (right panel). As the COSMOS sample is selected at higher redshift, we convert the $SFR$ to redshift $z = 0$ assuming that $SFR \approx (1+z)^{3}$. Black lines show the boundaries of the bins used in our analyses.}
    \label{fig:mu_SFR}
\end{figure}

After we have fitted the $\tau_{B}^{f}$, $F$, and bulge fraction of the entire sample, we now fit these parameters for different subsamples of galaxies separated by their physical properties. We aim to investigate how the magnitude-inclination relations, and thereby the T04 parameters describing global dust properties, change as a function of physical galaxy properties related to their star-formation history. We bin the galaxies in either $M_{*}$, $\mu_{*}$, $SFR$, $sSFR$, $dMS$, or $\Sigma_{SFR}$, and apply the T04 model fitting in each bin to obtain the best-fit parameters for galaxy samples with varying properties. We choose the bins such that they each cover a third of the physical property range constrained from Figure \ref{fig:optical_bias} after making the selection cuts.
The $M_{*}$ and $SFR$ properties are sensitive to systematic differences between the samples as the $M_{*}$ and $SFR$ are derived using different SED-fitting techniques for each survey. In Figure \ref{fig:mu_SFR} we show that the property distributions are comparable between the three samples when making our selection cuts, which should mean that the effects of the systematic differences on our reported trends can be neglected.

\subsubsection{Stellar mass}\label{Sec:FittingMass}
\begin{figure}[t]
    \centering
    \includegraphics[width = \columnwidth, trim = 0cm 1cm 0cm 0cm clip]{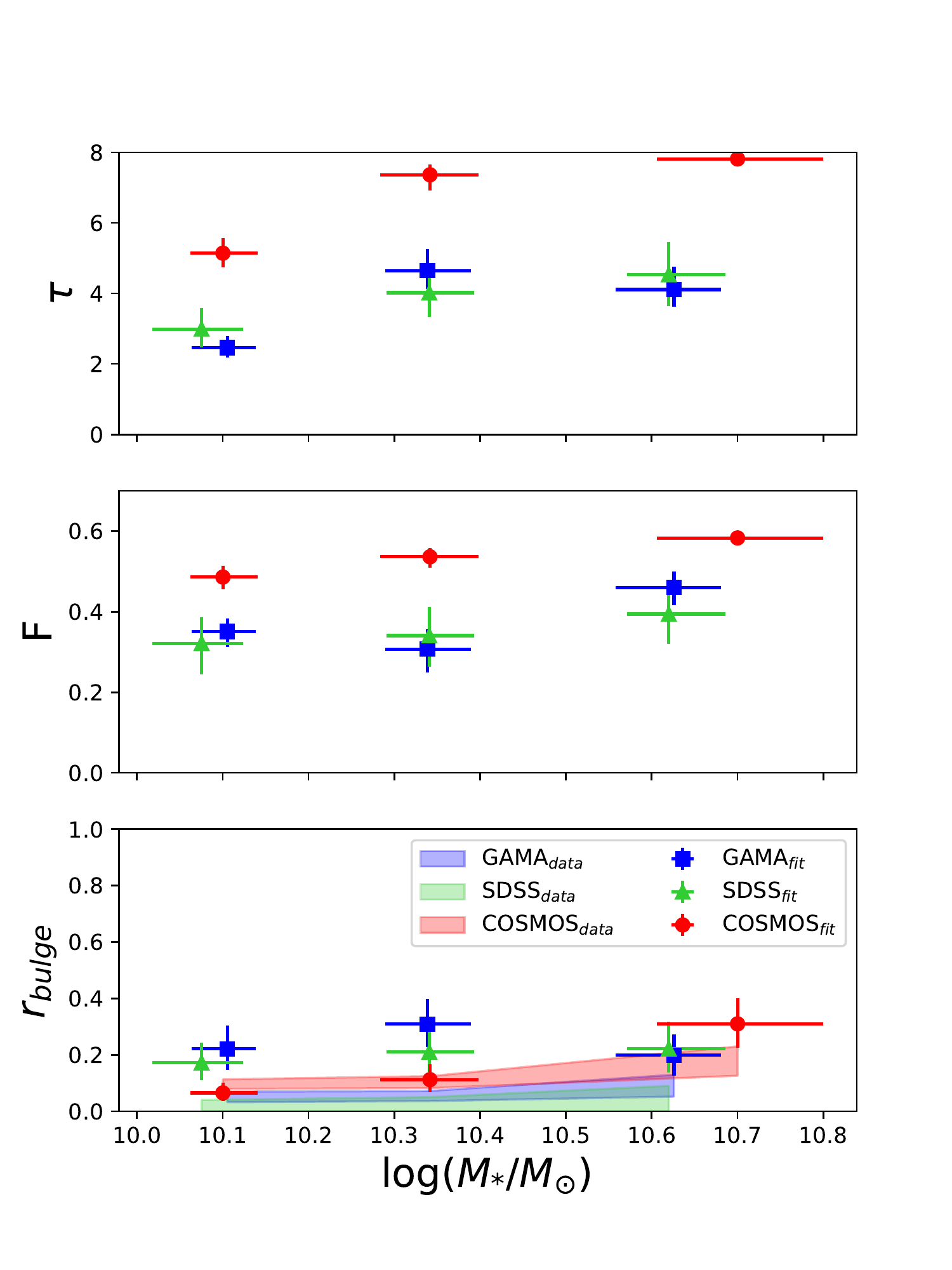}
    \caption{Best-fit values for $\tau_{B}^{f}$, $F$ and $r_{bulge}$ for galaxies in our datasets, divided in bins of $\log(M_{*}[\mathrm{M}_{\odot}])$. We compare the fitted values for $r_{bulge}$ to the average and 32-68th percentile of the bulge-to-total distribution of the selected samples, labeled with the subscript "data".}
    \label{fig:gal_mass}
\end{figure}

The first galaxy property we use for the binning is the $M_{*}$. Various studies have suggested that the $M_{*}$ and the dust mass $M_{dust}$ are positively correlated \citep[e.g.,][]{Groot13, DeVi17, Pas20}, with the $M_{dust}$ - $M_{*}$ ratio depending on redshift. Because $M_{dust}$ can be traced by the $\tau_{B}^{f}$, these results lead us to expect variations in our best-fit parameters as a function of $M_{*}$. Fig. \ref{fig:gal_mass} and Table \ref{tab:gal_mass} show the best-fit parameters in three bins of $M_{*}$: $9.0 \leq \log(M_{*}[\mathrm{M}_{\odot}) \leq 10.2$, $10.2 \leq \log(M_{*}[\mathrm{M}_{\odot}]) \leq 10.5$, and $10.5 \leq \log(M_{*}[\mathrm{M}_{\odot}]) \leq 12.0$.\\

\begin{table}[t]
\centering
\caption{Median $M_{*}$ and the best-fit values for the different datasets in each bin.}
\label{tab:gal_mass}
\resizebox{\columnwidth}{!}{%
\begin{tabular}{ll|l|l|l|l}
\hline\hline
 &  & $\log(M_{*}[\mathrm{M}_{\odot}])_{med}$ & $\tau_{B}^{f}$         & $F$         & $r_{bulge}$    \\ \hline
    &low& $10.08^{+0.06}_{-0.05}$   & $2.99^{+0.59}_{-0.52}$   & $0.32^{+0.08}_{-0.07}$ & $0.18^{+0.07}_{-0.06}$ \\
SDSS&mid& $10.34^{+0.084}_{-0.084}$   & $4.02^{+0.81}_{-0.69}$   &
    $0.34^{+0.07}_{-0.08}$ & $0.21^{+0.07}_{-0.07}$ \\
    &high& $10.65^{+0.13}_{-0.13}$   & $4.53^{+0.93}_{-0.89}$   & $0.39^{+0.06}_{-0.07}$ & $0.22^{+0.10}_{-0.08}$ \\ \hline
    &low& $10.10^{+0.04}_{-0.04}$   & $2.46^{+0.33}_{-0.28}$   & $0.35^{+0.03}_{-0.04}$ & $0.22^{+0.08}_{-0.07}$ \\
GAMA&mid& $10.34^{+0.085}_{-0.085}$   & $4.65^{+0.61}_{-0.51}$   & $0.31^{+0.05}_{-0.05}$ & $0.31^{+0.09}_{-0.08}$ \\
    &high& $10.65^{+0.13}_{-0.13}$   & $4.11^{+0.65}_{-0.49}$   & $0.46^{+0.04}_{-0.04}$ & $0.20^{+0.07}_{-0.07}$ \\ \hline
    &low& $10.10^{+0.04}_{-0.04}$   & $5.14^{+0.42}_{-0.39}$   & $0.49^{+0.03}_{-0.03}$ & $0.07^{+0.04}_{-0.03}$ \\
COSMOS &mid& $10.34^{+0.087}_{-0.087}$   & $7.36^{+0.29}_{-0.45}$   & $0.54^{+0.02}_{-0.03}$ & $0.11^{+0.06}_{-0.04}$ \\
    &high& $10.74^{+0.18}_{-0.18}$   & $7.81^{+0.09}_{-0.12}$   & $0.58^{+0.01}_{-0.01}$ & $0.31^{+0.09}_{-0.08}$\\
    \hline
\end{tabular}%
}
\end{table}

In Fig. \ref{fig:gal_mass} and Table \ref{tab:gal_mass}, we see that $\tau_{B}^{f}$ and $F$ increases with $M_{*}$. The fitted values for COSMOS are consistently higher compared to SDSS and GAMA, but the overall trend is the same. The increased $\tau_{B}^{f}$ in COSMOS relative to the low-z sample could mean that the $M_{dust}$ - $M_{*}$ ratio changes with redshift. Trends with $r_{bulge}$ and $M_{*}$ are inconsistent across the three samples and $r_{bulge}$ is significantly higher than what is constrained from observations in SDSS and GAMA. However, the fitted $r_{bulge}$ values are consistent within 2$\sigma$ with those inferred from our Sérsic index model in the COSMOS sample.  The COSMOS sample also shows a significant trend between $r_{bulge}$ and $M_{*}$, with more massive galaxies being more bulge-dominated as expected (e.g., \citealt{Lang2014}). This could indicate that higher resolution imaging data is required to constrain accurate structural parameters; COSMOS I-band imaging probes and average of $\sim$ 0.7 kpc resolution, compared to SDSS and GAMA g band-imaging at $\sim$ 2 kpc.
The lack of trend might be due to our choice of normalization by the observed magnitude value in the second inclination bin of the optical and NIR bands, where $\tau_{B}^{f}$ and $r_{bulge}$ are the free parameters. These bands are less affected by the attenuation and, therefore, the attenuation - inclination is more shallow, which would mean that the vertical offset would have the most influence on the $r_{bulge}$ and $\tau_{B}^{f}$. Since we cannot estimate the intrinsic emission in these bands and are only left with the inclination dependence, it would decrease the accuracy of these parameters. As the attenuation is better constrained by the inclusion of the FUV and NUV bands, where the effects are strongest and where $\tau_{B}^{f}$ is the only free parameter for the inclination dependent attenuation, we can still trust all our best-fit results of the $\tau_{B}$. Therefore, we do not have to worry that the inconsistency in $r_{bulge}$ will have a large impact on the other results.\\
We note that SDSS and GAMA vary slightly in the best-fit values. The reason for this effect has to do with the difference in the sample size. As our GAMA sample contains fewer galaxies than SDSS, it is more sensitive to detection bias. Our code will indicate whether the samplers of the fits converge and result in a trustworthy fit. As long as SDSS and GAMA have best-fit values within the uncertainties, we are confident in the best-fit results and trends we find for galaxies at redshift $z\sim 0.1$. 

\subsubsection{Stellar mass surface density}\label{Sec:FittingDens}
\begin{figure}[t]
    \centering
    \includegraphics[width = \columnwidth, trim = 0cm 1cm 0cm 0cm clip]{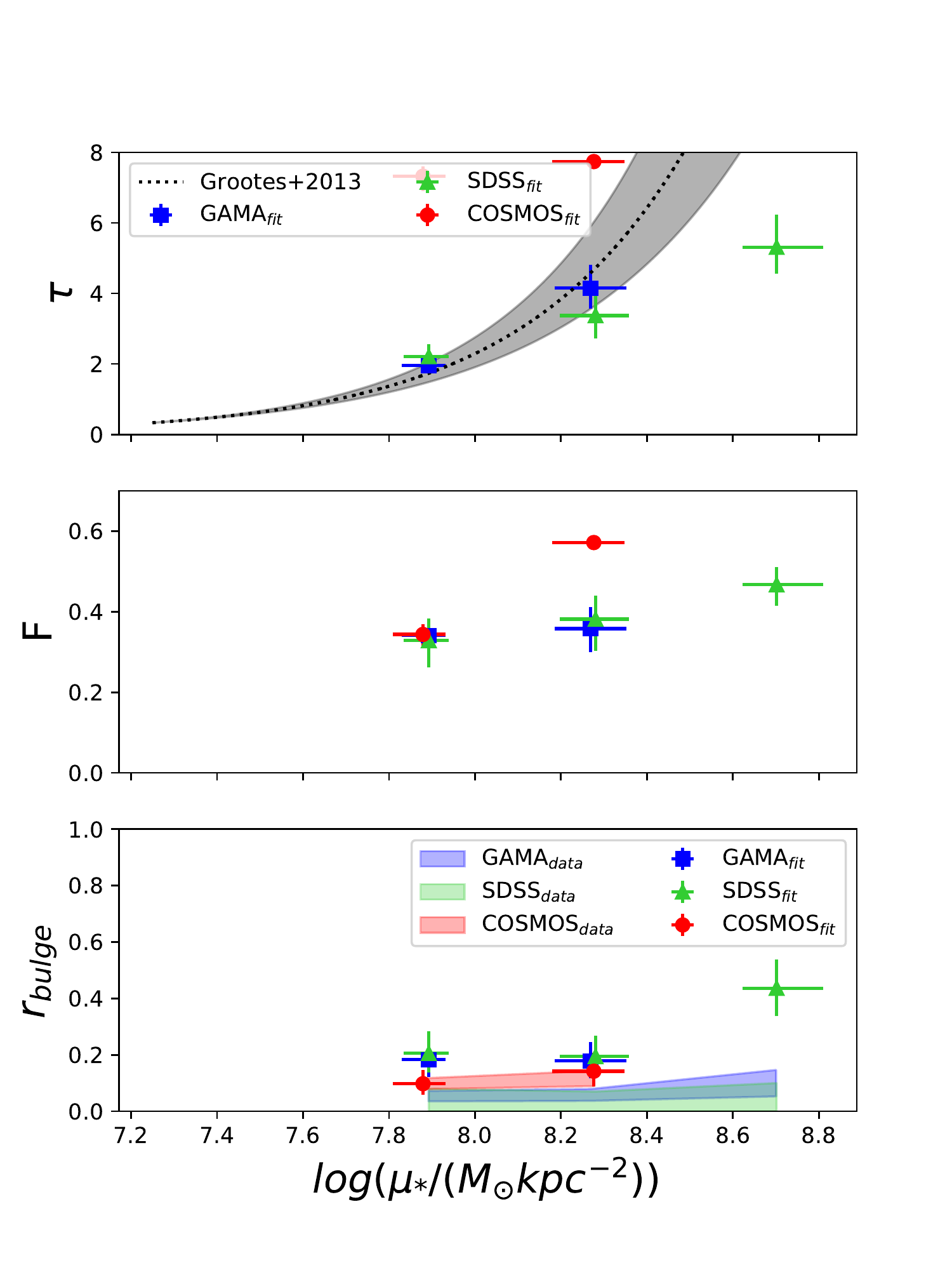}
    \caption{Best-fit values for $\tau_{B}^{f}$, $F$ and $r_{bulge}$ for galaxies in our datasets, divided in bins of $\mu_{*}$. We compare the fitted values for $r_{bulge}$ to the average and 32-68th percentile of the bulge-to-total distribution of the selected samples, labeled with the subscript "data". The results show an increase $\tau_{B}^{f}$ for an increase in $\mu_{*}$, similar to \citet{Groot13}, given by the black dashed line and gray. The GAMA and COSMOS surveys do not contain enough high surface density galaxies to describe a clear magnitude-inclination relation, resulting in the model not obtaining fitting results at $\log(\mu_{*}) \sim 8.7\mathrm{M}_{\odot}\mathrm{kpc}^{-2}$.}
    \label{fig:gal_mu}
\end{figure}

Next, we divide the galaxies into bins of $\mu_{*}$ calculated using Eq. (\ref{eq:dens}) before fitting the model. The fitting results of the three samples are given in Figure \ref{fig:gal_mu} and Table \ref{tab:gal_mu} in bins of $\mu_{*}$: $7.5 \leq \log(\mu_{*}[\mathrm{M}_{\odot}\mathrm{kpc}^{-2}]) \leq 8$, $8.0 \leq \log(\mu_{*}[\mathrm{M}_{\odot}\mathrm{kpc}^{-2}]) \leq 8.5$, and $8.5 \leq \log(\mu_{*}[\mathrm{M}_{\odot}\mathrm{kpc}^{-2}]) \leq 10.0$. The GAMA and COSMOS surveys do not contain enough high surface density galaxies to describe a clear magnitude-inclination relation, resulting in the model not obtaining fitting results.

\begin{table}[t]
\centering
\caption{Median $\mu_{*}$ and the best-fit values for the different datasets in each bin.}
\label{tab:gal_mu}
\resizebox{\columnwidth}{!}{%
\begin{tabular}{ll|l|l|l|l}
\hline\hline
       &      & $\log(\mu_{*,med}$ & $\tau_{B}^{f}$         & $F$         & $r_{bulge}$   \\
              &      & $\quad[\mathrm{M}_{\odot}\mathrm{kpc}^{-2}])$ &         &      &  \\ \hline
       & low  & $7.85^{+0.11}_{-0.11}$   & $2.21^{+0.36}_{-0.38}$   & $0.33^{+0.05}_{-0.07}$ & $0.21^{+0.08}_{-0.07}$ \\
SDSS   & mid  & $8.26^{+0.14}_{-0.14}$   & $3.37^{+0.86}_{-0.65}$   & $0.38^{+0.06}_{-0.08}$ & $0.19^{+0.07}_{-0.07}$ \\
       & high & $8.75^{+0.23}_{-0.23}$   & $5.31^{+0.92}_{-0.76}$   & $0.47^{+0.04}_{-0.05}$ & $0.44^{+0.10}_{-0.10}$ \\ \hline
       & low  & $7.85^{+0.11}_{-0.11}$   & $1.95^{+0.21}_{-0.20}$   & $0.34^{+0.03}_{-0.03}$ & $0.18^{+0.07}_{-0.06}$ \\
GAMA   & mid  & $8.26^{+0.14}_{-0.14}$   & $4.15^{+0.65}_{-0.57}$   & $0.36^{+0.05}_{-0.06}$ & $0.18^{+0.07}_{-0.06}$ \\
       & high & $8.75^{+0.19}_{-0.19}$   & ...   & ... & ... \\ \hline
       & low  & $7.84^{+0.13}_{-0.13}$   & $7.32^{+0.29}_{-0.42}$   & $0.34^{+0.03}_{-0.03}$ & $0.10^{+0.05}_{-0.04}$ \\
COSMOS & mid  & $8.26^{+0.14}_{-0.14}$   & $7.74^{+0.11}_{-0.16}$   & $0.57^{+0.01}_{-0.02}$ & $0.14^{+0.06}_{-0.05}$ \\
       & high & $8.82^{+0.24}_{-0.24}$   & ...   & ... & ...\\
       \hline
\end{tabular}%
}
\end{table}

We see in Figure \ref{fig:gal_mu} that $\tau_{B}^{f}$ increases when $\mu_{*}$ increases, with the increase being similar for SDSS and GAMA. The literature also suggests a positive $\tau - \mu_{*}$ correlation. For example, \citet{Groot13} computed the $\tau_{B}^{f}$ of low-z galaxies based on the dust mass derived from infrared emission and fitted an empirical relation between the $\tau_{B}^{f}$ of the galaxies and their $\mu_{*}$:
\begin{equation}\label{Grootes_t_mu}
    \log(\tau_{B}^{f}) = 1.12(\pm0.11)\cdot\log\left(\frac{\mu_{*}}{\mathrm{M}_{\odot}\mathrm{kpc}^{-2}}\right) - 8.6(\pm0.8).
\end{equation}
This \citet{Groot13} relation is shown in black in Figure \ref{fig:gal_mu} and aligns with the lowest and intermediate bin for SDSS and GAMA, given the uncertainties. The highest bin for SDSS is outside of the \citet{Groot13} curve that suggests a $\tau_{B}^{f}$ beyond the model limits, implying that the T04 model cannot reproduce the attenuation-inclination relation in the highest surface density bin. \\
We also see that $F$ slightly increases with the increase in $\mu_{*}$, with COSMOS having the steepest increase. The increase in $F$ for COSMOS could also be due to the model $\tau_{B}^{f}$ limit. The best-fit $\tau_{B}^{f}$ for the COSMOS sample is close to the maximum value allowed in the T04 model. If there is an increase in attenuation with $\mu_{*}$, but the model is already at the maximum allowed $\tau_{B}^{f}$, the lack of modeled attenuation will be compensated by artificially having a higher $F$. We could try to extrapolate the model for $\tau_{B}^{f} > 8$, but the current model assumptions might not hold at higher $\tau_{B}^{f}$. The T04 model is not calibrated for higher $\tau_{B}^{f}$ due to the increased likelihood of selecting starburst galaxies with irregular structures that are not expected to follow the same attenuation-inclination relation.

\subsubsection{Measures of star-formation activity: SFR, sSFR, and dMS}\label{Sec:FittingSFR}
\begin{figure}[t]
    \centering
    \includegraphics[width = \columnwidth]{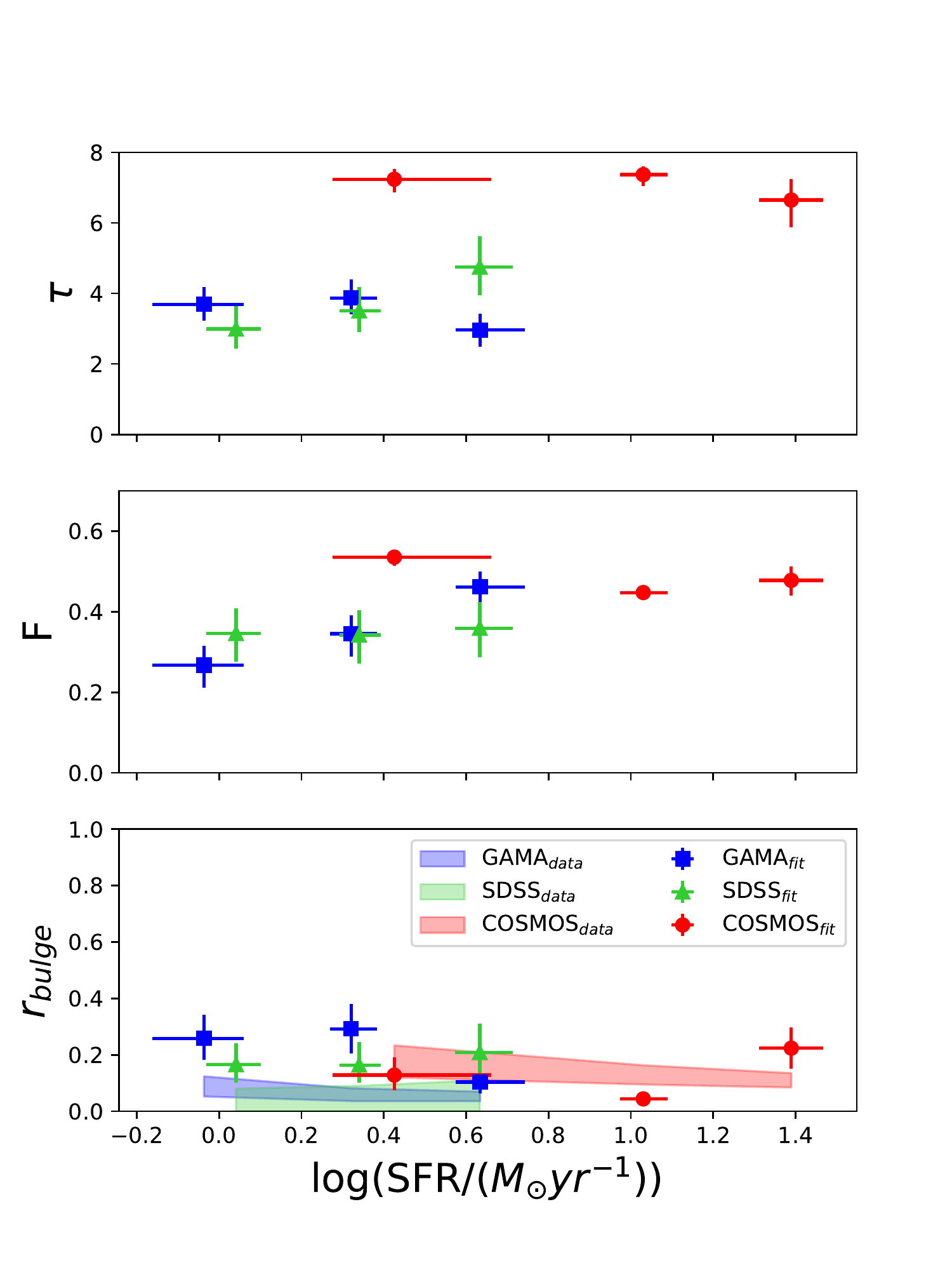}
    \caption{Best-fit values for $\tau_{B}^{f}$, $F$ and $r_{bulge}$ for galaxies in our datasets, divided in bins of $SFR$. We compare the fitted values for $r_{bulge}$ to the average and 32-68th percentile of the bulge-to-total distribution of the selected samples, labeled with the subscript "data". The best-fit parameters do not show a consistent correlation overall.}
    \label{fig:gal_SFR}
\end{figure}

\begin{figure}[t]
    \centering
    \includegraphics[width = \columnwidth]{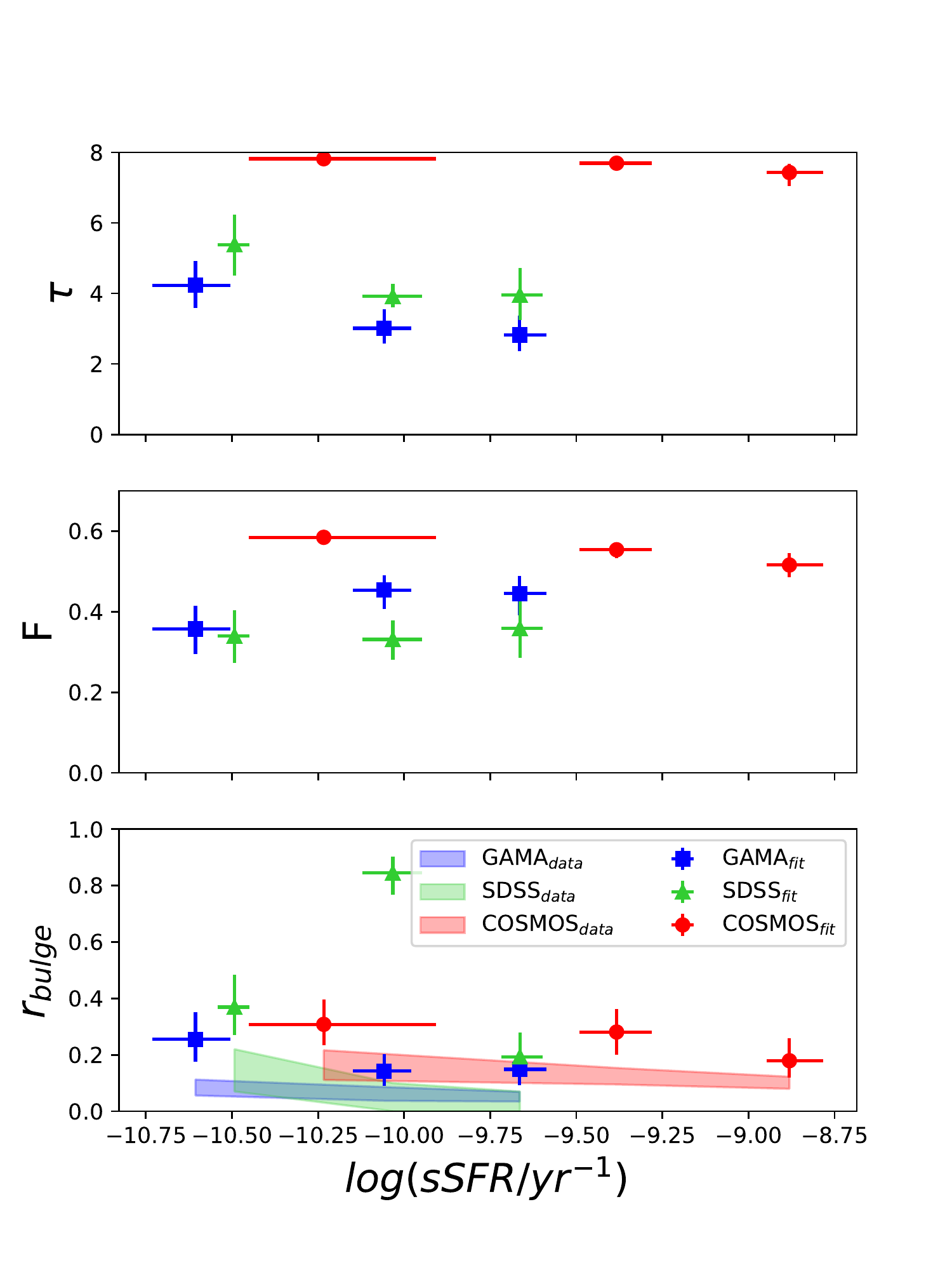}
    \caption{Best-fit values for $\tau_{B}^{f}$, $F$ and $r_{bulge}$ for galaxies in our datasets, divided in bins of $sSFR$. We compare the fitted values for $r_{bulge}$ to the average and 32-68th percentile of the bulge-to-total distribution of the selected samples, labeled as data.}
    \label{fig:gal_sSFR}
\end{figure}

The T04 model parameter $F$ traces the star-forming regions of a galaxy. If the $SFR$ or $sSFR$, changes, we might expect changes in these star-forming regions and, therefore, their attenuation.
Figure \ref{fig:gal_SFR} and Table \ref{tab:gal_SFR} illustrate the best-fit model parameters in our three datasets for different bins in SFR retrieved as explained in Section \ref{Sec:Data}. It is well known that out to redshift $z \leq$ 2 the $SFR$ and $sSFR$ scales approximately with $(1+z)^{3}$ \citep[e.g.,][]{Sargent2012, Il15, Tasc15, Pope19}. We scale the COSMOS $SFR$ and $sSFR$ by a factor of $(1+z)^{3}$ when binning in $SFR$ and $sSFR$ to ease comparison with the SDSS and GAMA samples. The ranges of the (scaled) $SFR$ bins are $-\infty \leq \log(\mathrm{SFR}[\mathrm{M}_{\odot} \mathrm{yr}^{-1}]) \leq 0.18$, $0.18 \leq \log(\mathrm{SFR}[\mathrm{M}_{\odot} \mathrm{yr}^{-1}]) \leq 0.48$, and $0.48 \leq \log(\mathrm{SFR}[\mathrm{M}_{\odot} \mathrm{yr}^{-1}]) \leq 1$. The ranges of the sSFR bins are: $-15. \leq \log(\mathrm{sSFR}[\mathrm{yr}^{-1}]) \leq -10.4$, $-10.4 \leq \log(\mathrm{sSFR}[\mathrm{yr}^{-1}]) \leq -9.8$, and $-9.8 \leq \log(\mathrm{sSFR}[\mathrm{yr}^{-1}]) \leq -9.0$. Figure \ref{fig:gal_SFR} and Table \ref{tab:gal_SFR}, and Figure \ref{fig:gal_sSFR} and Table \ref{tab:gal_sSFR} show the results for $SFR$ and $sSFR$ respectively without the redshift-scaling.

\begin{table}[t]
\centering
\caption{Median star-formation rate $SFR_{med}$ and the best-fit values for the different datasets in each bin.}
\label{tab:gal_SFR}
\resizebox{\columnwidth}{!}{%
\begin{tabular}{ll|l|l|l|l}
\hline\hline
       &      & $\log(\mathrm{SFR}_{med}$ & $\tau_{B}^{f}$         & $F$         & $r_{bulge}$    \\ 
       &    & $\quad[\mathrm{M}_{\odot} \mathrm{yr}^{-1}])$& & & \\ \hline
       & low  & $0.02^{+0.13}_{-0.13}$   & $2.99^{+0.64}_{-0.55}$   & $0.35^{+0.06}_{-0.07}$ & $0.17^{+0.08}_{-0.06}$ \\
SDSS   & mid  & $0.34^{+0.09}_{-0.09}$   & $3.50^{+0.67}_{-0.60}$   & $0.34^{+0.06}_{-0.07}$ & $0.16^{+0.08}_{-0.06}$ \\
       & high & $0.68^{+0.14}_{-0.14}$   & $4.75^{+0.88}_{-0.80}$   & $0.36^{+0.07}_{-0.07}$ & $0.21^{+0.10}_{-0.07}$ \\ \hline
       & low  & $-0.02^{+0.21}_{-0.21}$   & $3.69^{+0.49}_{-0.46}$   & $0.27^{+0.05}_{-0.06}$ & $0.26^{+0.08}_{-0.08}$ \\
GAMA   & mid  & $0.33^{+0.09}_{-0.09}$   & $3.87^{+0.53}_{-0.47}$   & $0.35^{+0.05}_{-0.06}$ & $0.29^{+0.08}_{-0.08}$ \\
       & high & $0.69^{+0.16}_{-0.16}$   & $2.96^{+0.46}_{-0.47}$   & $0.46^{+0.04}_{-0.04}$ & $0.10^{+0.05}_{-0.04}$ \\ \hline
       & low  & $0.50^{+0.29}_{-0.29}$   & $7.24^{+0.29}_{-0.37}$   & $0.54^{+0.02}_{-0.02}$ & $0.13^{+0.06}_{-0.05}$ \\
COSMOS & mid  & $1.04^{+0.09}_{-0.09}$   & $7.37^{+0.24}_{-0.32}$   & $0.45^{+0.02}_{-0.02}$ & $0.04^{+0.02}_{-0.02}$ \\
       & high & $1.41^{+0.14}_{-0.14}$   & $6.64^{+0.60}_{-0.77}$   & $0.48^{+0.03}_{-0.04}$ & $0.22^{+0.07}_{-0.07}$\\
       \hline
\end{tabular}%
}
\end{table}

The best-fit results indicate that $F$ has inconsistent trends in the three samples: it remains constant with $SFR$ for SDSS, it slightly increases for GAMA, and it decreases for COSMOS. Figure \ref{fig:gal_SFR} also shows that $\tau_{B}^{f}$ slightly increases with $SFR$ for SDSS, but the GAMA sample does not show this trend, indicating that any trend of our best-fit parameters with $SFR$ is not robust. The inconsistency is not a result of using different SED fitting methods to derive the $SFR$ because the samples have similar wavelength coverage and similar assumptions where made by the different studies, each reporting a $SFR$ averaged over the last 100 Myr. Therefore, the inconsistencies are driven by the uncertainty in the best-fit results. The trends are different when we fit in bins of $sSFR$; Table \ref{tab:gal_sSFR} and Figure \ref{fig:gal_sSFR} show that there is a negative correlation between $sSFR$ and $\tau_{B}^{f}$, but trends in other parameters remain unclear. The trend with $\tau_{B}^{f}$ and $sSFR$ is most likely driven by the $M_{*}$ because the high $sSFR$ bin could be dominated by low-mass galaxies.

\begin{table}[t]
\centering
\caption{Median specific star-formation rate $sSFR_{med}$ and the best-fit values for the different datasets in each bin.}
\label{tab:gal_sSFR}
\resizebox{\columnwidth}{!}{%
\begin{tabular}{ll|l|l|l|l}
\hline\hline
       &      & $\log(\mathrm{sSFR}_{med}$ & $\tau_{B}^{f}$         & $F$         & $r_{bulge}$    \\ 
       &    & $\quad[\mathrm{yr}^{-1}])$& & & \\ \hline
       & low  & $-10.5^{+0.04}_{-0.05}$   & $5.38^{+0.86}_{-0.87}$   & $0.34^{+0.06}_{-0.07}$ & $0.37^{+0.11}_{-0.10}$ \\
SDSS   & mid  & $-10.0^{+0.08}_{-0.08}$   & $3.91^{+0.36}_{-0.32}$   & $0.33^{+0.05}_{-0.05}$ & $0.84^{+0.06}_{-0.08}$ \\
       & high & $-9.7^{+0.07}_{-0.06}$   & $3.79^{+0.86}_{-0.76}$   & $0.37^{+0.07}_{-0.08}$ & $0.16^{+0.08}_{-0.06}$ \\ \hline
       & low  & $-10.6^{+0.10}_{-0.13}$   & $5.88^{+0.79}_{-0.81}$   & $0.34^{+0.06}_{-0.07}$ & $0.34^{+0.10}_{-0.09}$ \\
GAMA   & mid  & $-10.1^{+0.08}_{-0.09}$   & $5.05^{+0.82}_{-0.83}$   & $0.35^{+0.06}_{-0.08}$ & $0.20^{+0.09}_{-0.07}$ \\
       & high & $-9.65^{+0.07}_{-0.06}$   & $5.17^{+0.82}_{-0.77}$   & $0.34^{+0.07}_{-0.07}$ & $0.22^{+0.08}_{-0.08}$ \\ \hline
       & low  & $-10.23^{+0.33}_{-0.22}$   & $7.82^{+0.08}_{-0.11}$   & $0.58^{+0.01}_{-0.01}$ & $0.31^{+0.09}_{-0.07}$ \\
COSMOS & mid  & $-8.88^{+0.09}_{-0.09}$   & $7.43^{+0.25}_{-0.38}$   & $0.52^{+0.03}_{-0.03}$ & $0.18^{+0.08}_{-0.06}$ \\
       & high & $1.41^{+0.14}_{-0.14}$   & $6.64^{+0.60}_{-0.77}$   & $0.48^{+0.03}_{-0.04}$ & $0.22^{+0.07}_{-0.07}$\\
       \hline
\end{tabular}%
}
\end{table}

\begin{figure}[t]
    \centering
    \includegraphics[width = \columnwidth]{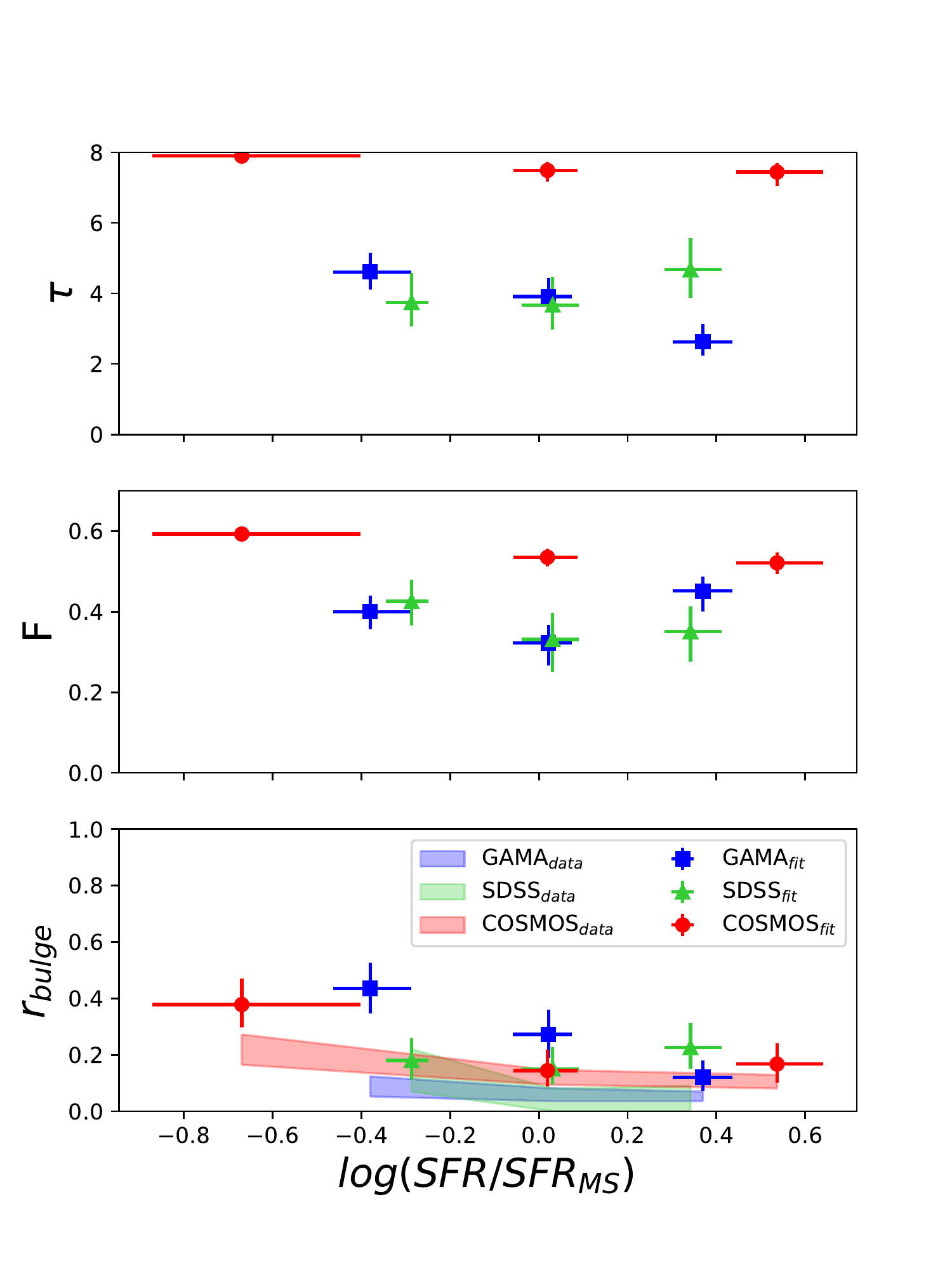}
    \caption{Best-fit values for $\tau_{B}^{f}$, $F$ and $r_{bulge}$ for galaxies in our datasets, divided in bins of dMS. We compare the fitted values for $r_{bulge}$ to the average and 32-68th percentile of the bulge-to-total distribution of the selected samples, labeled as data.}
    \label{fig:gal_dMS}
\end{figure}

As the best-fit results still show varying trends between the samples, we investigate what might drive the observed variations by using other parameters that depend on the $SFR$. The first parameter we test is the star-formation main-sequence offset $dMS$. We define the star-formation main sequence using the relation in \citet{Les18}:
\begin{equation}\label{eq:MS_SFR}
\log\left(\frac{SFR_{MS}}{M_{\odot \mathrm{yr}^{-1}}}\right) = 0.816\log\left(\frac{M_{*}}{\mathrm{M}_{\odot}}\right) - 8.248 + 3\log(1+z),
\end{equation}
with $SFR_{MS}$ the star-formation rate of a main-sequence galaxy, and $z$ the corresponding redshift. 
The star-formation main sequence offset or the logarithmic difference between the measured $SFR$ and the $SFR$ derived from the star-formation main-sequence relation is given as:
\begin{equation}\label{eq:dMS}
    dMS = \log(SFR) - \log(SFR_{MS}).
\end{equation}
Similar to our results for $SFR$, Fig. \ref{fig:gal_dMS} shows no clear trends of $\tau_{B}^{f}$ with $dMS$ and reveals no clear trends with $F$.

\begin{table}[t]
\centering
\caption{Median star-formation main-sequence offset $dMS_{med}$ and the best-fit values for the different datasets in each bin.}
\label{tab:gal_dMS}
\resizebox{\columnwidth}{!}{%
\begin{tabular}{ll|l|l|l|l}
\hline\hline
       &      & $\log(\mathrm{dMS}_{med}$ & $\tau_{B}^{f}$         & $F$         & $r_{bulge}$    \\ 
       &    & $\quad[\mathrm{dex}])$& & & \\ \hline
       & low  & $-0.29^{+0.06}_{-0.04}$   & $3.75^{+0.83}_{-0.68}$& $0.43^{+0.05}_{-0.06}$& $0.18^{+0.08}_{-0.07}$ \\
SDSS   & mid  & $0.03^{+0.07}_{-0.06}$   & $3.67^{+0.80}_{-0.69}$ & $0.33^{+0.07}_{-0.08}$ & $0.15^{+0.08}_{-0.06}$ \\
       & high & $0.34^{+0.06}_{-0.07}$   & $4.67^{+0.90}_{-0.79}$& $0.35^{+0.06}_{-0.08}$& $0.23^{+0.09}_{-0.07}$ \\ \hline
       & low  & $-0.38^{+0.08}_{-0.09}$   & $3.88^{+0.66}_{-0.52}$& $0.34^{+0.05}_{-0.06}$ & $0.22^{+0.08}_{-0.07}$ \\
GAMA   & mid  & $0.02^{+0.08}_{-0.05}$   & $3.91^{+0.53}_{-0.46}$ & $0.32^{+0.05}_{-0.05}$ &$0.28^{+0.08}_{-0.08}$ \\
       & high & $0.37^{+0.07}_{-0.07}$   & $2.11^{+0.41}_{-0.33}$ & $0.48^{+0.03}_{-0.04}$ & $0.10^{+0.05}_{-0.04}$ \\ \hline
       & low  & $-0.67^{+0.20}_{-0.27}$   & $7.90^{+0.05}_{-0.07}$ & $0.59^{+0.01}_{-0.01}$ & $0.38^{+0.09}_{-0.08}$ \\
COSMOS & mid  & $0.02^{+0.08}_{-0.07}$   & $7.49^{+0.23}_{-0.32}$ & $0.54^{+0.02}_{-0.02}$ & $0.14^{+0.07}_{-0.06}$ \\
       & high & $0.54^{+0.09}_{-0.10}$   & $7.44^{+0.25}_{-0.40}$ & $0.52^{+0.03}_{-0.03}$ & $0.17^{+0.07}_{-0.07}$\\
       \hline
\end{tabular}%
}
\end{table}

However, our trends of $\tau_{B}^{f}$ and $F$ with both $SFR$ and $sSFR$ are highly dependent on the assumed intrinsic UV emission. We discuss this further in Appendix \ref{Ap:IntrUV} and the implications of the lack of correlation between $F$ and $SFR$ in Section \ref{Sec:DiscussionGalProp}. 

\subsection{Star-formation rate surface density}
\begin{figure}[t]
    \centering
    \includegraphics[width = \columnwidth]{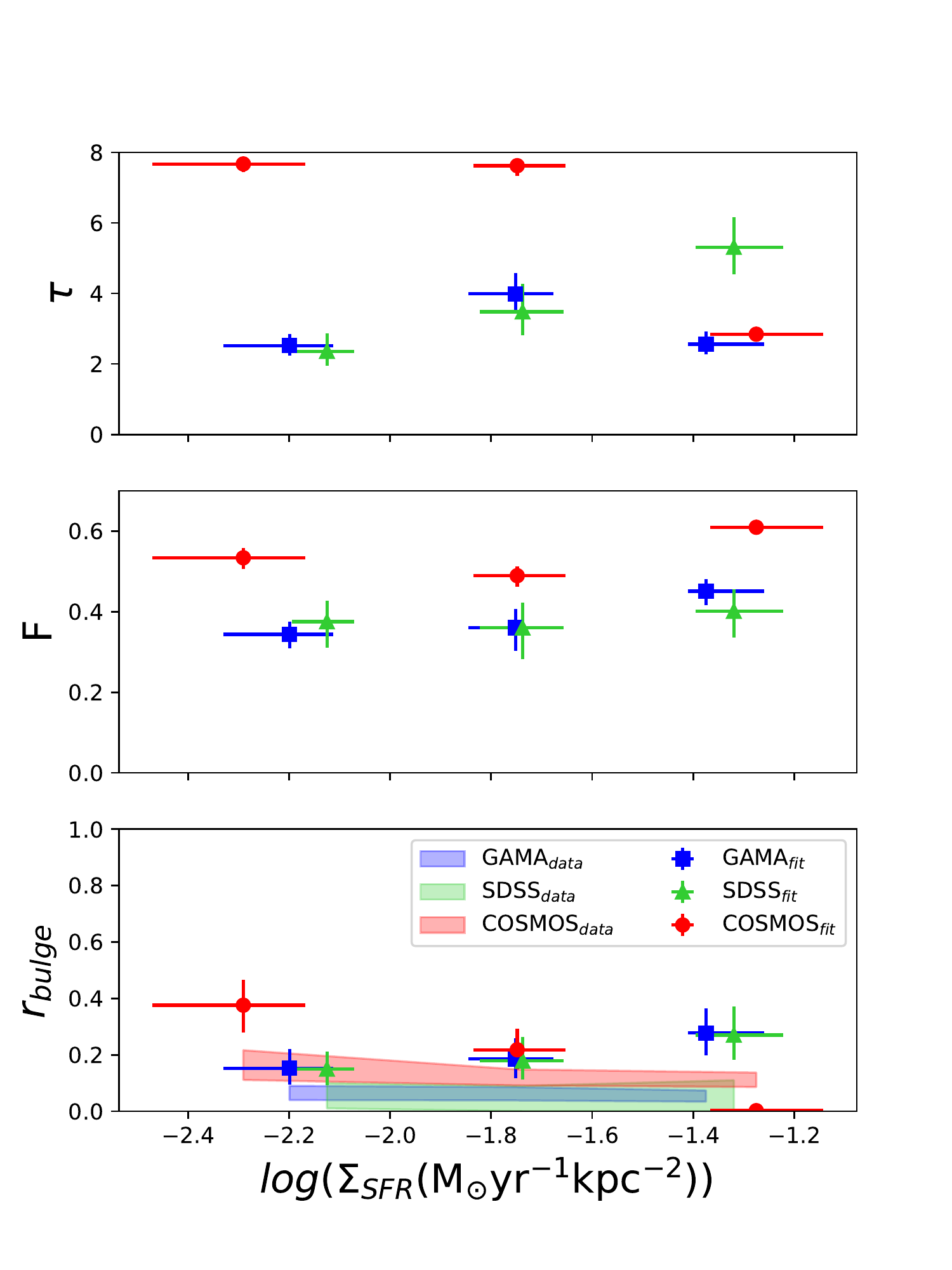}
    \caption{Best-fit values for $\tau_{B}^{f}$, $F$ and $r_{bulge}$ for galaxies in our datasets, divided in bins of $\Sigma_{SFR}$. We compare the fitted values for $r_{bulge}$ to the average and 32-68th percentile of the bulge-to-total distribution of the selected samples, labeled as data.}
    \label{fig:gal_SFRD}
\end{figure}

An important factor for star-formation is the fuel or the molecular gas. As our data do not directly measure the molecular gas, we use the star-formation rate surface density $\Sigma_{SFR}$ as a probe for the molecular gas mass surface density \citep{Schmidt59, Kennicutt98, Leroy2008, Bigi08}. Table \ref{tab:gal_SFRD} and Figure \ref{fig:gal_SFRD} show again that there is no clear correlation between $\Sigma_{SFR}$ and any of the best-fit parameters. We note that all results showing no relation between the fitted parameters and the binned galaxy property involve the $SFR$. We discuss what this means in Section \ref{Sec:DiscussionGalProp}. 

\begin{table}[t]
\centering
\caption{Median star-formation rate surface density $\Sigma_{SFR, med}$ and the best-fit values for the different datasets in each bin.}
\label{tab:gal_SFRD}
\resizebox{\columnwidth}{!}{%
\begin{tabular}{ll|l|l|l|l}
\hline\hline
&      & $\log(\mathrm{\Sigma_{SFR, med}}$ & $\tau_{B}^{f}$         & $F$         & $r_{bulge}$    \\ 
       &    & $\quad[\mathrm{M}_{\odot}\mathrm{yr}^{-1}\mathrm{kpc}^{-2}])$& & & \\ \hline
       & low  & $-2.12^{+0.05}_{-0.07}$   & $2.36^{+0.51}_{-0.42}$ & $0.38^{+0.05}_{-0.06}$ & $0.15^{+0.06}_{-0.06}$ \\
SDSS   & mid  & $-1.74^{+0.08}_{-0.09}$   & $3.48^{+0.79}_{-0.66}$ & $0.36^{+0.06}_{-0.08}$ & $0.18^{+0.08}_{-0.07}$ \\
       & high & $-1.32^{+0.10}_{-0.08}$   & $5.31^{+0.85}_{-0.76}$ & $0.40^{+0.05}_{-0.07}$ & $0.27^{+0.10}_{-0.09}$ \\ \hline
       & low  & $-2.20^{+0.09}_{-0.13}$   & $2.52^{+0.33}_{-0.29}$ & $0.34^{+0.03}_{-0.03}$ & $0.15^{+0.07}_{-0.06}$ \\
GAMA   & mid  & $-1.75^{+0.08}_{-0.09}$   & $3.99^{+0.60}_{-0.56}$ & $0.36^{+0.05}_{-0.06}$ & $0.19^{+0.07}_{-0.07}$ \\
       & high & $-1.37^{+0.12}_{-0.04}$   & $2.56^{+0.36}_{-0.30} $ & $0.45^{+0.03}_{-0.04}$ & $0.28^{+0.09}_{-0.08}$ \\ \hline
       & low  & $-2.29^{+0.12}_{-0.18}$   & $7.67^{+0.16}_{-0.23}$ & $0.53^{+0.02}_{-0.03}$ & $0.38^{+0.09}_{-0.10}$ \\
COSMOS & mid  & $-1.75^{+0.10}_{-0.09}$   & $7.62^{+0.17}_{-0.29} $ & $0.49^{+0.02}_{-0.03}$ & $0.22^{+0.07}_{-0.07}$ \\
       & high & $-1.28^{+0.13}_{-0.09}$   & $2.84^{+0.07}_{-0.07}$ & $0.61^{+0.01}_{-0.01}$ & $0.01^{+0.01}_{-0.01}$\\
       \hline
\end{tabular}%
}
\end{table}

\subsection{Balmer lines}\label{Sec:FittingBalmer}
The T04 model also allows us to describe relations between the observed $H\alpha$/$H\beta$ ratio and the inclination of the galaxy. These ionized emission lines are assumed to come purely from our dust-enshrouded star-forming regions and also suffer attenuation from both the thin and thick disk. In the T04 model, Hydrogen gas is ionized within HII regions, causing the Balmer recombination lines. Then the Balmer lines are either fully attenuated by optically thick fragments of the cloud or completely escape from the birth cloud unattenuated, similar to how the T04 model treats the attenuation of escaping stars (Section \ref{Sec:Methods}, Eq. (\ref{eq:T_UV})). $H\beta$ is more likely to be scattered and absorbed by dust particles in the disk components, increasing the $H\alpha/H\beta$ ratio with inclination. T04 modeled the ratio by using the radiative transfer predictions for the thin stellar disk at the wavelengths corresponding to the $H\alpha$ and $H\beta$ emission lines and fitted the ratio with a polynomial function similar to Eq. (\ref{Tuffs_model_power}). In this model, the ratio is independent of $F$ as a consequence of the assumed optical thickness and structure of the star-forming clouds. These assumptions result in the following function:
\begin{equation}\label{eq:H_ratio_T}
\begin{split}
    H\alpha/H\beta(\tau_{B}^{f}, i)
        &= H\alpha/H\beta_{model}(\tau_{B}^{f}, i)\\
        &= \Sigma_{j = 0}^{k} a_{j}(1 - \cos(i))^{j} .\\
\end{split}
\end{equation}

\begin{figure}[t]
    \centering
    \includegraphics[trim=2cm 2cm 3cm 3cm,clip, width = \columnwidth]{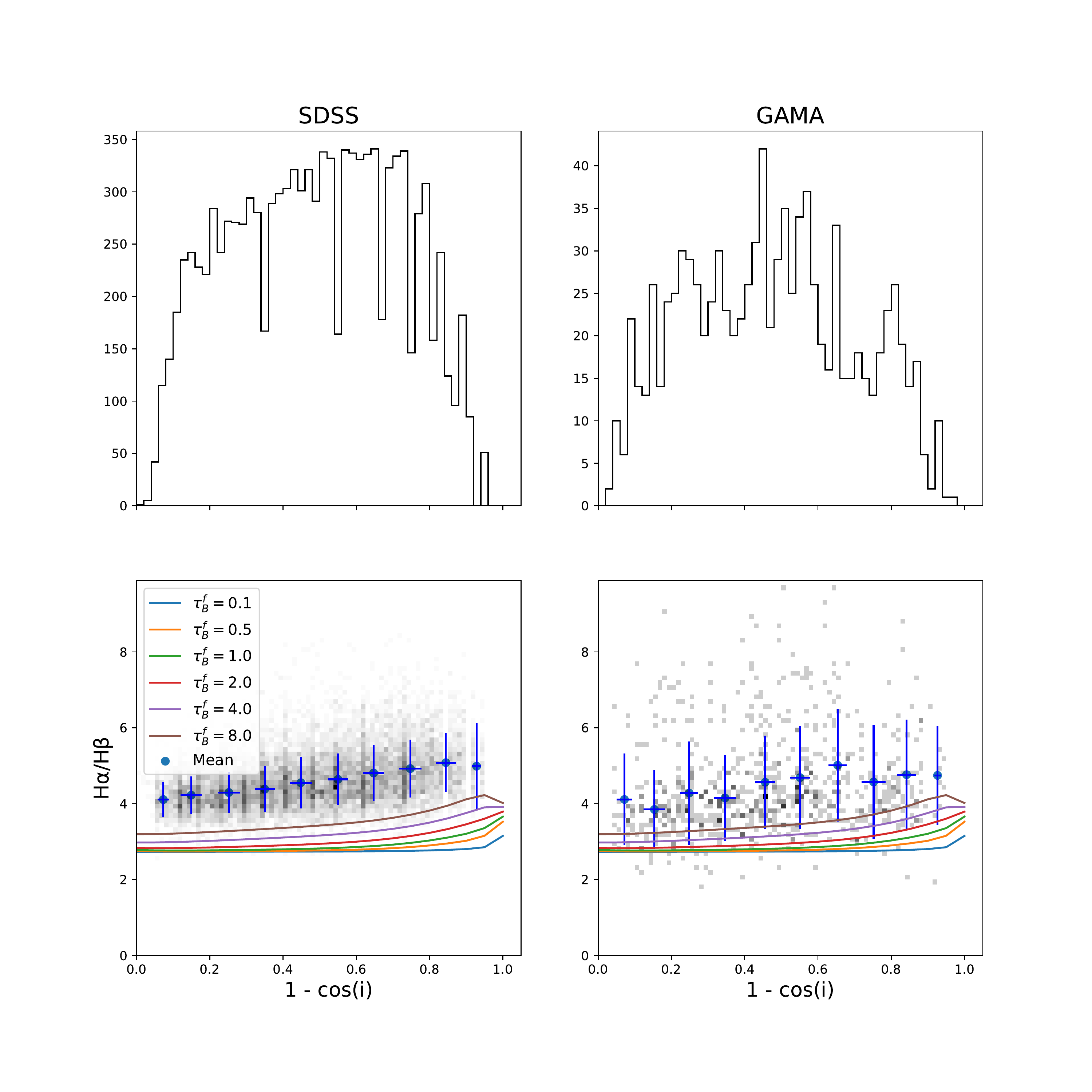}
    \caption{$H\alpha/H\beta$ ratio for the SDSS and GAMA sample for galaxies with varying inclination. The top two panels show the inclination distribution, and the bottom two panels show the $H\alpha/H\beta$ - inclination relation for the selected star-forming sample and the $H\alpha/H\beta$ ratio - inclination relation from the T04 model with varying $\tau_{B}^{f}$. We can see that none of the T04 models can describe the observed trend.}
    \label{fig:HaHb}
\end{figure}

\begin{figure}[t]
    \centering
    \includegraphics[scale = 0.5]{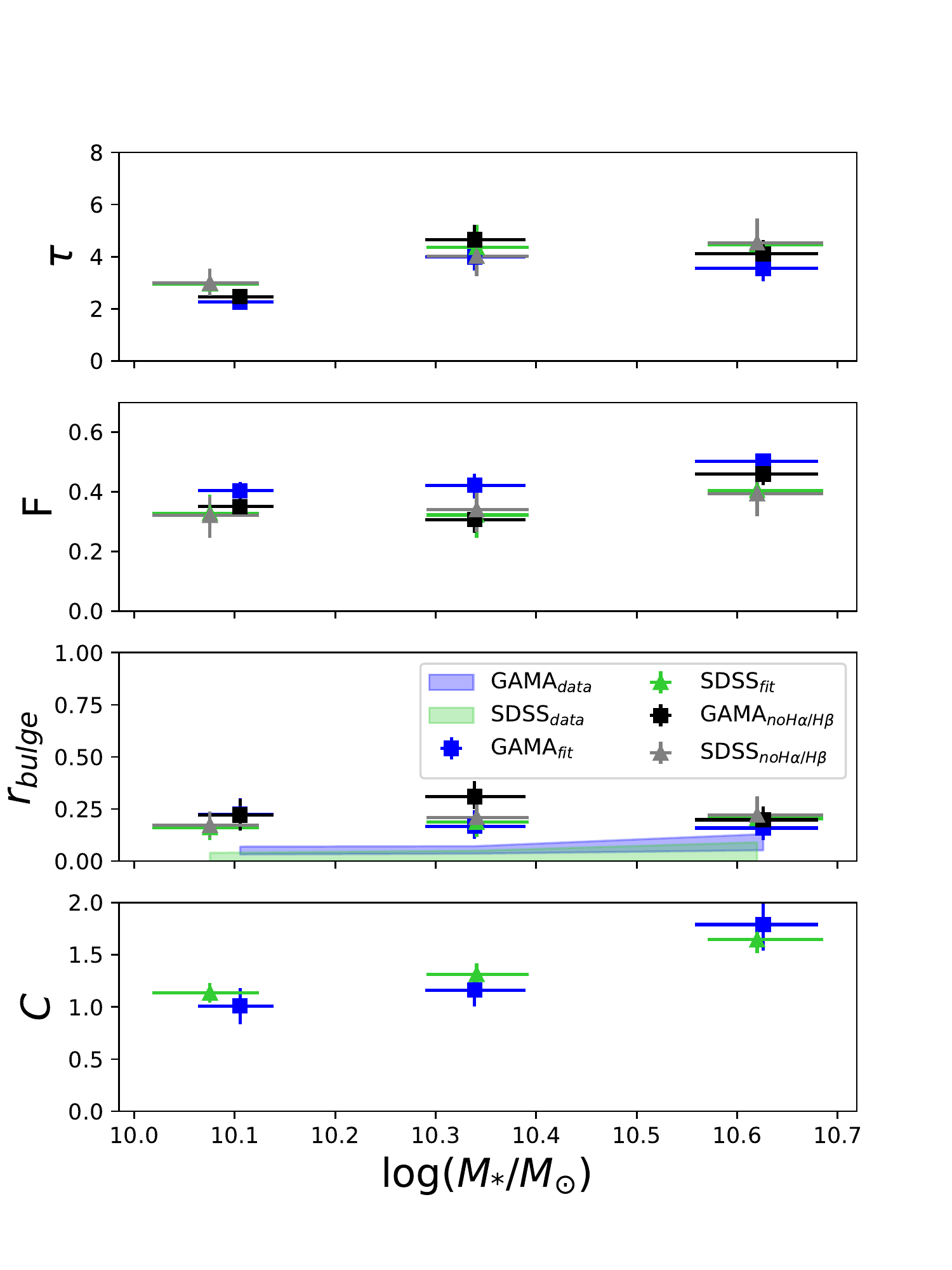}
    \caption{Best-fit values for $\tau_{B}^{f}$, $F$, $r_{bulge}$, and $C$ for galaxies in our datasets, divided in bins of mass $\log(M_{*}[\mathrm{M}_{\odot}])$.}
    \label{fig:gal_H}
\end{figure}

However, the existing T04 model could not reproduce the high ratios observed in our SDSS and GAMA samples, shown in Figure \ref{fig:HaHb}. This offset could mean that there is additional dust surrounding or inside the HII region influencing the transitions on such a small spatial scale that it does not affect the UV emission \citep{Yip10}. We use our fitting regime using the UV, optical, NIR, and $H\alpha$/$H\beta$ data for GAMA and SDSS and add a parameter $C$, describing the offset in the $H\alpha$/$H\beta$-inclination relation compared to the model:
\begin{equation}\label{eq:H_ratio_T2}
    H\alpha/H\beta(\tau_{B}^{f}, i)= \Sigma_{j = 0}^{k} a_{j}(1 - \cos(i))^{j} + C.
\end{equation}
After investigating the fitting of $C$ and how it varies with galaxy properties, we found consistent inter-sample trends for $\tau_{B}^{f}$, $F$, and $C$ with variation in $M_{*}$. We show the results in Table \ref{tb:gal_H} and Figure \ref{fig:gal_H}.

\begin{table}[t]
\centering
\caption{Median mass $M_{*,med}$ and the best-fit values for the different datasets in each bin including the $H\alpha/H\beta$-offset $C$.}
\label{tb:gal_H}
\resizebox{\columnwidth}{!}{%
\begin{tabular}{ll|l|l|l|l|l}
\hline\hline
 &  & $\log(M_{*}[\mathrm{M}_{\odot}])_{med}$ & $\tau_{B}^{f}$         & $F$         & $r_{bulge}$    & $C$\\\hline
    &low& $10.08^{+0.06}_{-0.05}$   & $2.62^{+0.54}_{-0.44}$   & $0.37^{+0.07}_{-0.06}$ & $0.15^{+0.07}_{-0.05}$ &$1.15^{0.09}_{0.08}$\\
SDSS&mid& $10.34^{+0.084}_{-0.084}$   & $3.97^{+0.74}_{-0.73}$   &
    $0.35^{+0.06}_{-0.08}$ & $0.18^{+0.07}_{-0.06}$ &$1.35^{0.12}_{0.12}$\\
    &high& $10.65^{+0.13}_{-0.13}$   & $3.99^{+0.84}_{-0.77}$   & $0.43^{+0.06}_{-0.07}$ & $0.19^{+0.08}_{-0.06}$ &$1.69^{0.12}_{0.14}$\\ \hline
    &low& $10.10^{+0.04}_{-0.04}$   & $2.49^{+0.52}_{-0.41}$   & $0.33^{+0.05}_{-0.06}$ & $0.25^{+0.09}_{-0.08}$ &$0.91^{0.18}_{0.18}$\\
GAMA&mid& $10.34^{+0.085}_{-0.085}$   & $4.55^{+0.64}_{-0.55}$   & $0.29^{+0.07}_{-0.07}$ & $0.34^{+0.08}_{-0.08}$ &$1.19^{0.17}_{0.18}$\\
    &high& $10.65^{+0.13}_{-0.13}$   & $4.35^{+0.52}_{-0.42}$   & $0.45^{+0.04}_{-0.05}$ & $0.18^{+0.05}_{-0.03}$ &$1.94^{0.22}_{0.24}$\\ \hline
\end{tabular}
}
\end{table}

In our new model shown in Table \ref{tb:gal_H} and Figure \ref{fig:gal_H}, we see similar trends with Table \ref{tab:gal_mass} and Figure \ref{fig:gal_mass}. The $\tau_{B}^{f}$ increases over bins of $M_{*}$, whereas the $F$ remains constant. Our results for $\tau_{B}^{f}$, $F$, and $r_{bulge}$ are consistent with what we found in Section \ref{Sec:FittingMass}, meaning the addition of the $H\alpha/H\beta$ ratio provides no additional constraints. We find that $C$ increases with the $M_{*}$ of the galaxy. This result means that the star-forming regions have a component not described in the T04 models, with a wavelength-dependent attenuation influencing the Balmer ratio. This component varies with the global stellar masses.

\section{Discussion}\label{Sec:Discussion}
With the fitting of the T04 parameters in bins of different galaxy properties, we now discuss the results of Section \ref{Sec:FittingBins} to gain insight into how the dust and galaxy properties are linked. In this section, we discuss the correlations found in more detail and describe their implications for dust formation and how different tracers could influence these implications.

\subsection{Dependence of model parameters on galaxy properties}\label{Sec:DiscussionGalProp}
We find an increase in $\tau_{B}^{f}$ with an increase in $M_{*}$ and $\mu_{*}$. \citet{Pop11} shows that $\tau_{B}^{f}$ can be linked to $M_{dust}$ based on the scale length of the stellar disk, the geometry of the dust, and dust properties (see \citealt{Pop11} Sec. 2.9, Eq. (44) for more information). Therefore, our results could imply that galaxies with higher $M_{*}$ have higher $M_{dust}$ \citep[e.g.,][]{Liu2019,Magn20,Koko21} hinting at the paired production of stars and dust \citep[e.g.,][]{DeVi17, Pas20}. The correlation with $\mu_{*}$  is similar to the \citet{Groot13} relation for low-$z$ galaxies but deviates for high-$\mu_{*}$ galaxies and high-$z$ because the T04 model is unable to cover the high $\tau_{B}$ estimated by the \citet{Groot13} relation for these bins.
The variation of $M_{dust}$ with galaxy properties is often traced using the ratio of the dust mass with the stellar mass $M_{dust}/M_{*}$. Studies report an anticorrelation between $M_{dust}/M_{*}$ with the $M_{*}$ both in the local universe \citep[e.g.,][]{Corte12, Clem13, Orel17, Casa20}, and out to $z\sim 2$ \citep[e.g.,][]{Calu17} with a slope ranging from -1 to 0, supporting our inferred positive relation between $M_{dust}$ and $M_{*}$.

\citet{DaCu10} investigated the relation between $M_{dust}$ of a galaxy and the $SFR$ using the model of \citet{DaCu08}. They used the two-screen attenuation relation from \citet{Char00} and separate attenuation from the interstellar medium and the birth clouds to calculate SED templates that best fit the observed galaxies in SDSS DR6. 
\citet{DaCu10} reported an increase in $M_{dust}$ with $SFR$ for SDSS galaxies with $-2.0 < \log(\mathrm{SFR}/(\mathrm{M}_{\odot}\mathrm{yr}^{-1})) <2.0 $. Although we find a positive correlation between $\tau_{B}^{f}$ (related to $M_{dust}$) and $SFR$ in the SDSS sample, we did not see this trend in the GAMA or COSMOS samples. Similarly, the three samples show inconsistent trends for $F$ and $r_{bulge}$ with $SFR$. Our investigation of $dMS$ and $\Sigma_{SFR}$ also resulted in inconsistent trends for all three samples.  

Our interpretation that the $\tau_{B}^{f}$, and therefore $M_{dust}$, is independent of $SFR$ is supported by the literature. For example, \citet{Casa17} investigated the radial distribution of the dust, gas, stars, and $SFR$ using data from DustPedia \citep{Dave17}. They found that the scale length for the dust-mass surface-density distribution is 1.8 times higher than for the $SFR$, assuming that both properties follow an exponential distribution. This is a consequence of the fact that the scale length of the stellar emissivity of the young stellar population is usually smaller than the scale length of the dust, a result derived from radiative transfer models of well-resolved galaxies \citep[e.g.,][]{Xil99, Pop00, Mis01, Popescu2017, Thir20, Natale21}. The difference in scale-length indicates that $M_{dust}$ and $SFR$ are not always spatially correlated, implying that the dust mass is not fully correlated with the $SFR$. The independence of $\tau_{B}^{f}$ on the $SFR$ could be due to a negative feedback mechanism, for example, radiative feedback, that regulates dust formation.

\citet{Casa17} also found that the dust-mass surface-density distribution differs from the stellar-mass surface-density distribution, which would imply that $M_{dust}$ and $M_{*}$ are not spatially correlated. This is again a consequence of the fact that the scale-length of the dust distribution is usually larger than the scale length of the NIR emissivity of the older stellar population that makes the bulk of the $M_{*}$, result also found in radiative transfer models of individual well-resolved galaxies \citep[e.g.,][]{Xil99, Pop00, Mis01, Popescu2017, Thir20, Natale21}. Other spatially resolved studies, such as \citet{Smith2016}, have found a spatial correlation between $M_{dust}$ and the $M_{*}$ up to twice the optical radius $r_{25}$. Whether or not $M_{dust}$ and $M_{*}$ are spatially correlated is still debated as the analysis of global properties indicate that there is a link between the $M_{dust}$ and the $M_{*}$ \citep[e.g.,][]{Corte12, Clem13, Orel17, Calu17, Casa20}. Considering the inconsistent trends with $SFR$, we interpret the anticorrelation found between $\tau_{B}^{f}$ and $sSFR$ to be driven by the $M_{*}$ rather than the star-formation timescale.

The SFR surface density $\Sigma_{SFR}$ can be used as a tracer for the total gas mass surface density following seminal work by \citet{Kennicutt98}. 
Several studies have shown that the dust and total gas are correlated \citep[e.g.,][]{Corb12, Sandstrom2013, Grov15, Casa20}, with the correlation dependent on the metallicity and the heating effects within a galaxy \citep[e.g.,][]{Lisen98, Draine_all2007, Gall08, Remy14, DeVi17}. Constraining the amount of total gas from $\Sigma_{SFR}$ is non trivial. Many studies of the Schmidt-Kennicutt relation have been performed on both galactic and subgalactic scales and find different results for the slope of the relation due to, for example, methodology, tracers used, sample selection, and influence of other galaxy properties such as $\Sigma_{*}$ \citep{Bigi08, Casa15, Elli2020, Mors20, Kenn21}.
However, it is now well established that $\Sigma_{SFR}$ is more closely correlated with the molecular gas mass surface density than total gas surface density \citep[e.g.,][]{Leroy2008, Schruba2011}. Although there are hints of a positive correlation between the dust mass and the molecular gas mass, this relationship comes with considerable scatter \citep[e.g.,][]{Orel17, Casa20}. Furthermore, the correlation between dust mass and $\tau_{B}^{f}$ is dependent on the scale height and relative distribution of stars and dust \citep{Pop11}, bringing uncertainties in relating our inferred $\tau_{B}^{f}$ with dust mass.
Given the systematic differences in Schmidt-Kennicutt relations, combined with the complex relationships between atomic gas, molecular gas, dust, and $SFR$, it is not surprising that we find inconsistent trends between $\Sigma_{SFR}$ and $\tau_{B}^{f}$ for our three samples. Future work could verify our results by using a more direct tracer of molecular gas mass, such as CO luminosity or dust emission in the Rayleigh-Jeans tail of the SED \citep[e.g.,][]{Corb12, Casa20}.

In Appendix \ref{Ap:IntrUV}, we show the results of Sect \ref{Sec:FittingZ} assuming different star-formation rate main-sequence relations and applying the \citet{Ken12} SFR-luminosity conversion to estimate the intrinsic emission in FUV and NUV. We find no trends of $F$ with $SFR$ when adopting different main-sequence normalizations. On the other hand, we find that any relations between $F$ and $M_{*}$ or $\mu_{*}$ vary depending on the main-sequence adopted. A possible explanation for this variation could be that the derived star-formation main-sequences are themselves affected by attenuation biases. The UV-SFR calibration from \cite{Ken12} assumes a constant star-formation history over the past 100 Myr, however galaxies above and below the main sequence have been shown to have rising and falling star-formation histories, respectively \citep[e.g.,][]{Leja19}, rendering the UV calibration inaccurate in particular bins. An incorrect $SFR$ calibration would cause a bias in our corrected FUV and NUV magnitudes, resulting in biased best-fitted values for $F$. Therefore, we chose not to use them to estimate intrinsic emissions. 
A different study would also be to check whether we need more accurate dust attenuation corrections than those derived from MIR.
We found that a change in the star-formation main-sequence adopted does not influence the relative relations between $\tau_{B}^{f}$ and $M_{*}$ or $\mu_{*}$, which is further evidence that the $M_{*}$ is strongly linked to the dust mass.

\subsection{Evolution of dust properties}\label{Sec:DiscussionZ}

\begin{figure}
    \centering
    \includegraphics[width=\columnwidth]{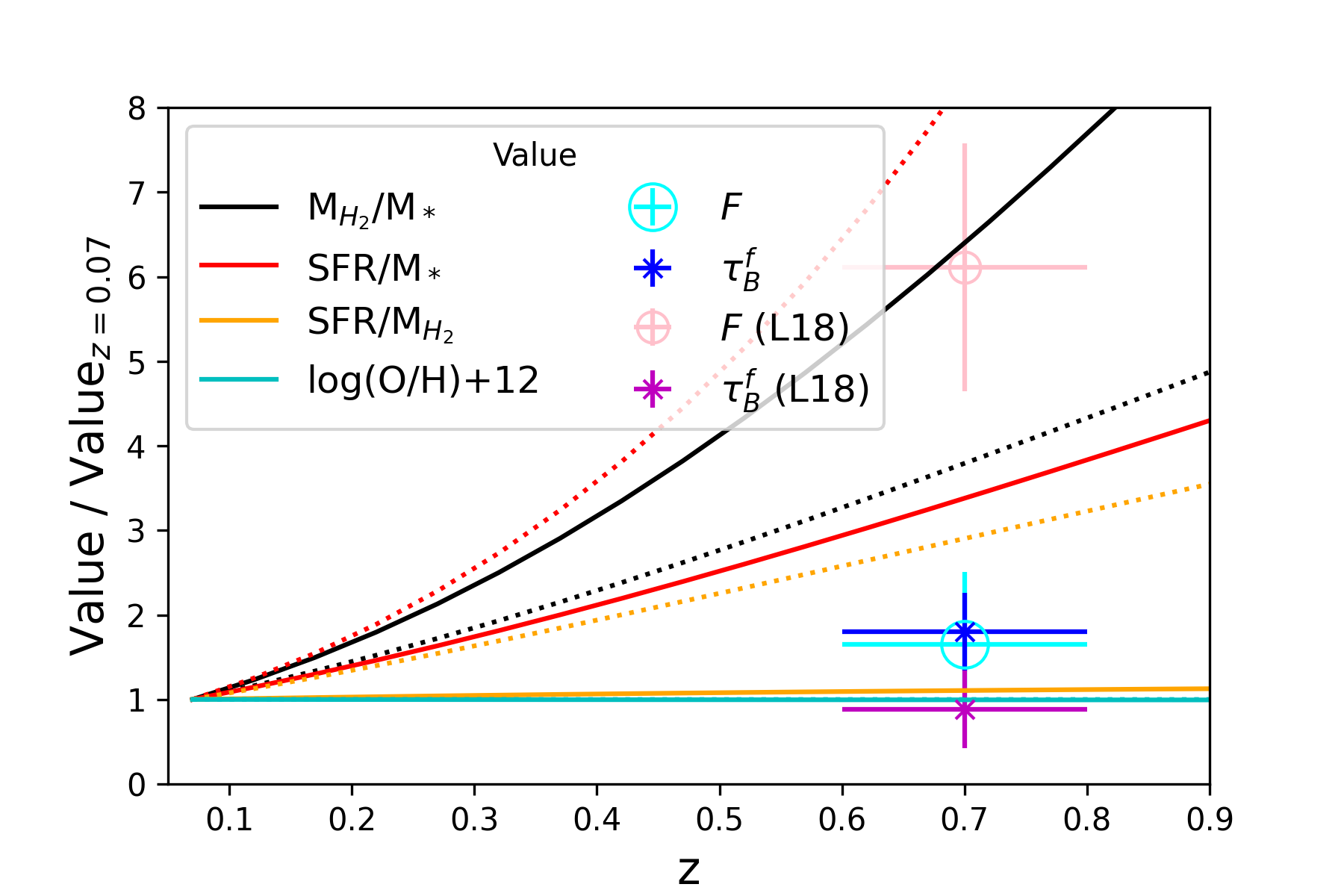}
    \caption{Evolution of massive galaxy properties from their low-z values out to $z\sim$1. Relationships for molecular gas to stellar mass ratio (M$_{H_2}$/M$_*$), specific SFR, SFR efficiency (SFR/M$_{H_2}$), and metallicity (log(O/H)+12) are from the A$^3$COSMOS scaling relations for main-sequence galaxies derived by \cite{Liu2019}, adopting the \cite{Schreiber2015} main-sequence. Solid lines show the relations at $\log(M_*/M_\odot)=10.2$ and dotted lines show the relations for galaxies with $\log(M_*/M_\odot)=11.5$.
Unlike \cite[][L18]{Les18}, who found that $F$ increases with redshift and the $\tau_{B}$ $\tau_B^f$ remains constant to $z\sim0.7$, we find that both these dust parameters evolve together increasing by a factor of $\sim$1.7 ($\tau_B^f$) and $\sim$1.8 ($F$) from low-z galaxies in GAMA and SDSS to intermediate-redshift galaxies in COSMOS. }
    \label{fig:evolution}
\end{figure}

Figure \ref{fig:evolution} summarizes how our model parameters $\tau_{B}^{f}$ and $F$, as well as galaxy molecular to stellar mass ratio (M$_{H_2}$/M$_*$), specific SFR, SFR efficiency (SFR/M$_{H_2}$), and metallicity (log(O/H)+12) increase relative to their values at $z=0.07$. For our $z=0.07$ reference, we have averaged the SDSS and GAMA values for $\tau_B^f$ and $F$ because they are consistent within the errors. We find that values of $\tau_{B}^{f}$ and $F$ are both a factor $\sim$1.75 higher in the COSMOS sample, implying these dust parameters scale roughly with $(1+z)$. Galaxy star-formation efficiency (inverse of depletion time), calculated for galaxies between $10.2<\log(M_*[\mathrm{M}_{\odot}])<11.5$ (the $M_{*}$ range of our samples), increases by a similar amount, according to the scaling relations reported by \citet[][see Fig. \ref{fig:evolution}]{Liu2019}.  

The $\tau_{B}^{f}$ for $z\sim0.7$ galaxies in COSMOS is higher than that found for $z\sim 0$ galaxies in SDSS and GAMA at fixed $M_{*}$, implying different relations between $M_{dust}$ and the $M_{*}$ for galaxies at different redshifts \citep[e.g.,][]{Bethermin2015, DeVi17, Calu17,  Pas20}. \citet{Bethermin2015} studied the mean $M_{dust}$ - $M_{*}$ ratio for star-forming galaxies with $M_{*}$ $>3\cdot10^{10}\mathrm{M}_{\odot}$ across redshift. They found that the $M_{dust}$ - $M_{*}$ ratio increased with redshift $M_{dust}/M_{*} \approx (1+z)^{x}$ with $x > 0.05$ \citep{Tan14} up to a redshift of $z = 1$ for star-formation main-sequence galaxies, whereafter it became constant. Because the COSMOS sample has higher fitted $\tau_{B}^{f}$, indicating a higher average $M_{dust}$, compared to SDSS and GAMA at similar $M_{*}$, our results imply a correlation between the $M_{dust}$ - $M_{*}$ ratio and redshift, which is consistent with the literature \citep[e.g.,][]{Liu2019,Magn20,Koko21}.

We found that the COSMOS galaxies have higher fitted $F$ than the low-z galaxies, indicating an increase in relative fraction of stars trapped in their birth clouds. This result hints at higher redshift galaxies having more optically thick star-forming regions than diffuse components compared to galaxies at lower redshift. 
\citet{Les18} found an increase in $F$ for the higher redshift galaxies with a factor of $\sim 5$, whereas we have an increase of $\sim 1.6$ with the difference caused by the inclusion of the NUV-band in the fitting. T04 suggested $F = 0.22$ for their low-z sample, close to our low-z sample results of $F = 0.19$, giving confidence in our fitting results. However, P11 suggested a higher value for typical low-z spiral galaxies using IR-submm wavelengths, $F = 0.35$. Because we constrain the effect of the $F$ with only the UV-emission, our result serves as a lower limit for $F$.

Changes in the galaxy metallicity can vary the overall attenuation, which would imply that $F$ does not only represent the clumpiness. Studies have shown that the metallicity of massive galaxies does not significantly evolve with redshift up to $z = 1$\citep[e.g.,][]{Cal01, Con10}. \citet{Con10} analyzed attenuation curves for galaxies with redshift $0.6 < z < 1.4$ from the DEEP2 Galaxy Redshift Survey and found the resulting attenuation curves to be similar to low-z star-forming galaxies. The similarity would suggest that the chemical composition of the diffuse dust is the same. Figure \ref{fig:evolution} shows the lack of metallicity evolution with redshift.

Our findings, that high redshift galaxies have higher values for $\tau_{B}^{f}$ and $F$ than their low redshift counterparts, are only valid if we assume no variation in the geometry of stars and dust with redshift. Observations found that galaxies at higher redshift have higher scale height and increased turbulent motion \citep[e.g.,][]{Burk16}. The higher scale height and increased turbulent motion cause galaxies to appear puffier, which will have two effects on constraining an attenuation-inclination trend. One effect is that the puffiness will result in incorrect estimates of the galaxy ellipticity. As we use the ellipticity to derive the inclination, the inclination might be wrongly estimated, resulting in an uncertain attenuation-inclination trend. Additionally, the increased puffiness could also mean that the relative distribution of the stars and dust varies with redshift, implying that the adopted model distribution is not applicable. We could confirm whether or not our results are affected by the increased puffiness by including IR data as an additional constraint on $\tau_{B}^{f}$, using a different measure for the inclination, or designing new attenuation-inclination models for galaxies with a puffier geometry.

Studies have tried to understand the increase in clumpiness for higher redshift galaxies using galaxy formation simulations \citep{Mand17, Soto17, In19}. \citet{Soto17} found evidence that the clumpiness is linked to the formation of the galaxies and their results align with clump migration theories. These theories suggest clumps form in turbulent disks due to gravitational instabilities and migrate to the center of galaxies to form the bulge. The increase in turbulent motion with redshift would suggest that it is more likely for star-forming clumps to be formed, resulting in higher clumpiness.

\subsection{Comparing optical depths derived from UV and optical}\label{Sec:DiscussionUVOpt}
Finding relations between the optical depth and physical properties is dependent on the definition of the optical depth. As we have explained in Section \ref{Sec:FittingBalmer}, the $H{\alpha}/H{\beta}$ ratio is dependent on the inclination, and with the assumptions in the T04 model, we could not recreate the observed ratios. We have adjusted the T04 model to describe the $H{\alpha}/H{\beta}$-inclination relation using $\tau_{B}^{f}$ as the contribution of the diffuse component and an inclination independent offset $C$ for the star-forming regions different from the $F$, allowing us to fit the offset in FUV and NUV separate from the found $H{\alpha}/H{\beta}$ ratio offset. 

The separation of $F$ and $C$ allows us to find correlations between physical properties and the newly introduced components using the $H\alpha/H\beta$ ratios. We found that the contribution of the new component of star-forming regions $C$ is also dependent on $M_{*}$. This new component could be diffuse dust within or surrounding the clumpy components that influence the emission when a stellar photon escapes, ionizing additional gas, contributing to the diffuse ionized gas (DIG) emission and not impacting the derived $F$. These effects could explain why studies on the UV-slope $\beta$ and Balmer optical depth $\tau_{Bal}$ find different dependencies on galaxy properties, as they trace different parts of the star-forming regions. \citet{Bat16} and \citet{Bat17} studied the dependence of $\beta$ and $\tau_{Bal}$ on inclination, $M_{*}$, and $SFR$. \citet{Bat16} found that $\beta$ and $\tau_{Bal}$ share positive correlations with the $M_{*}$ and SFR, but had different functional forms, and \citet{Bat17} found that the $\beta$-$\tau_{Bal}$ relation depends on the inclination, also hinting at the UV- and Balmer-derived parameters trace different regions. Studying the relations in more detail might link the number, size, and properties of clumpy regions to the properties of their host galaxy. 
We could analyze the existence of the additional component using HII region models where the escaped fraction of Balmer emission from the star-forming regions gets attenuated. We do note that the $H\alpha/H\beta$ ratios from the T04 model are integrated properties, whereas we use measurements from fibers, which only detect the Balmer lines over a small scale of the galaxy. Studies have shown that the $SFR$ and attenuation are dependent on location in galaxies \citep[e.g.,][]{Thir20}. Future work with observations of spatially resolved star-forming regions could allow us to better understand the attenuation of the Balmer lines occurring in the gaps between or around the optically thick birth clouds.


\section{Conclusion}\label{Sec:Conclusion}
The goal of this study is to constrain the dependence of attenuation on galaxy physical properties and redshift, and we use parameters as defined in the \citet{T04} model to track these dependencies. The \citet{T04} model contains the parameters optical depth $\tau_{B}^{f}$ describing the diffuse dust, clumpiness $F$ describing the optically thick star-forming regions, and bulge fraction $r_{bulge}$ describing the bulge-to-total luminosity ratio. We compare the results between low-z galaxies and galaxies at redshift $z \sim 0.7$. We find that galaxies at redshift $z \sim 0$ and $z \sim 0.7$ have a different $\tau_{B}^{f}$, $F$, and $r_{bulge}$. The average values for the low-z sample are $\tau_{B}^{f} \approx 4.1$, $F\approx0.33$, $r_{bulge} \approx 0.2$. We found, for galaxies at $z \sim 0.7$, that $\tau_{B}^{f} \approx 7.1$, $F\approx0.57$, $r_{bulge} \approx 0.1$. 

We found that $\tau_{B}^{f}$ increases with both $M_{*}$ and $\mu_{*}$. This dependence implies that galaxies with higher $M_{*}$ also have more dust. The $M_{dust}$ - $M_{*}$ ratio increases with redshift, as $\tau_{B}^{f}$ is higher at $z\sim 0.7$ than at $z\sim0$ for fixed $M_{*}$.
We find no robust trends for $\tau_{B}^{f}$ or $F$ varying with $SFR$, $\Sigma_{SFR}$, or $dMS$. Given the lack of trends with $SFR$, we conclude that the decrease of $\tau_{B}^{f}$ seen with $sSFR$ is driven by trends with $M_{*}$. 

There are additional effects we need to take into account when relating $\tau_{B}^{f}$ to the optical depth of HII regions, which we explore using the $H{\alpha}/H{\beta}$ ratio for the low-z galaxies. We found that, besides the predicted attenuation of the Balmer line emission escaping the star-forming regions, there is an additional wavelength-dependent attenuation, described in our study by the parameter $C$. Our findings are consistent with a more optically thin component of dust, perhaps filling between the otherwise optically thick cloud fragments. The increase in $C$ with $M_{*}$ could mean that the individual HII regions inside more massive galaxies have a higher fraction of optically thin dust within or around them.

Our results imply that dust properties are dependent on the global properties of a galaxy, the properties of the HII regions, and the redshift. This dependence means that the formation of galaxies and dust are linked, and we need more detailed dust models to calculate the photometric corrections at different redshifts. High-resolution dust continuum observations using ALMA \citep[e.g.,][]{Hodg19} could help determine the dust structures for different galactic components in galaxies at $z\sim0.7$, following analysis methods developed for local galaxies such as \citet{Thir20}.

\begin{acknowledgements}
We thank the referee for their positive report and constructive suggestions to improve this work. We would like to thank Sabine Bellstedt for discussing the updated GAMA photometric data. BG acknowledges the support of the Australian Research Council as the recipient of a Future Fellowship (FT140101202). JH gratefully acknowledges support of the VIDI research program with project number 639.042.611, which is (partly) financed by the Netherlands Organisation for Scientific Research (NWO). MTS acknowledges support from a Scientific Exchanges visitor fellowship (IZSEZO\_202357) from the Swiss National Science Foundation.

\end{acknowledgements}

\bibliographystyle{aa} 
  \bibliography{manuscript} 

\appendix
\section{Comparison of SDSS and GAMA photometry}\label{Ap:Photometry}
\begin{figure}[!ht]
    \centering
    \includegraphics[trim=0cm 1cm 1cm 0cm,clip, width = \columnwidth]{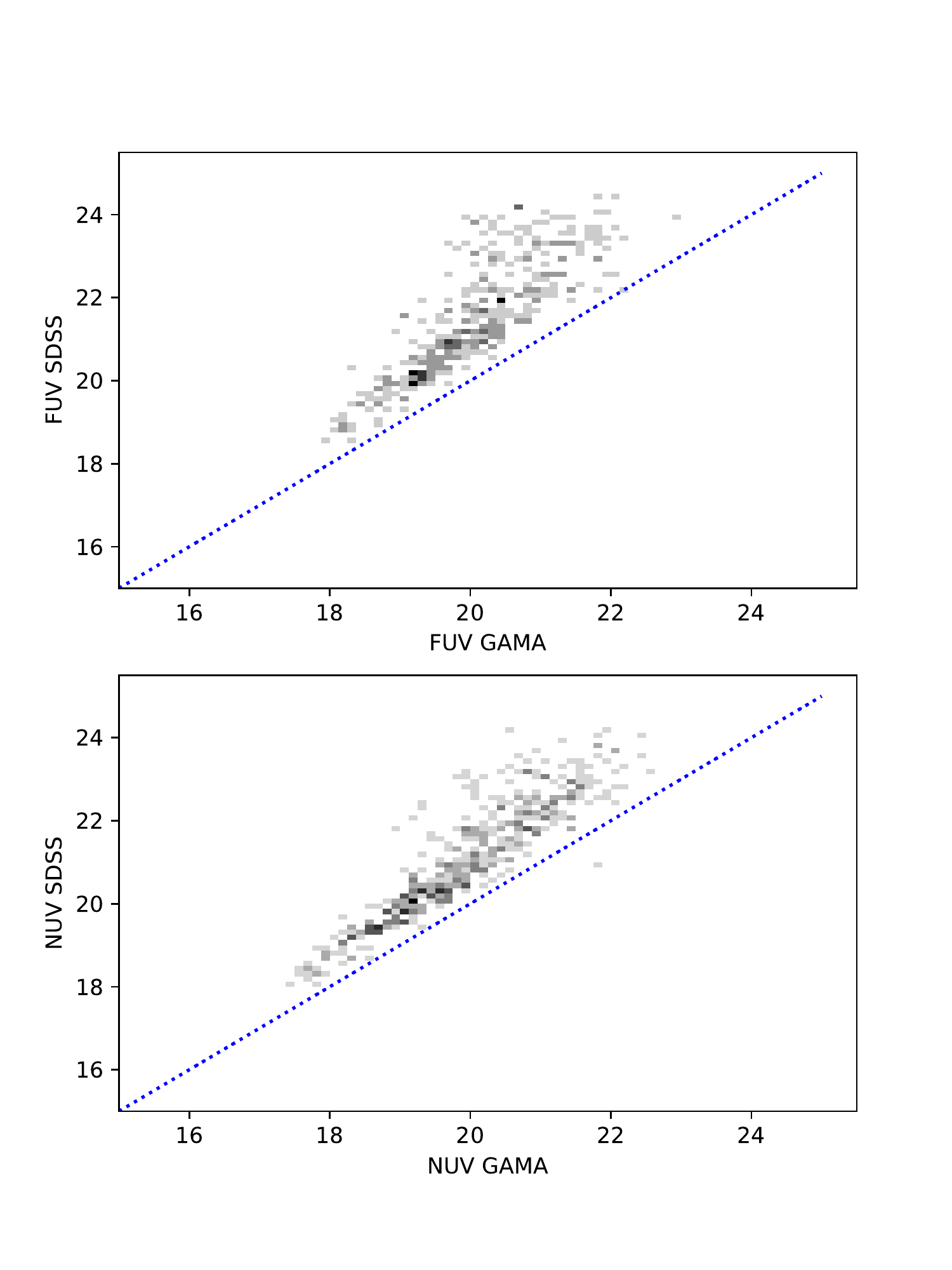}
    \caption{Comparison of the FUV and NUV measurements from \citet{Bell20} for GAMA and \citet{Bianchi11} for SDSS of galaxies appearing in both SDSS and GAMA. The blue line shows the one-to-one relation for the different bands. The figure shows that the measurements do not align, meaning we need to decide which photometry we trust the most for our calibrations.}
    \label{fig:UV_Sabine}
\end{figure}

For this study, we use the updated GAMA photometry catalog of \citet{Bell20}. They used a new method of deriving apertures needed for extracting galaxy photometry. The new technique results in different fluxes between GAMA and SDSS. We compare the photometry in UV, optical, and NIR bands by selecting galaxies that appear in both SDSS and GAMA illustrated in Figure \ref{fig:UV_Sabine}.

\begin{figure}[!ht]
    \centering
    \includegraphics[trim=0cm 1cm 1cm 0cm,clip, width = \columnwidth]{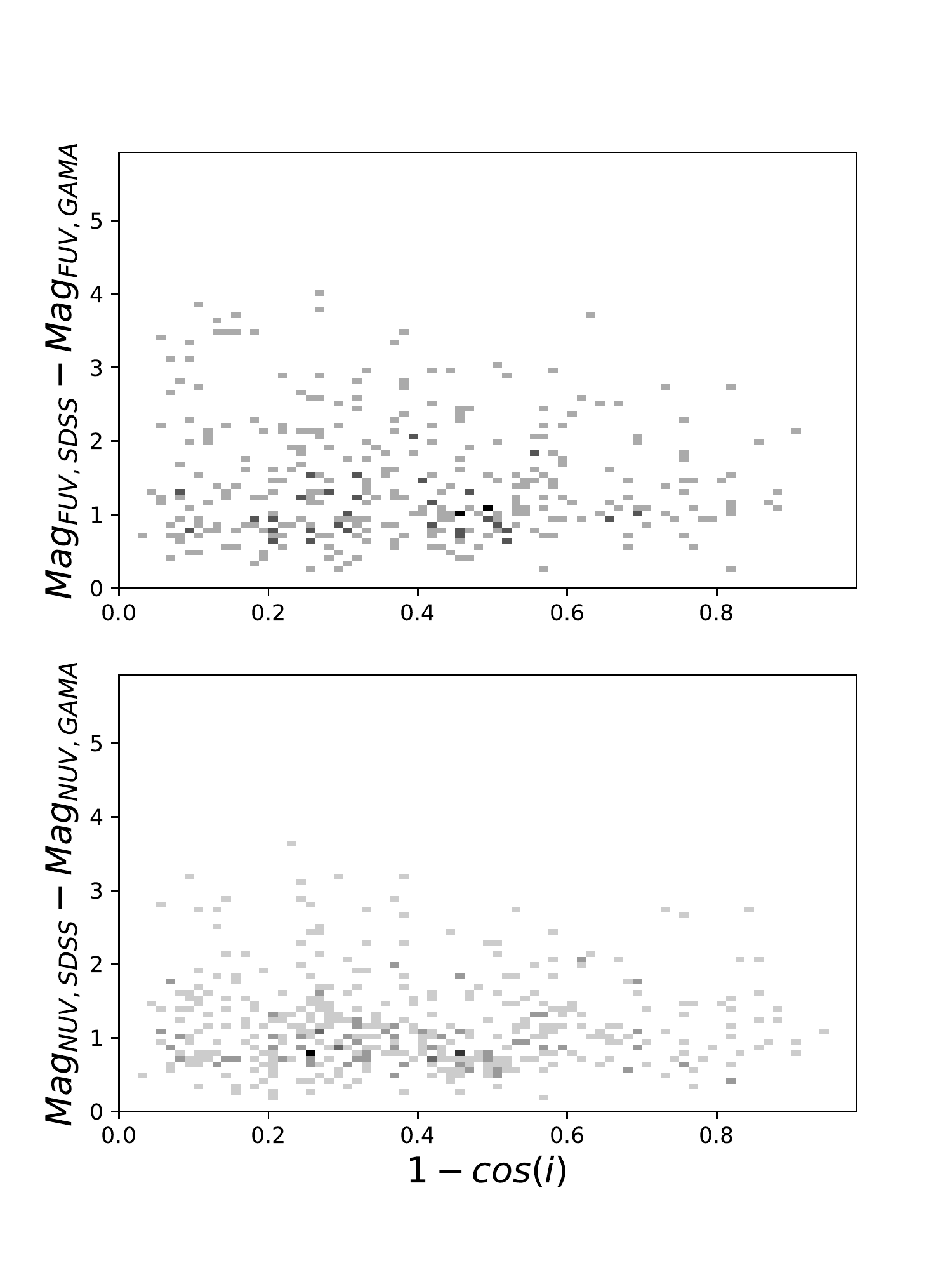}
    \caption{FUV (top) and NUV (bottom) measurement comparison of \citet[ GAMA]{Bell20} and \citet[ SDSS]{Bianchi11} photometry for galaxies appearing in both surveys as a function of inclination $1 -\cos(i)$. The figure shows that the difference in photometry is independent of inclination.}
    \label{fig:UV_Sabine_inc}
\end{figure}

\begin{figure}[!ht]
    \centering
    \includegraphics[trim=0cm 1cm 1cm 0cm,clip, width = \columnwidth]{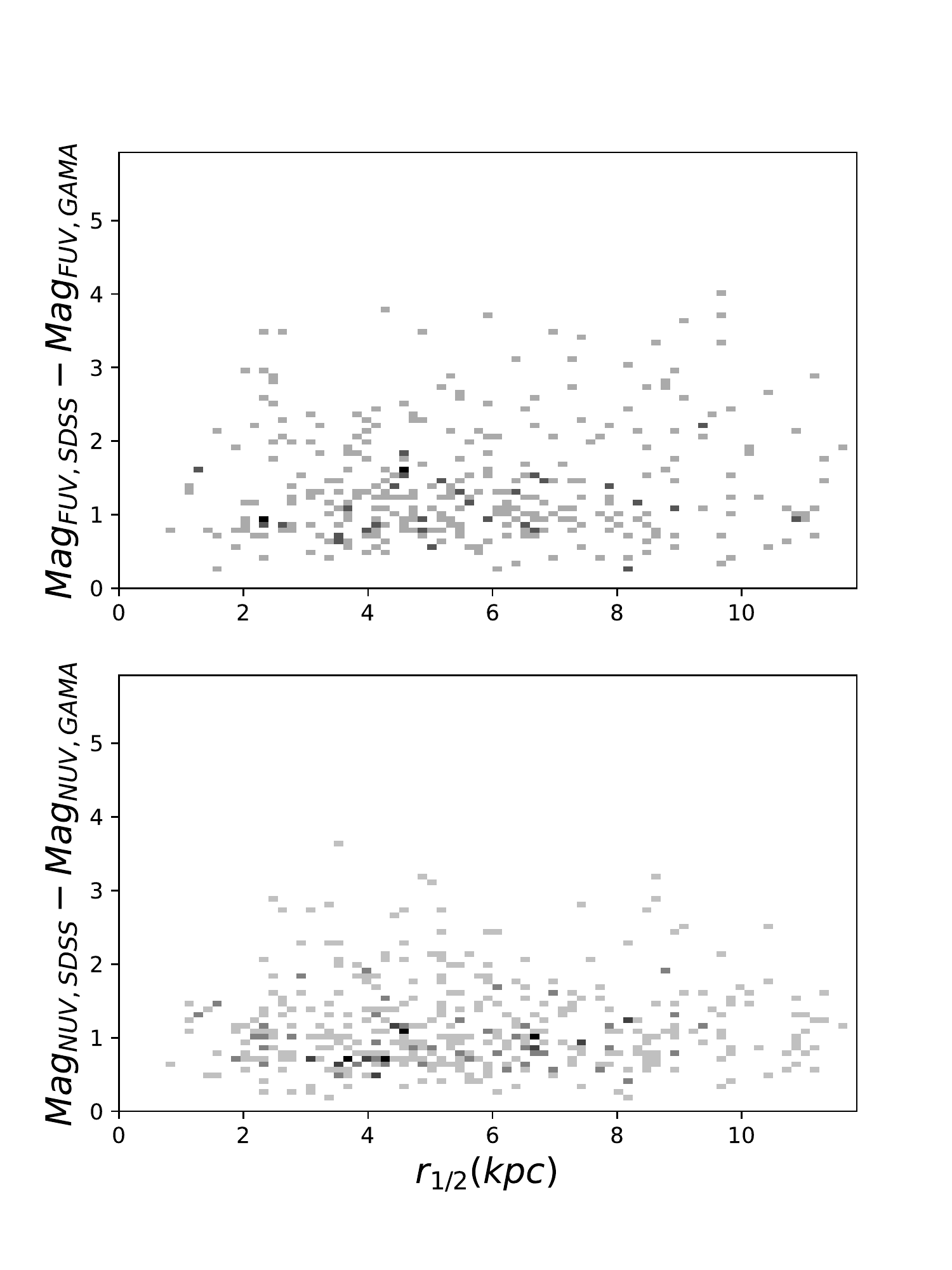}
    \caption{FUV and NUV measurement comparison from \citet{Bell20} for GAMA and \citet{Bianchi11} for SDSS of galaxies appearing in both SDSS and GAMA as a function of half-light radius $r_{1/2}$. The figure shows that the difference in photometry is independent of $r_{1/2}$.}
    \label{fig:UV_Sabine_r}
\end{figure}

The comparison shows that the FUV and NUV have a constant offset in magnitudes of $\sim1.2\pm 0.8$ and $\sim1.0\pm 0.6$ respectively, whereas the optical and NIR magnitudes (not shown) are in agreement with each other. The UV offset is independent of galaxy inclination (Fig. \ref{fig:UV_Sabine_inc}) and g-band half-light radius (Fig. \ref{fig:UV_Sabine_r}), meaning it is likely independent of the size and shape of the aperture used in the \citet{Bell20} derivations. Due to the updated method used in the new GAMA photometry, we have decided to correct the SDSS GALEX UV MIS magnitudes by subtracting the offset. 

\section{Fitting bulge-to-total ratio}\label{Ap:GalProp}
One of the model parameters from the T04 model is the bulge-to-total ratio $r_{bulge}$, which tells us about how the radiation is divided between the bulge and the disks. The impact of the parameter on $\Delta m$ is dependent on inclination, which could influence the accuracy of our results as such effects can also be caused by the $\tau_{B}^{f}$. Studies such as \cite{Haus13} and \citet{Sim02} used different methods to constrain the ratio. As mentioned in Section \ref{Sec:Methods}, we derive the distribution of the bulge-to-total ratio derived for our selected star-forming galaxies in GAMA and COSMOS to act as the prior for the T04 model fitting.

\begin{figure}[!ht]
    \centering
    \includegraphics[trim=3cm 1cm 3cm 0cm,clip, width = \columnwidth]{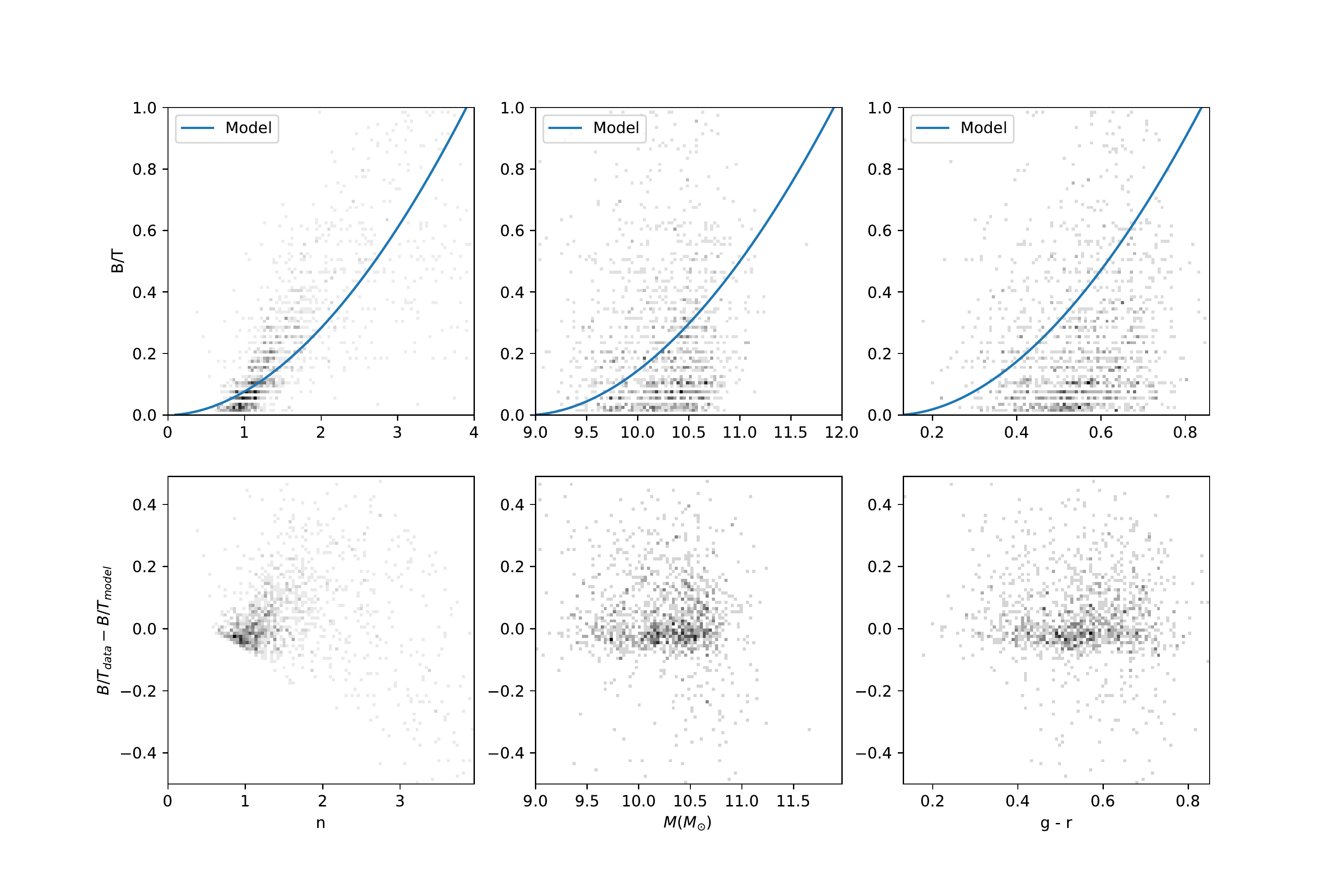}
    \caption{Bulge-to-total ratio $B/T$ of galaxies appearing in both SDSS and GAMA against the Sérsic index $n$, $M_{*}$ and g - r color. The blue line shows the best-fit model where we change the variable per plot: top-left is for varying $n$, top-middle is for varying $M_{*}$ and top right is for varying g-r. The bottom plots illustrate the residuals of the data - model for each data point.}
    \label{fig:BT_GAMA}
\end{figure}

\begin{figure}[!ht]
    \centering
    \includegraphics[trim=3cm 1cm 3cm 0cm,clip, width = \columnwidth]{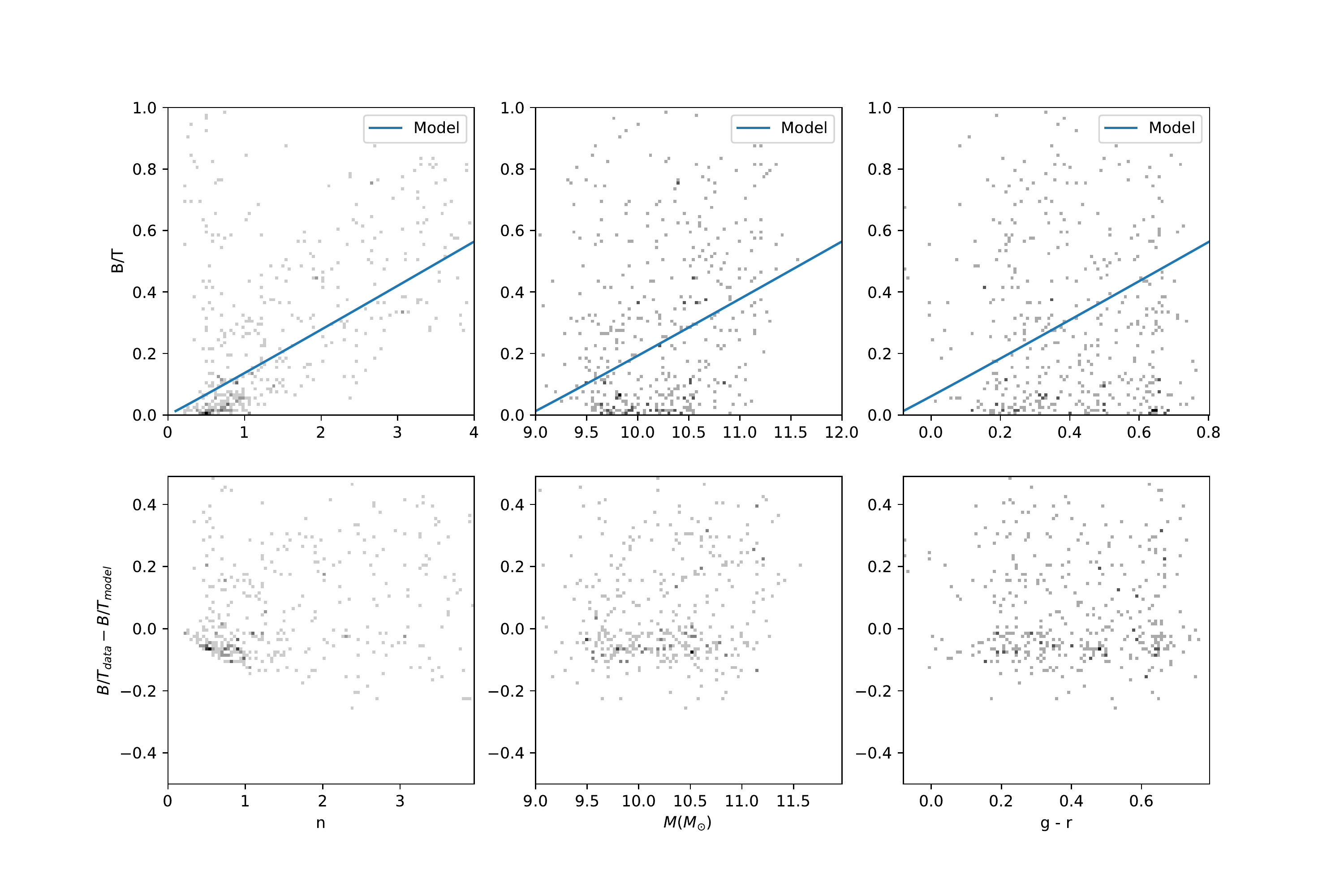}
    \caption{Bulge-to-total ratio $B/T$ of galaxies appearing in COSMOS and CANDELS against the Sérsic index $n$, $M_{*}$ and B - V color. The blue line shows the best-fit model where we change the variable per panel: top-left is for varying $n$, top-middle is for varying $M_{*}$ and top right is for varying B-V. The bottom panels illustrate the residuals of the data - model for each galaxy. The model seems to mostly agree but tends to underestimate $B/T$ for higher values of the galaxy properties.}
    \label{fig:BT_COSMOS}
\end{figure}

We first select star-forming galaxies from the GAMA sample following Section \ref{Sec:SFselec} and only use the galaxies which can be found in SDSS to use the derived bulge-to-total ratios from \citet{Sim11}. For COSMOS, we select star-forming galaxies following Section \ref{Sec:SFselec} and use the bulge-to-total ratios from the CANDELS data set \citep{Haus13}. We use scikit.learn to fit $r_{bulge}$ as a function of $M_{*}$, $r_{1/2}$, and g-r color. We assume that the ratio follows a power law in both parameters and apply the linear regression module to derive the relations. In figures \ref{fig:BT_GAMA} and \ref{fig:BT_COSMOS}, we show the fitting results and show that, together, these three variables give good results in predicting the $r_{bulge}$.

\section{Importance sampling}\label{Ap:ImpSamp}
Our analysis relies on unbiased photometry and galaxies with similar properties that do not vary in inclination. We have found in Section \ref{Sec:PhysSelec} that the physical properties derived from photometry have an implicit relation with inclination. The inclination dependence hints at an inclination bias in our photometry that might come from detection bias and could interfere with the attenuation-inclination trends we want to analyze. The inclination bias of the physical properties gives us insight into how the inclination biases our photometry and, therefore, the selected galaxy distribution. Importance sampling uses this information to correct the biases. We refer readers to \cite{Chev15} for an overview. In this section, we describe the basic principles needed to understand the implementation of the technique in our analyses.

Importance sampling is a statistical technique that corrects for implicit biases in datasets. This technique comes from the mathematical description of finding expected values with a distribution of observable $x$:
\begin{equation}\label{eq:av_stand}
    \bar{x} = \int x f(x)dx,
\end{equation}
where $f(x)$ is the probability function of observable $x$ and $\bar{x}$ the average value of the observable.
Importance sampling is a result of rewriting Eq. (\ref{eq:av_stand}) as:
\begin{equation}\label{eq:av_bias}
    \bar{x} = \int x \frac{f(x)}{g(x)}g(x)dx,
\end{equation}
with $f(x)$ and $g(x)$ both being different probability distributions, allowing us to transform between distributions of parameters. The transformation for our case is between the observed distribution, represented by $g(x)$ and the inclination unbiased distribution $f(x)$. Since we are dealing with discrete measurements, we rewrite Eq. (\ref{eq:av_bias}) as a sum instead of an integral:
\begin{equation}
    \bar{x} = \frac{\Sigma_{i} x \frac{f(x)}{g(x|y_{i})}}{\Sigma \frac{f(x)}{g(x|y_{i})}},
\end{equation}
with $x$ being the observed physical property and $g(x\mathrm{|}y_i)$ being the probability function for $x$, given a known $y$, in our case, the inclination. The ratio $\frac{f(x)}{g(x|y_{i})}$ represent weights that describes how much the measurement of distribution $f(x)$ would contribute to the mean if the data would be drawn from a different distribution $g(x)$. We need to calculate the weights to ensure that the averages we calculate in photometry are unbiased by the inclination before deriving the inclination dependent model parameters.

The physical properties that show inclination bias are $\log(M_{*})$, $r_{1/2}$, and $n$. We add redshift $z$ to the list to take into account possible cosmological volume corrections to create one weight per galaxy. For each property, we derive the distribution per inclination bin by separating the galaxies into ten bins of inclination angle. In each bin, we assume a Gaussian distribution for the physical properties and determine distributions using the Gaussian kernel density estimator provided in the scipy package. We refer to the distributions of single physical property as $h(x\mathrm{|} \mathrm{y})$.
The global probability distribution $g(x\mathrm{|} \mathrm{y})$ is found by multiplying the single probability distribution with each other:
\begin{equation}
    g(x|y) = \prod_{j}^{n} h_{j}(x_{j}|\mathrm{y}),
\end{equation}
with $\mathrm{h_{j}(x_{j}|\mathrm{y})}$ being the probability function of the physical properties, assuming the properties are independent. \\
We cannot define an inclination unbiased distribution $f(x)$ because each bin could be inclination biased. Therefore, we follow \citet{Chev15} for calculating the inclination unbiased distribution:
\begin{equation}
    f(x) = \prod_{i} g_{i}(x|y_i)^{0.1}.
\end{equation}
With this relation, we use both $f(x$) and $g(x|y)$ to calculate the weights $\frac{f(x)}{g(x|y_{i})}$ that we apply to the photometry to calculate unbiased mean values.

\section{Testing model fidelity}\label{Ap:Model}
The decisions made in Section \ref{Sec:Methods}, such as the face-on normalization of optical emission, could influence the accuracy of our fitting results, causing small offsets in fitting for the best-fit parameters. In this section, we investigate how any offset influences our results.
The analysis was done using simulated attenuation-inclination curves from the T04 model in the GALEX FUV \& NUV bands, B, V, and I bands, and 2MASS J and K bands without adding any uncertainty. We simulate attenuation-inclination for different values of $\tau_{B}^{f}$, $F$ and $r_{bulge}$ and apply our fitting regime on the models. When varying the one of the parameters, we fix the other two to default values of $\tau_{B}^{f} = 4.0$, $F = 0.3$, and $r_{bulge} = 0.5$.

\subsection{Face-on normalization}\label{Ap:ModelNorm}
\begin{figure}[!ht]
    \centering
    \includegraphics[trim=1cm 1cm 1.5cm 0cm,clip,width=\linewidth]{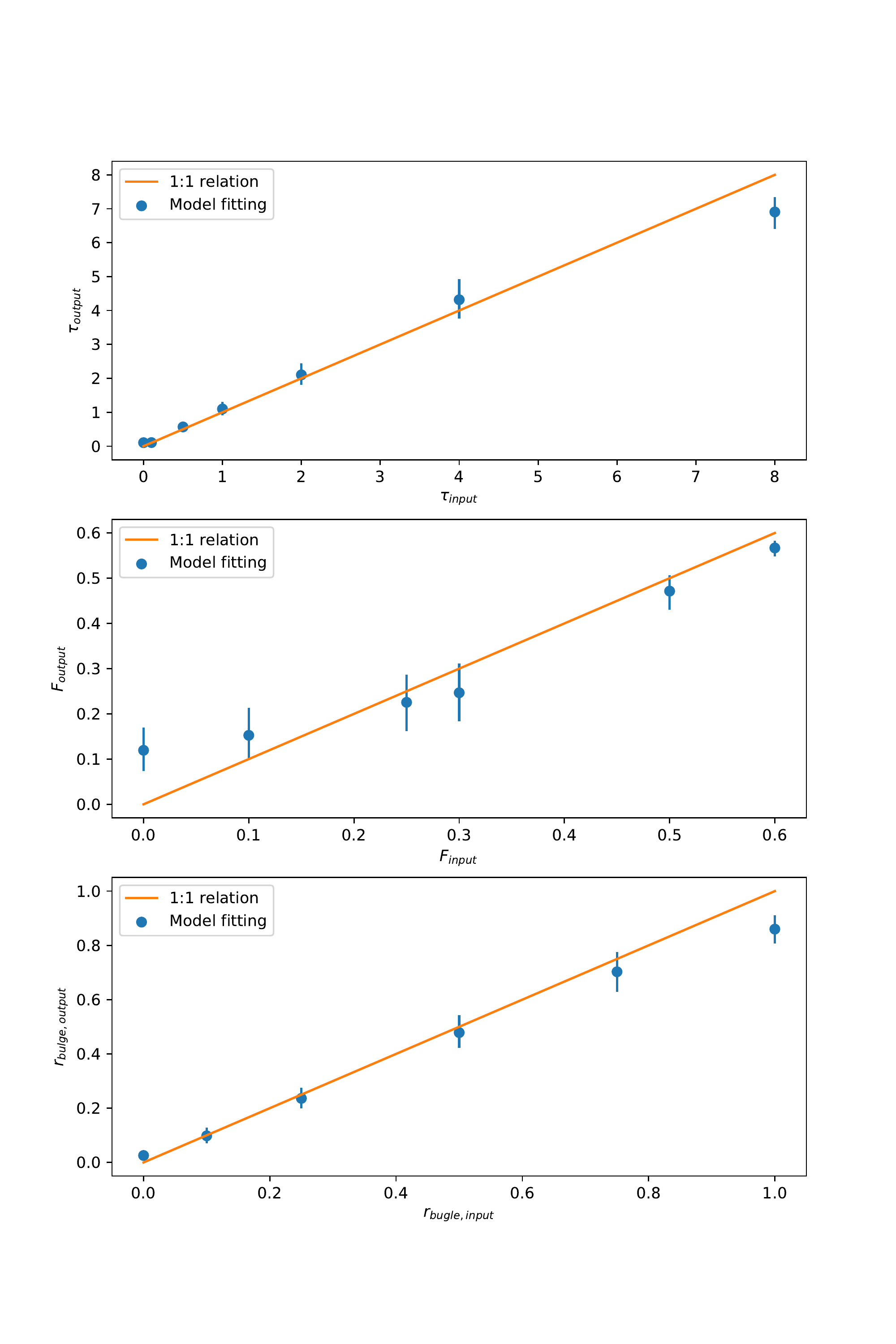}
    \caption{Comparison between the input parameters used to create the T04 model from which simulated NIR to UV attenuation-inclination relations were drawn, and the resulting best-fit parameter. The blue points show the results of the fitting using different values of $\tau_{B}^{f}$, $F$ and $r_{bulge}$, the orange line is the one$-$to$-$one relation. We see that the offset increases the closer the parameters. For each panel, we only vary one of the three parameters. The other two parameters are set to the values given in Section \ref{Ap:Model}. The figure shows that the fitting result will differ based on how close the model is to the boundaries defined by the prior.}
    \label{fig:prior}
\end{figure}

First, we show the impact of the normalization in the optical bands on our results. Since we cannot estimate the intrinsic emission in optical and NIR bands, we normalize these bands by the average value of galaxies with inclination $0.1 < 1 - \cos(i) < 0.2$. The results of the model fitting compared to our input are given in Figure \ref{fig:prior}. For each panel, we only vary one of the three parameters. The two parameters we leave as constant are set on the values given in Section \ref{Ap:Model}.

In Figure \ref{fig:prior} we see that there is an offset for the model fitted parameters at different input for $\tau_{B}^{f}$, $F$ and $r_{bulge}$. This offset is mostly driven by the priors and the distribution of samplers. The distribution of parameters has a Poisson distribution. The median value of the distribution is dependent on the boundaries, as the distribution becomes skewed at the boundaries, meaning the median value has a slight offset. For $F$, we also have the influence of $\tau_{B}^{f}$. The vertical offset in the attenuation-inclination relation in UV is dependent on $F$ and $\tau_{B}^{f}$. Low values of $F$ mean that the offset is very small. This causes the fitting for low values of $F$ to be very uncertain and is often paired with increased uncertainty in $\tau_{B}^{f}$ because the small vertical offset could also be explained by a change in $\tau_{B}^{f}$.\\

\subsection{Sensitivity of the bands}\label{Ap:ModelBands}
\begin{figure}[!ht]
    \centering
    \includegraphics[trim=1cm 1cm 1.5cm 0cm,clip,width=\linewidth]{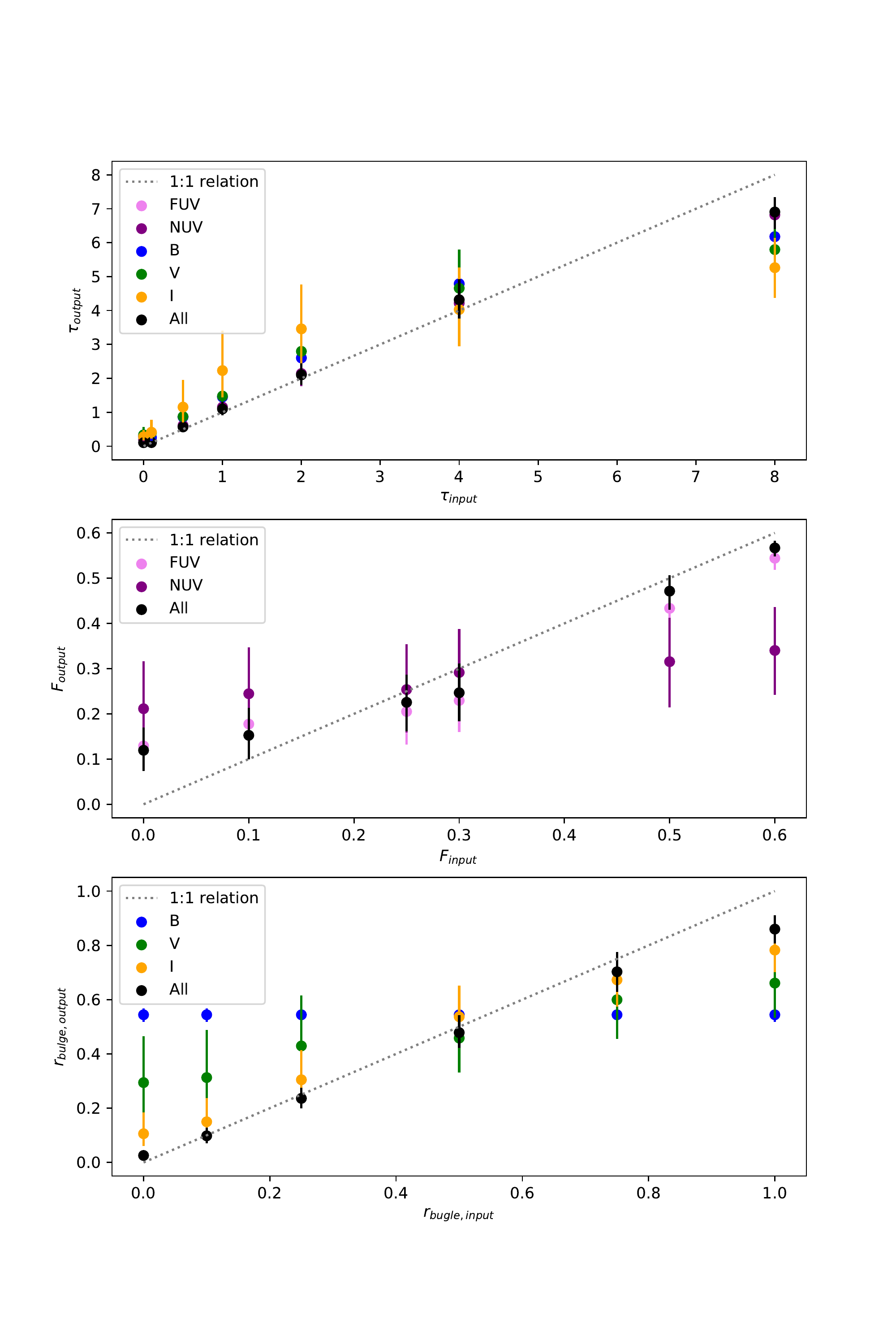}
    \caption{Comparison between the input parameters used to create the T04 model and the resulting best-fit parameter using the T04 model in different bands. The accuracy of the fitting is dependent on how much the band is affected by attenuation. Using all the bands gives the best results.}
    \label{fig:band}
\end{figure}

Next, we cover the number of photometric bands used, as this is a key difference between our work and \citet{Les18}. \citet{Les18} used only the GALEX FUV bands to find the best-fit value for $\tau_{B}^{f}$ and $F$, whereas we use GALEX FUV and NUV together with BVIJK for COSMOS and SDSS griJK for SDSS and GAMA. 
We model attenuation-inclination relations using the T04 model for several combinations of $\tau_{B}^{f}$, $F$ and $r_{bulge}$. We apply the model fitting using separate bands and compare this best-fit parameter from each band with the best-fit parameter using all the bands. The results are given in Figure \ref{fig:band}, where we only vary one of the three parameters for each panel.

In Figure \ref{fig:band} we see the fitting results for different values of $\tau_{B}^{f}$, $F$ and $r_{bulge}$ using different bands. We know that the UV bands are more affected by attenuation than the optical bands are. This difference in sensitivity means that the $\tau_{B}^{f}$ found from UV bands should be more accurate than the optical bands, which we see in the top panel. In the middle panel, we see the FUV results obtaining the $F$ very well, whereas the NUV results show a larger offset closer to the priors. This difference is due to the FUV being more sensitive to $F$ compared to NUV. In the bottom panel, we see that the I-band has the closest best-fit results to the input values of $r_{bulge}$. The I-band magnitude is not very affected by the inclination effects of the diffuse dust from the disks, making the inclination dependence primarily driven by $r_{bulge}$. This dependence should, in turn, mean that bands that are more affected by the diffuse dust will find less accurate results for the fitted $r_{bulge}$, which is also what we see. For both $\tau_{B}^{f}$, $F$ and $r_{bulge}$ we see that using all bands gives the best fit, as this takes all the previously mentioned effects into account, similar to what we found when comparing our results with \citet{Les18}. \\

\section{Estimating intrinsic UV emission}\label{Ap:IntrUV}
In this study, we derived intrinsic emission by correcting the GALEX FUV and NUV bands using MIR emission following \citet{Hao11}. In \citet{Les18}, they used different methods of deriving the intrinsic FUV emission based on the star-formation main-sequence. The star-formation main-sequence is highly dependent on sample selection, resulting in multiple papers with different $SFR-M_{*}$ relations. Here, we discuss the three different star-formation main-sequences used to derive intrinsic UV emission in \citet{Les18} and how they impact the model fitting results.

\subsection{Star-formation main-sequence}\label{Ap:IntrUVSFMS}
The main star-formation main-sequence (SF MS) used in \citet{Les18}, is derived from an updated sample presented in \citet{Sargent2014}:
\begin{equation}
\log(\frac{SFR_{MS}}{M_{\odot \mathrm{yr}^{-1}}}) = 0.816\log(\frac{M_{*}}{\mathrm{M}_{\odot}}) - 8.248 + 3\log(1+z),
\end{equation}
with $SFR_{MS}$ the star-formation rate of a main-sequence galaxy, and z the corresponding redshift.
Appendix A of \citet{Les18} also shows results using two additional main-sequences: the \citet{Speagle2014} best-fit main-sequence derived by compiling 25 different literary works, and the \citet{Schreiber2015} main-sequence, derived using a stacking analysis of star-forming galaxies, combining UV and FIR data. The \citet{Speagle2014} is given as:
\begin{equation}
\log(\frac{SFR_{MS}}{M_{\odot \mathrm{yr}^{-1}}}) = (0.84 - 0.026t)\log(\frac{M_{*}}{\mathrm{M}_{\odot}}) - (6.51 - 0.11t),
\end{equation}
with $t$ the age of the Universe in Gyr and assuming a Kroupa IMF. The \citet{Schreiber2015} MS is given as:
\begin{equation}
\begin{split}
\log(\frac{SFR_{MS}}{M_{\odot \mathrm{yr}^{-1}}})
    & = m - 0.5 + 1.5\log(1+z) \\
    & - 0.3[max(0, m - 0.36 - 2.5\log(1+z))]^{2},
\end{split}
\end{equation}
where $m = \frac{M_{*}}{10^{9}\mathrm{M}_{\odot}}$.

\subsection{From SFR to UV}\label{Ap:IntrIVConv}
We use the \citet{Ken12} conversions to derive the expected intrinsic emission:
\begin{equation}\label{eq:MS_L}
    \mathrm{SFR_{MS}} = C_{\nu} \nu L_{\nu}.
\end{equation}
The assumption needed to use the conversion in our analysis is that all galaxies in our sample are main-sequence galaxies. We derive the SFR from the different star-formation main-sequences and convert the $SFRs$ to FUV and NUV luminosities. Since the T04 models normalize by the intrinsic magnitudes rather than the intrinsic luminosities, we apply the conversion to magnitudes and follow the same procedure as Section \ref{Sec:Methods}. 

\subsection{Fitting comparison}\label{Ap:IntrUVFit}
\begin{figure}[!ht]
    \centering
    \includegraphics[width=\columnwidth]{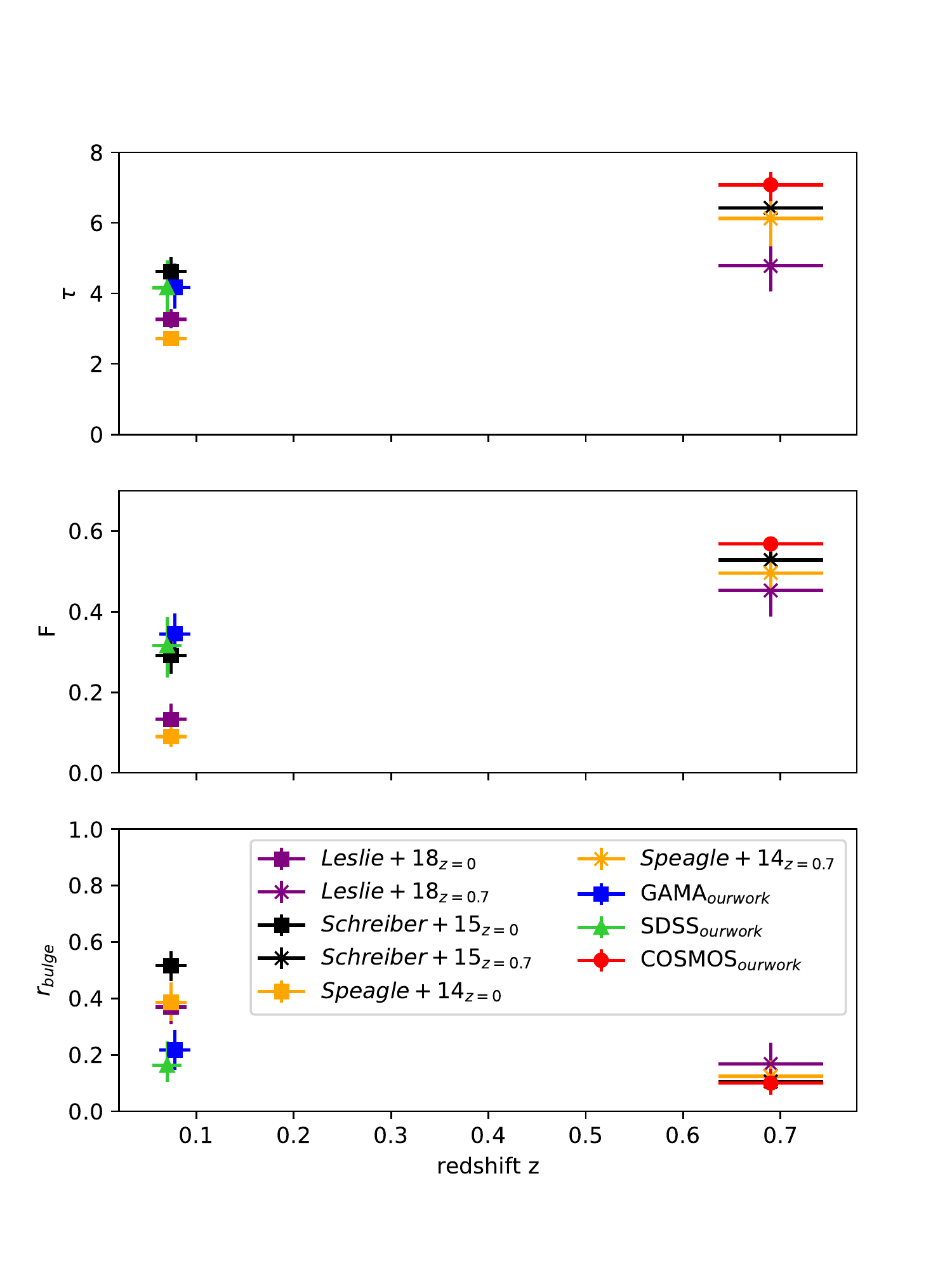}
    \caption{Results of fitting the T04 model for galaxies in the GAMA, SDSS and COSMOS datasets using different assumptions for the intrinsic UV emission. The fitted parameters are the $\tau_{B}^{f}$ (top), $F$ (middle), and $r_{bulge}$ (bottom). The assumptions for intrinsic emission are made using different star-formation main-sequences: \citet{Les18} (purple), \citet{Schreiber2015} (green), and \citet{Speagle2014} (orange), where square markers indicate the sample around $z\approx0.0$ and the crosses for the sample around $z\approx0.7$. We see that in all cases $\tau_{B}^{f}$ and $F$ increase with redshift, with the different assumptions only shifting the value.}
    \label{fig:MS_z_fit}
\end{figure}

\begin{figure}[!ht]
    \centering
    \includegraphics[width=\columnwidth]{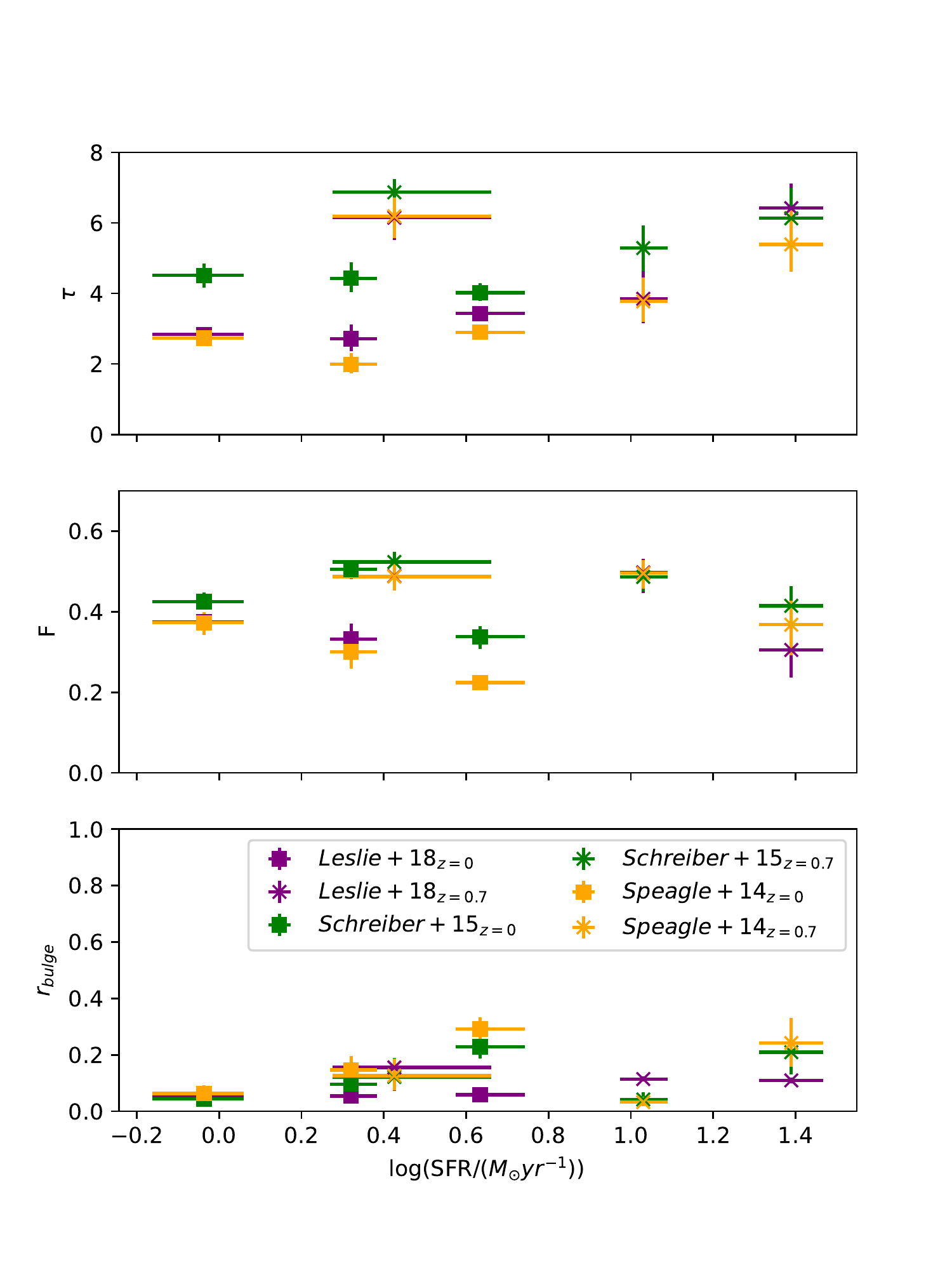}
    \caption{Results of fitting the T04 model for galaxies in the GAMA, SDSS and COSMOS datasets using different assumptions for the intrinsic emission separated in bins of $SFR$. The fitted parameters are the $\tau_{B}^{f}$ (top), $F$ (middle), and $r_{bulge}$ (bottom). The assumptions for intrinsic emission are made using different star-formation main-sequences: \citet{Les18} (purple), \citet{Schreiber2015} (green), and \citet{Speagle2014} (orange), where square markers indicate the sample around $z\approx0.0$ and the crosses for the sample around $z\approx0.7$. We see that none of the normalizations result in a consistent trend between $SFR$ and any of the fitted parameters.}
    \label{fig:MS_SFR_fit}
\end{figure}

We use the same fitting regime as in the main analysis to obtain the best-fit parameters for the $\tau_{B}$, $F$, and bulge fraction. We show the best-fit values from the magnitude-inclination relation of all galaxies in Figure \ref{fig:MS_z_fit} and for galaxies binned by SFR in Figure \ref{fig:MS_SFR_fit}.

Figure \ref{fig:MS_z_fit} shows that the results for the low-z galaxies are consistent across the different intrinsic corrections, whereas the fitting for the $z\sim0.7$ galaxies has different fitted values. Figure \ref{fig:MS_z_fit}, for instance, shows that \citet{Les18} obtains lower $\tau_{B}^{f}$ for $z\approx0.7$ galaxies, where our results in Figure \ref{fig:gal_z} are in line with the higher $\tau_{B}^{f}.$ The trends over redshift remain qualitatively the same, meaning that the star-formation main-sequence has a negligible impact on the redshift evolution.
The main-sequence starts to have more influence on the results when we study the dependence on $SFR$. The trends remain inconsistent, which could mean that either the uncertainty in $SFR$ causes the inconsistency or that the $SFR$ and dust parameters are independent. The inconsistent trends in these results may be an indirect result of us assuming intrinsic emission based on the $SFR$ following \citet{Ken12}. The relation between intrinsic UV emission and $SFR$ have all been estimated using the Starburst99 stellar evolution models \citep{Leitherer1999} assuming a Chabrier IMF, solar abundances, and constant star-formation on 100 Myr timescales. When we bin our galaxies based on a property, we confirm that the remaining properties investigated remain constant between the bins. However, we cannot check how the metallicity or star-formation histories vary between our sample bins. Therefore, we cannot confirm whether the assumptions made for the \citet{Ken12} calibration are valid, and this may add systematic uncertainties to our findings.
The inconsistent results when applying different main-sequence relations could also hint at the variation in the star-forming galaxy selection used to derive the different main-sequence relations. Thus, we can not confirm that the main-sequence combined with the \citet{Ken12} SFR calibration is an appropriate method for deriving the intrinsic UV emission of main-sequence galaxies and adopt the MIR correction for our main analyses.
\end{document}